\documentclass[final]{article}
\usepackage{longtable}
\usepackage{multirow}
\usepackage{booktabs}
\usepackage{longtable}
\usepackage{color}
\usepackage{graphicx}
\usepackage{amsmath}
\usepackage{amsfonts}
\renewcommand{\vec}[1]{\boldsymbol{#1}}

\begin{document}
%%%%% title : short title may not be used but TITLE is required.
% \title{TITLE}
% \title[short title]{TITLE}
\title{Mathematical modeling of tumor-immune interactions: methods, applications, and future perspectives}

%%%%% author(s) :
% single author:
% \author[name in running head]{AUTHOR\corrauth}
% [name in running head] is NOT OPTIONAL, it is a MUST.
% Use \corrauth to indicate the corresponding author.
% Use \email to provide email address of author.
% \footnote and \thanks are not used in the heading section.
% Another acknowlegments/support of grants, state in Acknowledgments section
% \section*{Acknowledgments}
%\author[O.~Author]{Only Author\corrauth}
%\address{School of Mathematical Sciences, Beijing Normal University,Beijing 100875, P.R. China}
%\email{{\tt author@email} (O.~Author)}

% multiple authors:
% Note the use of \affil and \affilnum to link names and addresses.
% The author for correspondence is marked by \corrauth.
% use \emails to provide email addresses of authors
% e.g. below example has 3 authors, first author is also the corresponding
%      author, author 1 and 3 having the same address.
% \author[Z. Zhang et~al.]{Zhengru Zhang\affil{1}\comma\corrauth,
%       Author Chan\affil{2}~and Author Zhao\affil{1}}
% \address{\affilnum{1}\ School of Mathematical Sciences,
%          Beijing Normal University,
%          Beijing 100875, P.R. China. \\
%           \affilnum{2}\ Department of Mathematics,
%           Hong Kong Baptist University, Hong Kong SAR.}
% \emails{{\tt zhang@email} (Z.~Zhang), {\tt chan@email} (A.~Chan),
%          {\tt zhao@email} (A.~Zhao)}
% \footnote and \thanks are not used in the heading section.
% Another acknowlegments/support of grants, state in Acknowledgments section
% \section*{Acknowledgments}
\author{Chenghang Li\thanks{School of Mathematical Sciences, Tiangong University, Tianjin, 300387, China (Email: lichenghang@tiangong.edu.cn)}  \and Jinzhi Lei\thanks{School of Mathematical Sciences, Center for Applied Mathematics, Tiangong University, Tianjin, 300387, China (Email: jzlei@tiangong.edu.cn).}}

%
%same address:
%\author[F. Author and A.~Co-Author,]{First Author and A.~Co-Author\corrauth}
%\address{address of First Author and His Best Friend}
%

\maketitle

%%%%% Begin Abstract %%%%%%%%%%%
\begin{abstract}

Mathematical oncology is a rapidly evolving interdisciplinary field that uses mathematical models to enhance our understanding of cancer dynamics, including tumor growth, metastasis, and treatment response. Tumor-immune interactions play a crucial role in cancer biology, influencing tumor progression and the effectiveness of immunotherapy and targeted treatments. However, studying tumor dynamics in isolation often fails to capture the complex interplay between cancer cells and the immune system, which is critical to disease progression and therapeutic efficacy. Mathematical models that incorporate tumor-immune interactions offer valuable insights into these processes, providing a framework for analyzing immune escape, treatment response, and resistance mechanisms. In this review, we provide an overview of mathematical models that describe tumor-immune dynamics, highlighting their applications in understanding tumor growth, evaluating treatment strategies, and predicting immune responses. We also discuss the strengths and limitations of current modeling approaches and propose future directions for the development of more comprehensive and predictive models of tumor-immune interactions. We aim to offer a comprehensive guide to the state of mathematical modeling in tumor immunology, emphasizing its potential to inform clinical decision-making and improve cancer therapies. 

\end{abstract}
%%%%% end %%%%%%%%%%%

%%%%% AMS/PACs/Keywords %%%%%%%%%%%
%\pac{}


\vspace{0.5cm}
\textbf{MSC Codes:\ \ } 92C42, 92B05, 92B10%The information of the AMS subject classification can be found in http://mathscinet.ams.org/msc/msc2010.html

%%%% maketitle %%%%%

%%%% Start %%%%%%

\section{Introduction}

\begin{figure}[htp!]
	\centering
	\includegraphics[width=13cm]{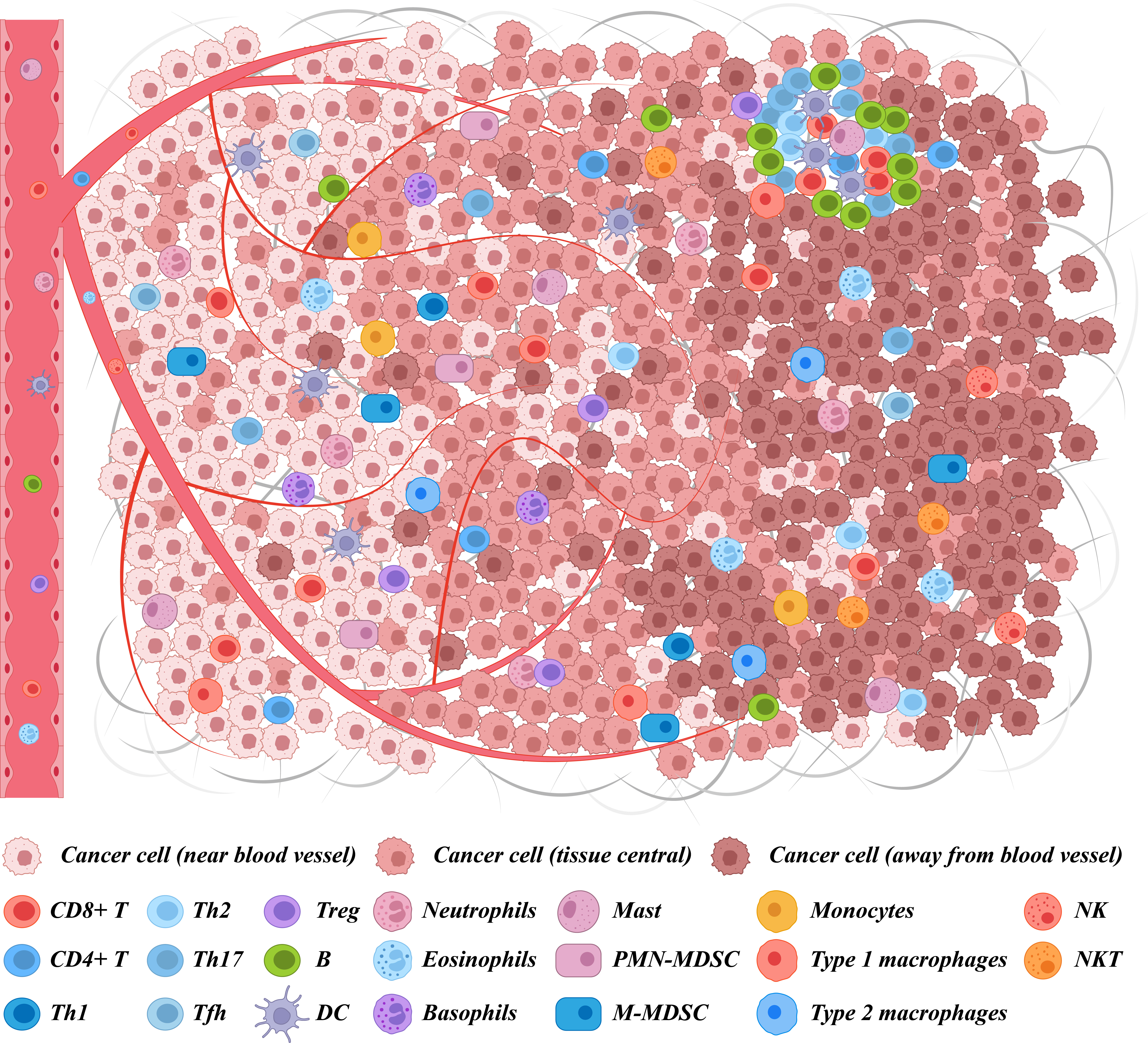}
	\caption{ A global map of the tumor microenvironment.}
	\label{1}
\end{figure}

Cancer, often described as a malignant tumor, represents a complex and dynamic ecosystem \cite{Chen.CancerCommun.2022,Anderson.CurrBiol.2020,Visser.CancerCell.2023}. This ecosystem, known as the tumor microenvironment (TME) (Figure \ref{1}), comprises not only malignant tumor cells capable of rapid proliferation and metastasis but also includes various non-cancerous components such as immune cells, stromal cells, fibroblasts, and vascular endothelial cells \cite{Anderson.CurrBiol.2020,Visser.CancerCell.2023,Gajewski.NatImmunol.2013,Hinshaw.CancerRes.2019}. The TME plays a pivotal role in the processes of tumor growth, progression, metastasis, and drug resistance \cite{Visser.CancerCell.2023,Gajewski.NatImmunol.2013,Hinshaw.CancerRes.2019,Meads.NatRevCancer.2009,Quail.NatMed.2013}. Within this environment, tumors actively shape conditions favorable to their survival and proliferation through mechanisms such as the secretion of cytokines, immune-modulating factors, and the expression of immune checkpoint molecules \cite{Rabinovich.AnnuRevImmunol.2007,Joyce.Science.2015}. Meanwhile, immune cells infiltrate tumor tissue via migration, chemotaxis, and recruitment, influencing tumor development \cite{Joyce.Science.2015,Ozga.Immunity.2021,Nagarsheth.NatRevImmunol.2017}.

Tumor-immune system interactions are marked by a dynamic and complex interplay of mutual promotion, competition, and adaptation \cite{Prendergast.CancerRes.2007,Galon.Immunity.2020}. These interactions not only influence tumor growth, metastasis, and regression but also modulate the immune system's composition, function, and responsiveness \cite{Dunn.NatImmunol.2002,Dunn.AnnuRevImmunol.2004,Dunn.Immunity.2004,Schreiber.Science.2011}. Recent advances in single-cell sequencing and other biotechnological tools have significantly enhanced our understanding of these tumor-immune interactions \cite{Ren.AnnuRevImmunol.2021,Xu.SignalTransductTargetTher.2021,Kashyap.TrendsBiotechnol.2022}. However, the inherent complexity of these interactions poses challenges that experimental techniques alone cannot fully address, necessitating the use of mathematical modeling as a powerful complementary approach to uncover underlying patterns and mechanisms. 

Mathematical models provide a framework for describing and simulating complex biological systems, allowing researchers to abstract and quantify interactions with the tumor-immune landscape \cite{Gatenby.Nature.2003,Anderson.NatRevCancer.2008,Byrne.NatRevCancer.2010,Altrock.NatRevCancer.2015,Rockne.PhysBiol.2019}. These models offer several key advantages in studying tumor-immune dynamics: (1) \textbf{Quantitative Description}: Mathematical models enable the quantitative analysis of tumor-immune interactions through differential equations and algorithms, offering new perspectives on the complex processes underlying these interactions. (2) \textbf{Systematic Analysis}: By modeling tumor-immune interactions as integrated systems, these approaches capture feedback loops and multicomponent interactions, providing insights into the regulation of tumor growth, immune evasion, and immune cell dynamics. (3) \textbf{Multi-Scale Simulation}: Mathematical models can simulate biological processes across multiple scales, from molecular and cellular to tissue levels, facilitating a comprehensive understanding of the dynamic nature of tumor-immune interactions. (4) \textbf{Treatment Predictions}: These models are also valuable tools for predicting the effects of various treatment strategies, aiding in the design of personalized therapies, and supporting clinical decision-making through the simulation of therapeutic outcomes. 

Despite their potential, mathematical models of tumor-immune interactions face significant challenges \cite{Michor.Cell.2015,Konstorum.JRSocInterface.2017,Clarke.NatRevCancer.2020,Butner.NatComputSci.2022}. The complexity of tumor-immune dynamics involves multiple time scales, diverse cellular components, and intricate regulatory networks, requiring an interdisciplinary approach that integrates knowledge from applied mathematics, computational science, tumor immunology, and clinical medicine. 
Additionally, the acquisition and processing of multi-source data are critical yet challenging aspects of model development, necessitating robust data integration and validation methods to ensure model reliability. Finally, interpatient variability in tumor types and immune characteristics adds another layer of complexity, underscoring the need for adaptable modeling approaches that can account for individualized tumor behavior and biomarker variability. 

In this review, we comprehensively analyze the current landscape of mathematical models in tumor immunology, focusing on their methodologies, applications, and impact on understanding tumor dynamics and treatment responses. In Section \ref{sec:2}, we discuss key immunological mechanisms and recent research. Section \ref{sec:3} delves into modeling approaches and regulatory networks of tumor-immune interactions. In Section \ref{sec:4}, we explore the application of these models to various cancer treatment strategies. Finally, we discuss current limitations and propose future directions for the advancement of mathematical models in the study of tumor-immune systems.

\section{Biological background of immunological mechanisms}
\label{sec:2}
\subsection{Hallmarks of cancer}

\begin{figure}[htp!]
	\centering
	\includegraphics[width=13cm]{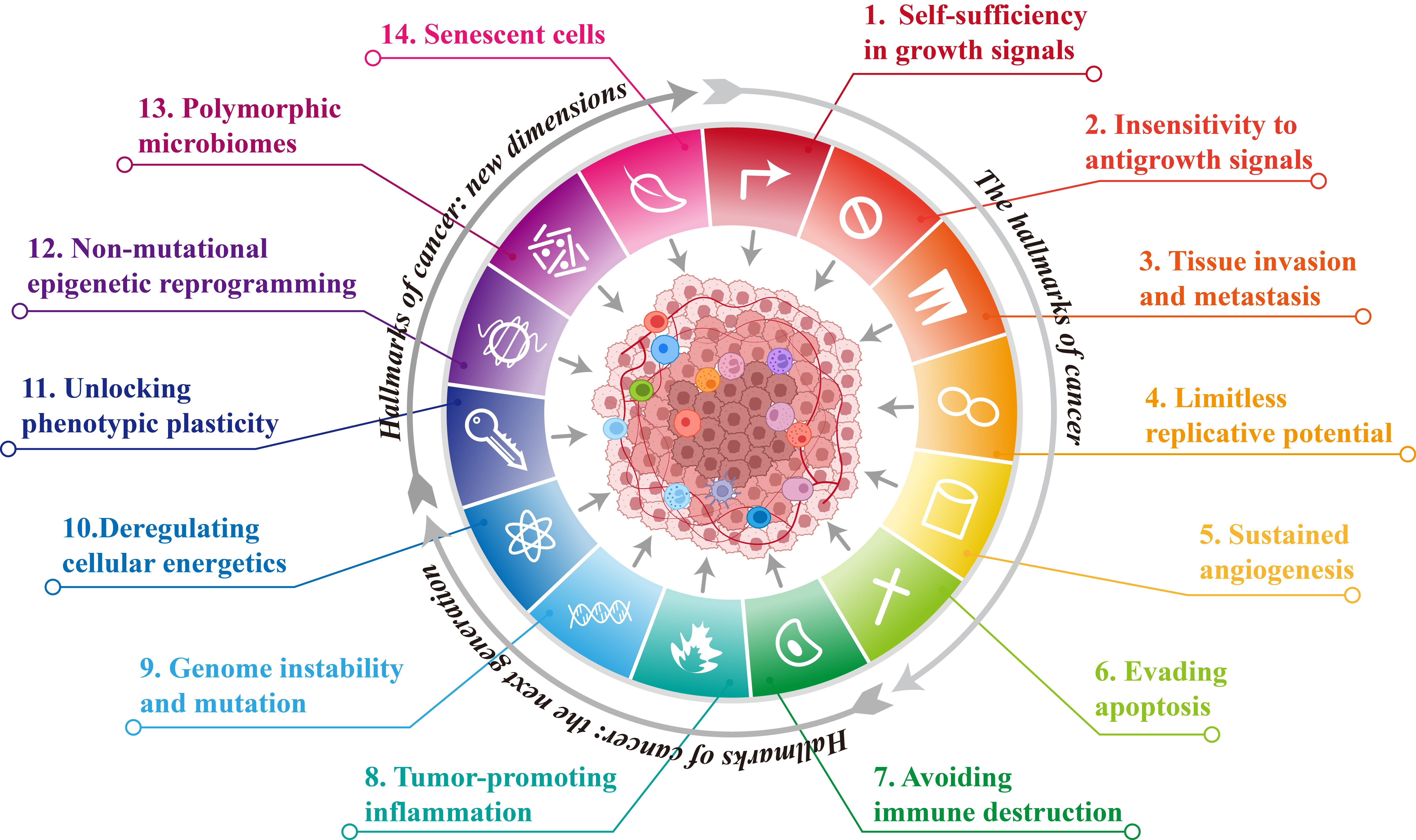}
	\caption{ The hallmarks of cancer\cite{Hanahan.Cell.2000,Hanahan.Cell.2011,Hanahan.CancerDiscov.2022}.}
	\label{2-1}
\end{figure}

The hallmarks of cancer define the fundamental characteristics that drive cancer development and progression (Figure \ref{2-1}) \cite{Hanahan.Cell.2000,Hanahan.Cell.2011,Hanahan.CancerDiscov.2022}. In 2000, Douglas Hanahan and Robert A. Weinberg identified six original hallmarks: self-sufficiency in growth signals, insensitivity to anti-growth signals, evasion of apoptosis, limitless replicative potential, sustained angiogenesis, and tissue invasion and metastasis \cite{Hanahan.Cell.2000}. In 2011, four additional hallmarks were introduced: avoiding immune destruction, tumor-promoting inflammation, genome instability and mutation, and deregulating cellular energetics \cite{Hanahan.Cell.2011}. By 2022, four more hallmarks were recognized: unlocking phenotypic plasticity, non-mutational epigenetic reprogramming, polymorphic microbiomes, and the influence of senescent cells \cite{Hanahan.CancerDiscov.2022}. 

These hallmarks provide a comprehensive framework for understanding the progression and evolution of cancer. Recently, Joshua Adam Bull and Helen Mary Byrne proposed the ``hallmakers'' of mathematical oncology, which define how mathematical models can help elucidate the complex processes of tumor initiation and progression \cite{Bull.PIEEE.2022}. The integration of mathematics, oncology, and immunology is driving new advances in cancer research.

The hallmarks of cancer emphasize the unique distinctions between tumor cells and normal cells, many of which are closely linked to the immune system. For example, immunosuppressive cells and tumor-associated fibroblasts contribute to the formation of pre-metastatic niches, facilitating tumor invasion and metastasis \cite{Hanahan.Cell.2000,Visser.CancerCell.2023}. Tumor-promoting inflammation is driven by the infiltration of inflammatory cells and cytokines, which significantly impact tumor-immune interactions \cite{Grivennikov.Cell.2010,Hanahan.Cell.2011}. The interplay between polymorphic microbiomes, tumors, and the immune system forms a cancer-immune-microbiome axis that influences tumor progression and therapeutic response \cite{Yang.SignalTransductTargetTher.2023,Hanahan.CancerDiscov.2022}. Mathematical modeling of tumor-immune interactions is central to mathematical oncology, providing quantitative insights into the dynamics of cancer development and progression.

\subsection{Immune cells}

\begin{figure}[htp!]
	\centering
	\includegraphics[width=13cm]{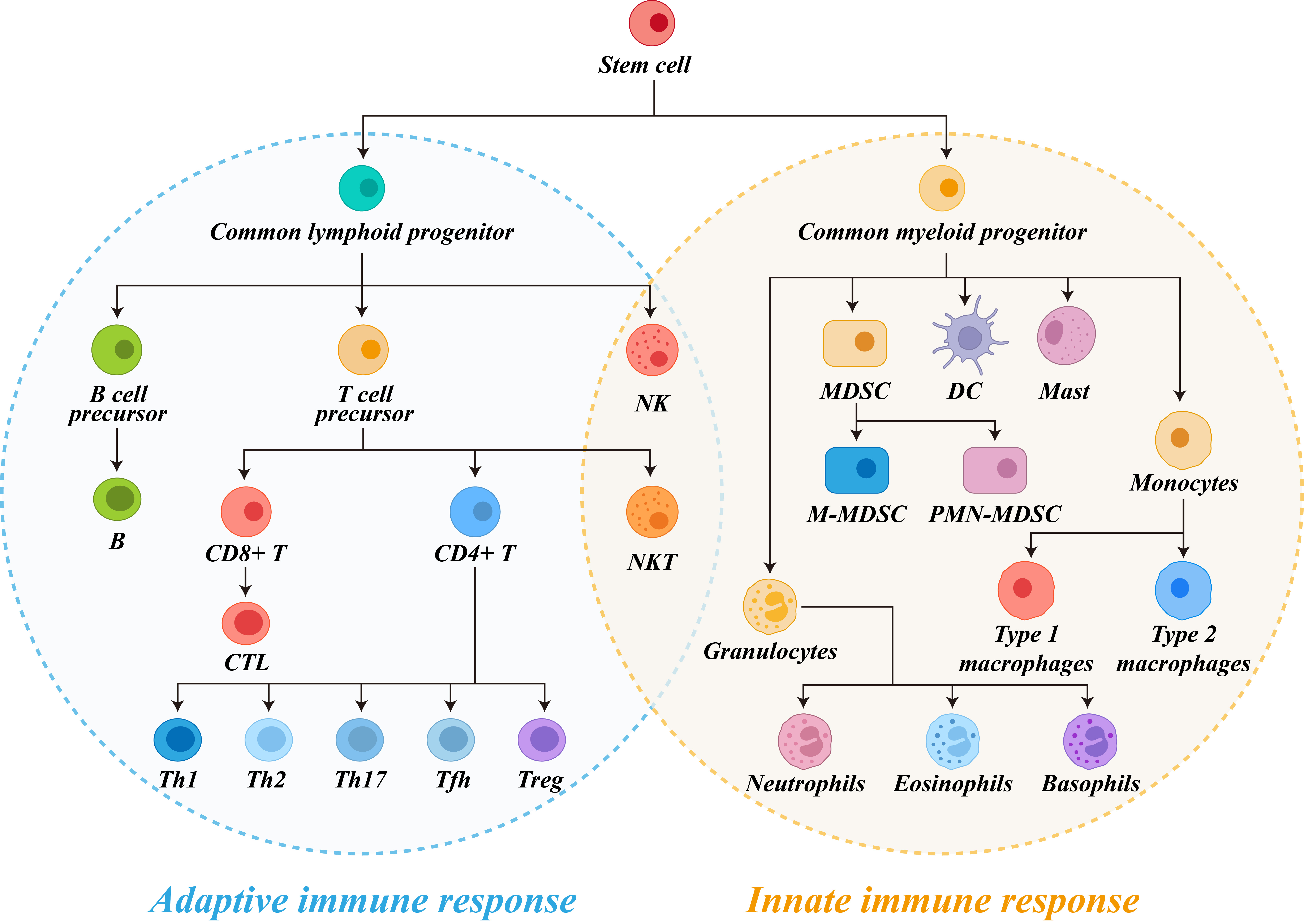}
	\caption{Immune cell lineage.}
	\label{2-2}
\end{figure}

The immune system is a complex and highly coordinated defense network that safeguards the body against infections, diseases, and abnormal cells, including cancer \cite{Parkin.Lancet.2001,Delves.NEnglJMed.2000}. It consists of various cells, tissues, and organs that collaborate to identify and eliminate harmful pathogens as well as damaged or malignant cells. The system's core consists of immune cells, primarily lymphocytes and myeloid cells \cite{Gajewski.NatImmunol.2013,Parkin.Lancet.2001,Delves.NEnglJMed.2000} (Figure \ref{2-2}). Lymphocytes, which originate from lymphoid organs, are key players in the adaptive immune response against tumors. They are further classified based on their distinct functions and surface markers into T lymphocytes, B lymphocytes, natural killer (NK) cells, and natural killer T (NKT) cells. Myeloid cells, a crucial component of innate immunity, include granulocytes, myeloid-derived suppressor cells (MDSCs), dendritic cells (DCs), monocytes, and macrophages, all of which play vital roles in the body's immediate response to threats.

T lymphocytes are primarily involved in cellular immunity, recognizing and binding to specific antigens to initiate immune responses \cite{Wculek.NatRevImmunol.2020}. Due to their robust tumor-killing abilities, T cells have become a central focus in contemporary tumor immunology research. Naive T cells can differentiate into effector T cells under the influence of various cytokines, resulting in distinct subtypes such as helper T cells (Th), regulatory T cells (Treg), and cytotoxic T lymphocytes (CTL) \cite{Zhu.Blood.2008,Zhou.Immunity.2009,OShea.Science.2010}. 

Th cells are a subset of T lymphocytes that play a crucial role in regulating and coordinating the immune response by aiding in the activation and function of other immune cells through the secretion of specific cytokines. They are further divided into subtypes, including Th1, Th2, Th17, and follicular helper T cells (Tfh), based on their transcription factors and cytokine profiles:
\begin{itemize}
\item Th1 cells differentiate from naive CD4+ T cells under the influence of IL-12 and primarily secrete IL-2, IFN-$\gamma$, and TNF-$\alpha$ \cite{Zhu.Blood.2008,Zhou.Immunity.2009,OShea.Science.2010,Liew.NatRevImmunol.2002}. Th1 cells enhance CTL expansion through IL-2 and exert direct anti-tumor effects by secreting IFN-$\gamma$ and TNF-$\alpha$, which contributed to the killing of cancer cells.
\item Th2 cells can promote tumor growth by secreting cytokines such as IL-4, IL-5, and IL-10 \cite{Zhu.Blood.2008,Zhou.Immunity.2009,OShea.Science.2010,Liew.NatRevImmunol.2002}. The differentiation of naive CD4+ T cells into the Th2 subtype is driven by IL-4, produced by granulocytes, mast cells, and already differentiated Th2 cells.  
\item Th17 cells are a subpopulation of effector CD4+ T cells known for secreting IL-17. Recent research indicates that TGF-$\beta$, IL-6, and IL-23 promote the differentiation of naive CD4+ T cells into the Th17 cells, whereas IFN-$\gamma$ and IL-4 inhibit this process \cite{Zhu.Blood.2008,Stadhouders.JAutoimmun.2018}. 
\item Tfh cells are primarily located in peripheral immune organs and play a crucial role in the formation of germinal centers \cite{Crotty.AnnuRevImmunol.2011}.
\end{itemize}

Tregs are a subset of T cells with potent immunosuppressive functions, known for secreting high levels of immunosuppressive cytokines such as IL-10 and TGF-$\beta$ \cite{Beyer.Blood.2006,Zou.NatRevImmunol.2006}. Tregs can be classified into two main types: naturally occurring Treg (nTreg) derived from the thymus, and induced adaptive Treg (iTreg). Within the TME, Tregs predominantly refer to iTreg, which facilitates tumor immune evasion, suppresses anti-tumor immune responses, and contributes to the establishment of an immunosuppressive microenvironment.

CTLs are derived from naive CD8+ T cells and are central to the anti-tumor immune response. CTLs specifically recognize cancer cells through the interaction of their T-cell receptors (TCRs) with major histocompatibility complex (MHC) expressed on the surface of cancer cells \cite{Wculek.NatRevImmunol.2020}. CTLs directly induce tumor cell death via the FasL-Fas signaling pathway and can also trigger apoptosis indirectly by secreting granzymes and perforin \cite{Barry.NatRevImmunol.2002}. However, recent studies have shown that intratumoral CTLs often display an exhausted phenotype, marked by impaired immune function \cite{Wherry.NatImmunol.2011,McLane.AnnuRevImmunol.2019}. Addressing CTL exhaustion presents a significant therapeutic opportunity in cancer treatment.

B cells primarily contribute to humoral immunity. Within the tumor-immune system, plasma cells derived from B cells secrete antibodies that recognize and bind to tumor antigens, facilitating the immune system's ability to target and eliminate cancer cells \cite{Sharonov.NatRevImmunol.2020}. Additionally, B cells act as antigen-presenting cells (APCs), presenting tumor antigens to other immune cells and thereby initiating immune responses. Recent studies have shown that B cells support the maintenance of secondary lymphoid organ structures and promote the formation of intratumoral tertiary lymphoid structures (TLS) \cite{Schumacher.Science.2022,Fridman.Immunity.2023}. TLS are clusters of immune cells that develop in non-lymphoid tissues and are typically found in chronically inflamed areas of cancers. Their presence is often associated with better survival outcomes for patients \cite{Schumacher.Science.2022,Fridman.Immunity.2023}. 

NK cells are the archetypal innate immune cells, capable of recognizing and destroying tumor cells in a non-specific manner. NK cells eliminate cancer cells by releasing cytolytic mediators such as perforin and granzyme \cite{Chiossone.NatRevImmunol.2018,Fridman.Immunity.2023}. Another typical function of NK cells is their ability to kill tumor cells by CD16 receptor-mediated antibody-dependent cell-mediated cytotoxicity (ADCC) \cite{Chiossone.NatRevImmunol.2018,Fridman.Immunity.2023}. Although they are traditionally considered part of the innate immune system, some NK cells display adaptive-like traits, including clone specificity and memory. Additionally, activated NK cells can secrete a range of cytokines and chemokines, further regulating the immune response \cite{Chiossone.NatRevImmunol.2018,Fridman.Immunity.2023}.

DC cells, as the most potent professional APCs, play a crucial role in mediating innate immune responses and inducing adaptive immunity \cite{Wculek.NatRevImmunol.2020,Eisenbarth.NatRevImmunol.2019}. They are central to initiating, regulating, and sustaining anti-tumor immune responses. Immature DCs efficiently capture, process, and present tumor-associated antigens (TAAs) released by cancer cells. Once activated, DCs upregulate MHC molecules, which present antigens to T cell receptors (TCRs) on naive T cells, providing the first signal required for T cell activation. Simultaneously, DCs deliver the second activation signal through costimulatory molecules. Moreover, activated DCs secrete chemokines that promote T cell recruitment and cytokines such as IL-12, which drive the differentiation of Th1 and CTLs, providing the third signal for effective immune responses \cite{Wculek.NatRevImmunol.2020,Eisenbarth.NatRevImmunol.2019}. Together, these mechanisms orchestrate a robust anti-tumor immune response.

Tumor-associated macrophages (TAMs) are classified into two main types, M1 and M2, based on their functional roles and activation states within the TME \cite{Biswas.NatImmunol.2010}. The differentiation of macrophages into these phenotypes is known as polarization. M1 macrophages are generally considered anti-tumor, as they secrete pro-inflammatory cytokines like IL-12, IFN-$\gamma$, and TNF-$\alpha$ \cite{Biswas.NatImmunol.2010,Galli.NatImmunol.2011,Noy.Immunity.2014}. On the other hand, M2 macrophages are linked to tumor progression, producing anti-inflammatory cytokines such as IL-4, IL-6, and CCL7 \cite{Biswas.NatImmunol.2010,Galli.NatImmunol.2011,Noy.Immunity.2014}. Macrophage polarization demonstrates significant plasticity: factors like M-CSF and TGF-$\beta$ promote the transition of TAMs from the M1 to the M2 phenotype, while TNF-$\alpha$ and IL-12 drive the reverse transition from M2 to M1 \cite{Biswas.NatImmunol.2010,Galli.NatImmunol.2011,Noy.Immunity.2014,Basak.FrontImmunol.2023}. Understanding this bidirectional polarization is crucial for unraveling the complexities of tumor-immune interactions and their regulatory mechanisms.

Neutrophils, the most abundant granulocytes, serve as the body's first line of defense against infections. Within the TME, neutrophils can adopt distinct phenotypes \cite{Giese.Blood.2019,Shaul.NatRevClinOncol.2019,Sansores-Espana.IntJMolSci.2022}. N1 neutrophils exhibit anti-tumor properties, while N2 neutrophils promote tumor progression \cite{Giese.Blood.2019,Shaul.NatRevClinOncol.2019,Sansores-Espana.IntJMolSci.2022}. TGF-$\beta$ is a key driver of neutrophil polarization towards the tumor-promoting N2 phenotype \cite{Sansores-Espana.IntJMolSci.2022}, while type I interferons facilitate polarization towards the anti-tumor N1 phenotype \cite{Sansores-Espana.IntJMolSci.2022}. N1 neutrophils combat tumors by releasing reactive oxygen species (ROS) to kill cancer cells, and by promoting T cell activation and macrophage recruitment \cite{Giese.Blood.2019,Shaul.NatRevClinOncol.2019,Sansores-Espana.IntJMolSci.2022}. In contrast, N2 neutrophils contribute to tumor growth through angiogenesis, suppression of NK cell activity, and recruitment of Tregs \cite{Giese.Blood.2019,Shaul.NatRevClinOncol.2019,Sansores-Espana.IntJMolSci.2022}.

MDSCs are a heterogeneous group of myeloid cells with strong immunosuppressive capabilities \cite{Hegde.Immunity.2021}. They are categorized into two major subtypes: polymorphonuclear MDSCs (PMN-MDSCs), which resemble neutrophils, and monocytic-MDSCs (M-MDSCs), which are more akin to monocytes \cite{Groth.BrJCancer.2019}. In the TME, MDSCs exert potent pro-tumor and immunosuppressive effects through various mechanisms, including the introduction of immunosuppressive cells, inhibition of lymphocyte trafficking, production of reactive oxygen species, and expression of immune checkpoint molecules \cite{Hegde.Immunity.2021,Groth.BrJCancer.2019,Li.SignalTransductTargetTher.2021}. Emerging evidence suggests that MDSCs are a hallmark of malignant tumors and represent a promising target of cancer immunotherapy \cite{Hegde.Immunity.2021,Groth.BrJCancer.2019,Li.SignalTransductTargetTher.2021}.

\subsection{Cancer immunoediting}

\begin{figure}[htp!]
	\centering
	\includegraphics[width=13cm]{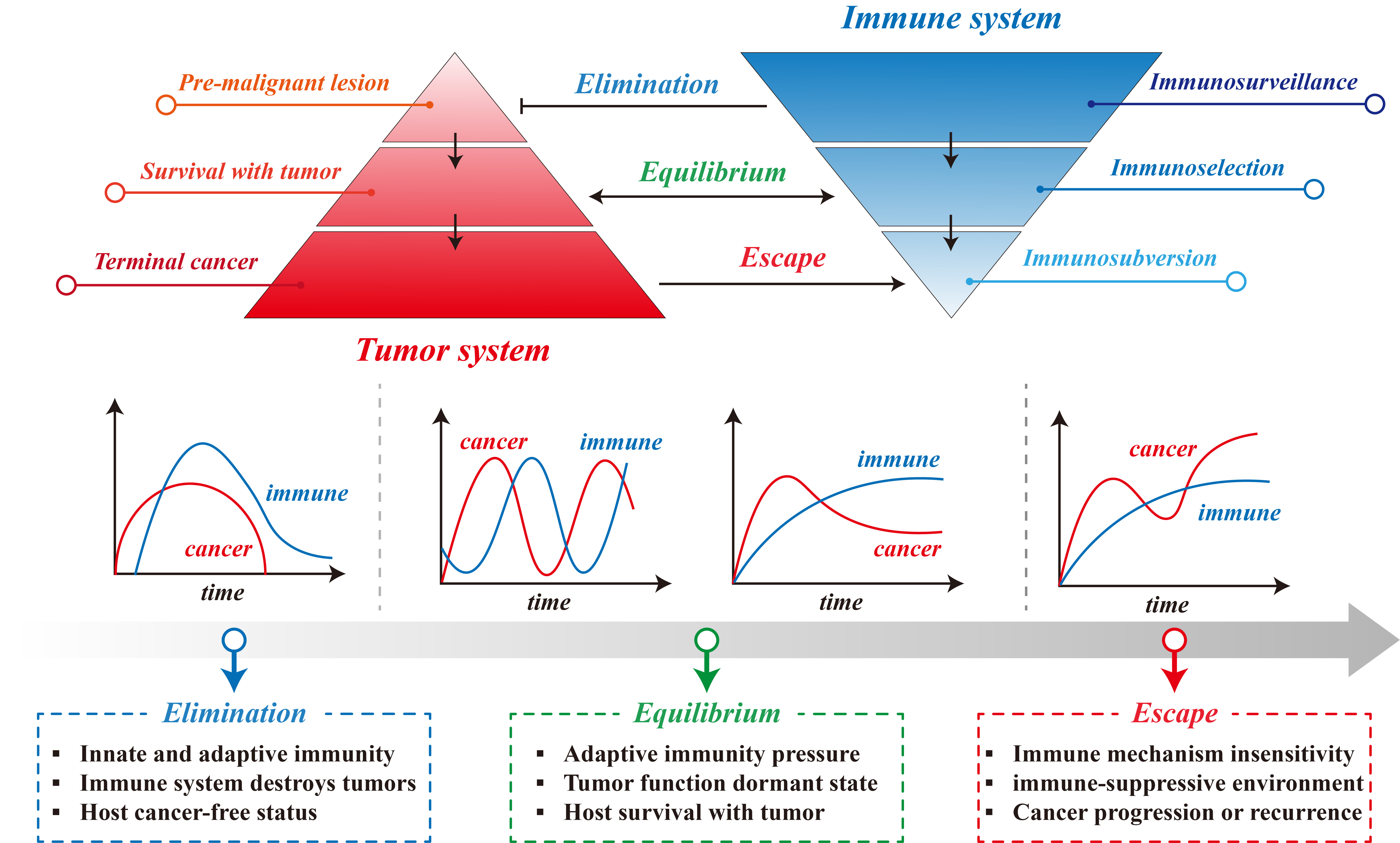}
	\caption{Mechanistic framework and dynamic perspectives on cancer immunoediting \cite{Zitvogel.NatRevImmunol.2006,Kareva.FrontImmunol.2021,Gubin.ClinCancerRes.2022}.}
	\label{2-3}
\end{figure}

Cancer Immunoediting describes the dynamic interplay between tumors and the immune system, evolving across three distinct phases: elimination, equilibrium, and escape (Figure \ref{2-3}) \cite{Dunn.NatImmunol.2002,Dunn.AnnuRevImmunol.2004,Dunn.Immunity.2004,Schreiber.Science.2011}. These phases capture the dynamic struggle between tumor growth and immune surveillance, highlighting the interactions between the immune system and cancer progression.

The elimination phase marks the onset of immune surveillance, where the immune system identifies and attacks developing tumors (Figure \ref{2-3}). DCs detect TAAs released by tumor cells and present them to T lymphocytes, initiating an immune response \cite{Wculek.NatRevImmunol.2020}. Upon antigens recognition, naive T cells differentiate into effector T cells, which target and destroy tumor cells by engaging specific antigens on the tumor surface. In addition, innate immune cells such as NK cells contribute to this phase by directly identifying and eliminating cancer cells using their inherent cytotoxic abilities \cite{Gubin.ClinCancerRes.2022}. This phase is characterized by the coordinated actions of both innate and adaptive immune systems, aiming to eliminate tumor cells at an early stage. While successful completion of this phase can result in the clearance of tumors, factors such as tumor heterogeneity, the complexity of the TME, and immune system limitations often allow for the survival of residual cancer cells \cite{Zitvogel.NatRevImmunol.2006,Kareva.FrontImmunol.2021,Gubin.ClinCancerRes.2022}.

The equilibrium phase is a critical stage in cancer immunoediting marking a prolonged standoff between the tumor and immune system (Figure \ref{2-3}). This phase is characterized by a sustained balance, where tumor cells enter a dormant state to evade immune detection and destruction \cite{Zitvogel.NatRevImmunol.2006,Kareva.FrontImmunol.2021,Gubin.ClinCancerRes.2022}. During equilibrium, tumors continue to evolve, accumulating mutations that promote immune escape and modulate tumor antigen expression \cite{Aguirre-Ghiso.NatRevCancer.2007,Manjili.CancerRes.2017,Santos-de-Frutos.CommunBiol.2021}. Although the immune system persists in eliminating detectable tumor cells, only the most immunogenic subsets are cleared. If this phase is prolonged, tumors may accumulate enough genetic alterations to evade immune control, setting the stage for eventual immune escape and recurrence \cite{Zitvogel.NatRevImmunol.2006,Kareva.FrontImmunol.2021,Gubin.ClinCancerRes.2022}. Under this continuous immune pressure, the tumor evolves through mutation and selection, progressively developing traits that enable immune evasion. 

The escape phase is the final stage of cancer immunoediting (Figure \ref{2-3}). At this point, the tumor gains the ability to evade immune destruction, leading to clinical progression and malignancy. Tumor immune escape is driven by two main mechanisms \cite{Zitvogel.NatRevImmunol.2006,Kareva.FrontImmunol.2021,Gubin.ClinCancerRes.2022}. First, tumors reduce their immunogenicity by downregulating antigen expression, allowing them to slip past immune surveillance. Second, tumors enhance immune suppression by upregulating immune checkpoints, which induces T cell apoptosis or impairs their function, weakening immune attacks. Tumor cells also secrete cytokines and chemokines to limit lymphocyte infiltration into the TME, while promoting the recruitment of MDSCs and Tregs. This creates an immune-privileged niche that supports tumor growth and survival.

\subsection{Cancer-immunity cycle}

\begin{figure}[htp]
	\centering
	\includegraphics[width=13cm]{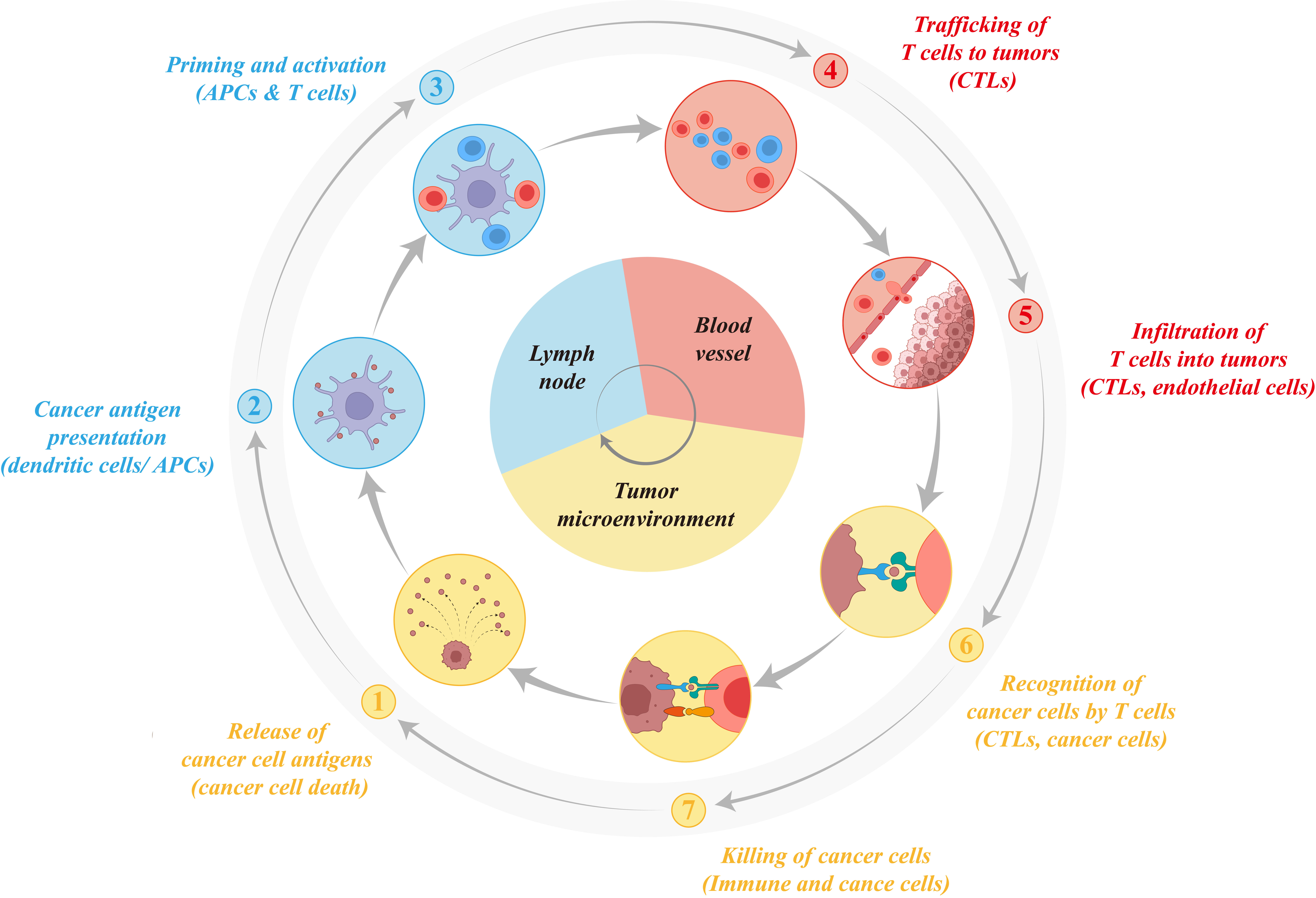}
	\caption{ Cancer-immunity cycle\cite{Chen.Immunity.2013}. }
	\label{2-4}
\end{figure}

The cancer-immunity cycle is a mechanistic model that outlines the sequential events between tumors and the immune system, providing a framework for understanding tumor immunology \cite{Chen.Immunity.2013,Mellman.Immunity.2023}. This cycle consists of seven key steps, each contributing to the initiation and amplification of the immune response against cancer (Figure \ref{2-4}): 
\begin{enumerate}
\item[(1)] \textbf{Release of cancer antigens:} Genetic alterations in cancer cells lead to the production of TAAs, which are immunogenic proteins specific to the tumor \cite{Lewis.IntRevImmunol.2003}. As tumors grow and undergo apoptosis, these antigens are released into the TME, serving as signals to initiate a tumor-specific immune response (Step 1 in Figure \ref{2-4}). 
\item[(2)] \textbf{Cancer antigen presentation:} APCs capture and process TAAs, displaying them on their surface via MHC molecules. These APCs then travel through the lymphatic system to tumor-draining lymph nodes, where antigen presentation occurs, initiating an immune response (Step 2 in Figure \ref{2-4}) \cite{Wculek.NatRevImmunol.2020}. This step is crucial for triggering a T-cell response against the tumor. 
\item[(3)] \textbf{Priming and activation of T cells:} In the tumor-draining lymph nodes, naive T cells recognize the peptide-MHC complexes on APCs through their TCRs. This recognition activates the T cells, causing them to differentiate into effector T cells capable of targeting tumor cells (Step 3 in Figure \ref{2-4}) \cite{Wculek.NatRevImmunol.2020}. The priming and activation of T cells are critical to the immune system's ability to fight the tumor. 
\item[(4)] \textbf{Trafficking of T cells to tumors:} Once activated, effector T cells exit the lymph nodes and travel the bloodstream towards the tumor site, guided by various chemotactic signals (Step 4 in Figure \ref{2-4}) \cite{Visser.CancerCell.2023,Quail.NatMed.2013}. 
\item[(5)] \textbf{Infiltration of T cells into tumors:} Effector T cells infiltrate the tumor tissue in response to chemokine signals, dispersing throughout the TME to locate tumor cells (Step 5 in Figure \ref{2-4}) \cite{Joyce.Science.2015,Ozga.Immunity.2021,Nagarsheth.NatRevImmunol.2017}. 
\item[(6)] \textbf{Recognition of cancer cells by T cells:} Effector T cells identify tumor cells by recognizing specific antigens on their surface via their TCRs (Step 6 in Figure \ref{2-4}) \cite{Gajewski.NatImmunol.2013}. This recognition step is essential for the immune system to selectively target and destroy cancer cells.  
\item[(7)] \textbf{Killing of cancer cells:} After recognizing the tumor cells, T cells release cytotoxic molecules such as granzyme and perforin, which induce apoptosis in the tumor cells (Step 7 in Figure \ref{2-4}). The death of tumor cells releases additional antigens, which continue to fuel the cancer-immunity cycle, creating a feedback loop (Step 1 in Figure \ref{2-4})\cite{Chen.Immunity.2013,Mellman.Immunity.2023}.
\end{enumerate}

\subsection{Cancer immunotype}

\begin{figure}[htp!]
	\centering
	\includegraphics[width=13cm]{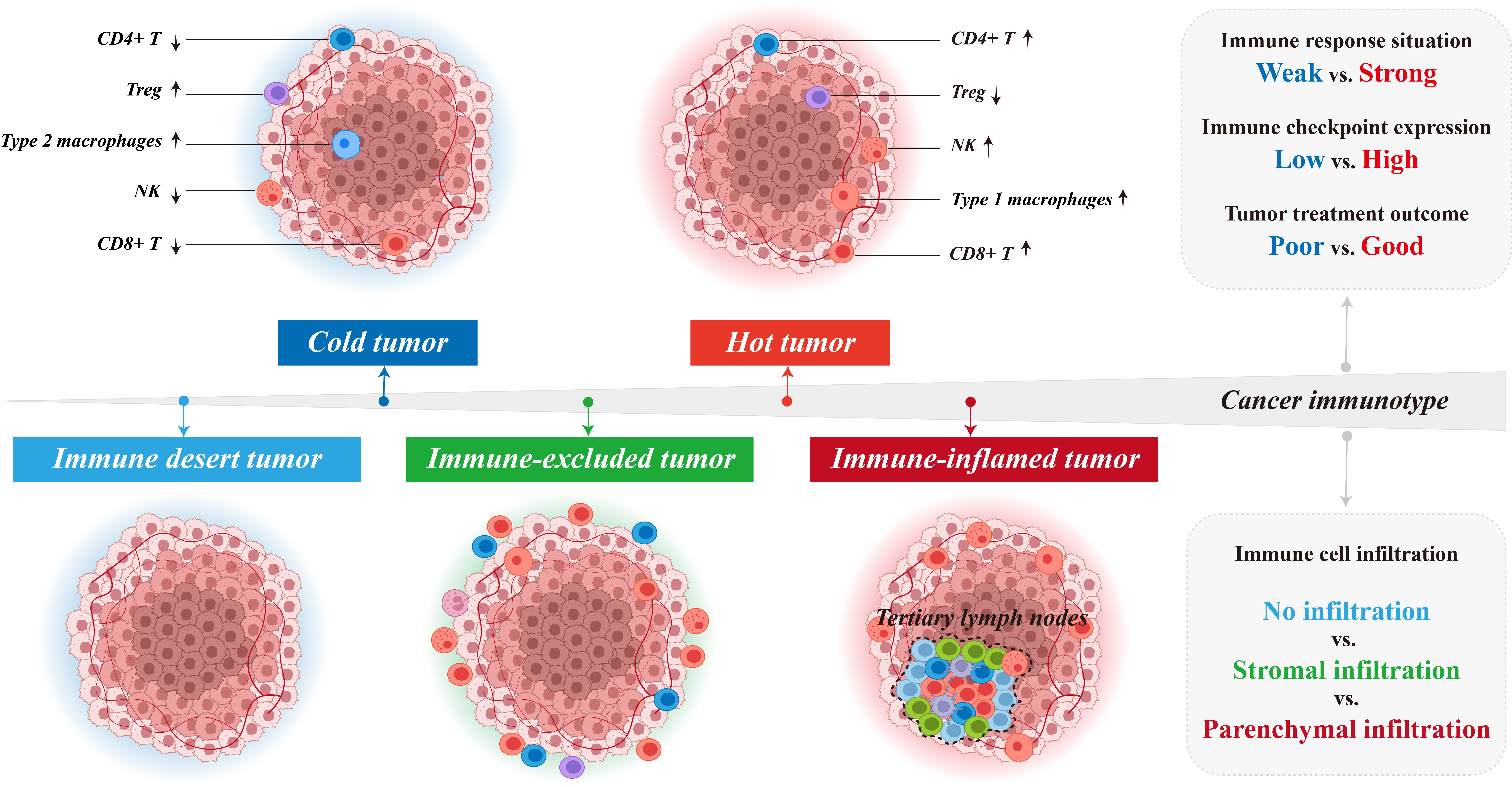}
	\caption{Cancer immunotype \cite{Chen.Nature.2017}.}
	\label{2-5}
\end{figure}

The term ``cancer immunotype'' refers to the distinct patterns of interaction between tumors and the immune system \cite{Nagarsheth.NatRevImmunol.2017,Duan.TrendsCancer.2020}. One of the most common ways to classify cancer immunotypes is by distinguishing between ``cold'' and ``hot'' tumors (Figure \ref{2-5}) \cite{Duan.TrendsCancer.2020,Liu.Theranostics.2021,Zhang.TrendsImmunol.2022}. Cold tumors are characterized by weak or absent immune responses, with three defining features: (1) minimal immune cell infiltration, (2) low expression of immune checkpoint molecules, and (3) poor response to treatment. In contrast, hot tumors exhibit strong immune activity, with high levels of immune cell infiltration and immune checkpoint expression. These tumors generate a robust anti-tumor immune response, often leading to more favorable treatment outcomes. The primary difference between cold and hot tumors lies in the degree of immune system engagement. 

Based on the biological mechanisms of the cancer-immunity cycle, tumors can also be classified into three categories: immune-desert, immune-excluded tumors, and immune-inflamed (Figure \ref{2-5}) \cite{Chen.Nature.2017,Anandappa.CancerDiscov.2020,Mellman.Immunity.2023,Khosravi.CancerCommunLond.2024}. Immune desert tumors lack immune cell infiltration in the TME, resulting in minimal response. Immune-excluded tumors display immune cells that surround the tumor but fail to penetrate its interior, leading to ineffective immune surveillance and action. Immune-inflamed tumors feature substantial immune cell infiltration, particularly T cells, which are crucial for anti-tumor responses. These tumors are associated with elevated IFN-$\gamma$ signaling and high tumor mutational burden, both of which enhance immune activity against the tumor. Additionally, immune-inflamed tumors often develop TLS within the TME, which are linked to better clinical outcomes for patients \cite{Schumacher.Science.2022,Fridman.Immunity.2023}.

\section{Mathematical formulations of tumor-immune interactions}
\label{sec:3}

\begin{figure}[htp!]
	\centering
	\includegraphics[width=13cm]{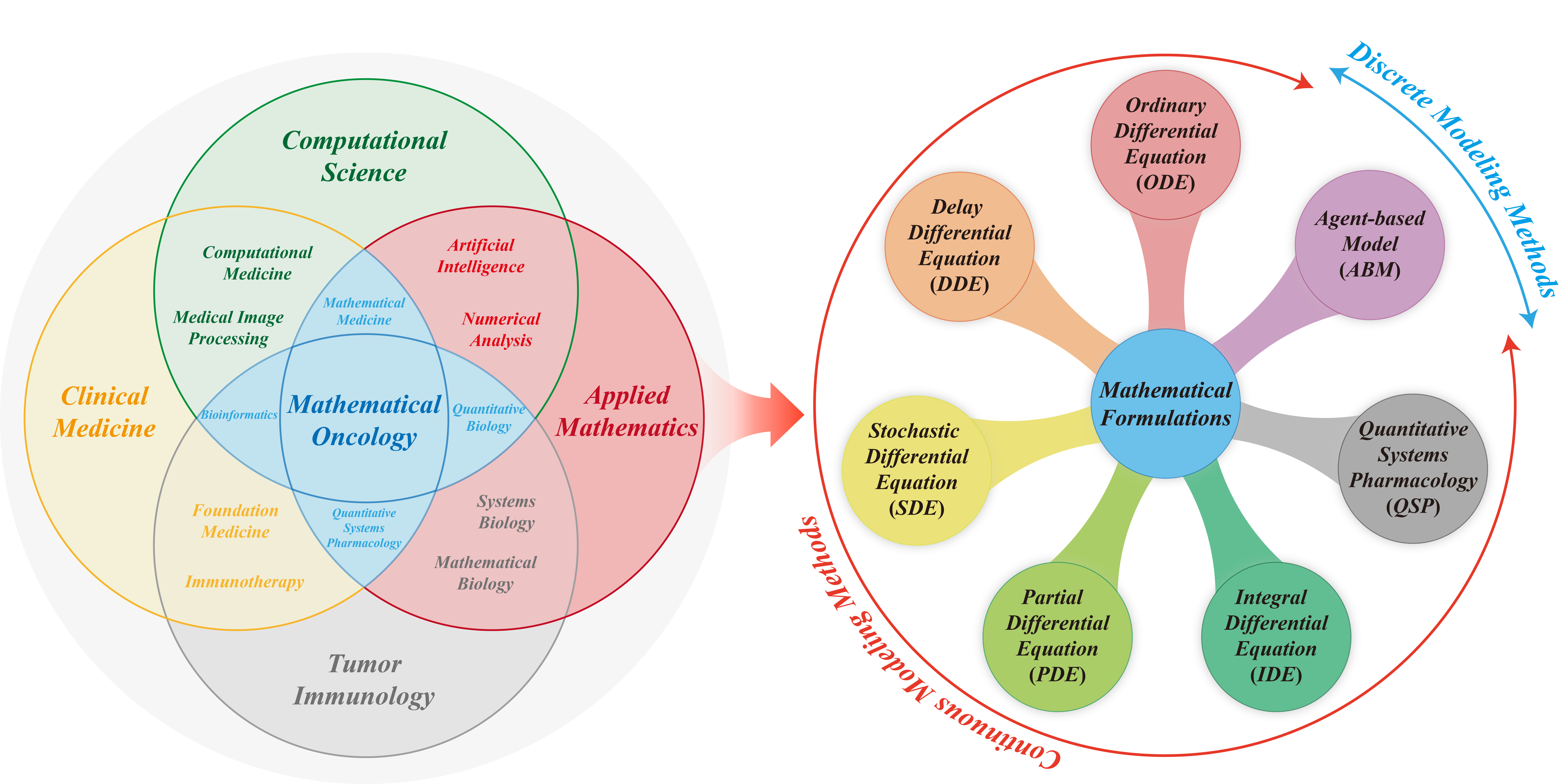}
	\caption{Venn diagrams and related professional keywords of interdisciplinary and intersectional research methods in the field of mathematical oncology.}
	\label{3-1}
\end{figure}

Mathematical Oncology is an emerging interdisciplinary field (Figure \ref{3-1}) that leverages foundational knowledge in tumor immunology and real clinical data to explore cancer dynamics. By applying mathematical models and computational methods, it investigates key aspects of cancer such as tumor evolution, metastasis, drug resistance, prognosis prediction, and optimized treatment strategies \cite{Gatenby.Nature.2003,Anderson.NatRevCancer.2008,Byrne.NatRevCancer.2010,Altrock.NatRevCancer.2015,Rockne.PhysBiol.2019}. This approach provides valuable insights into cancer behavior, helping to refine therapeutic approaches and enhance patient outcomes.

Mathematical models of tumor-immune interactions offer powerful tools and analytical frameworks for exploring key dynamics in tumor-immune systems \cite{Eftimie.BullMathBiol.2011,Arabameri.MathBiosci.2018,Mahlbacher.JTheorBiol.2019}. In this review, we present two primary categories of modeling approaches for mathematically representing these interactions. The first category is equation-based models (EBMs), which use differential equations to capture the temporal and spatial dynamics of genes, cells, and molecules. These models are grounded in principles such as mass action laws, enzyme reaction kinetics, and fluid dynamics. EBMs, which are typically continuous models, include various formulations: ordinary differential equations (ODEs), delayed differential equations (DDEs), stochastic differential equations (SDEs), partial differential equations (PDEs), integral differential equations (IDEs), and quantitative systems pharmacology (QSP). The second category is rule-based models (RBMs), also known as agent-based models (ABMs). These models describe system dynamics by simulating interactions between individual entities, such as protein molecules or cells, with rules derived from experimental data and biological mechanisms. ABMs are generally discrete models. While continuous models focus on the macroscopic interactions between tumors and the immune system, discrete models emphasize the stochasticity and uncertainty present at the microscopic level.

\subsection{Ordinary differential equation (ODE) model}

The ODE model is a fundamental mathematical tool for describing tumor cell interactions with the immune system, providing a strong framework for analyzing tumor dynamics over time. By applying ODEs, researchers can thoroughly explore how tumors interact with various immune system components, such as immune cells, receptors, and cytokines. In this review, we present a unified framework for the ODE model of the tumor-immune system, represented as:
\begin{equation}\label{eq:ODE}
	\frac{\mathrm{d} X_i(t) }{\mathrm{d} t}  = G_{i}+\sum_{j=1}^n F_{i,j}(\mathbf{X}(t);\Theta) + D_{j},
\end{equation}
where $\mathbf{X}=(X_1, X_2, \dots, X_n)$ represents the cell numbers of different components in the tumor immune system, $F_{i,j}$ captures the interactions between components $j$ and $i$, $\Theta=(\theta_1,\theta_2,\dots,\theta_m)$ denotes the set of parameters. Additionally,  $G_{i}$ and $D_{i}$ represent the dynamic behaviors of cell growth and death, respectively. 

The growth rate term $G_i$ in equation \eqref{eq:ODE} can be expressed using several well-known growth models, classified into six types: exponential, power law, logistic, Hill function, Gompertzian, and von Bertalanffy models \cite{Gerlee.CancerRes.2013,Sarapata.BullMathBiol.2014,Benzekry.PLoSComputBiol.2014,Talkington.BullMathBiol.2015,Lei.JTheorBiol.2020}. These models are detailed below:
\begin{itemize}
\item \textbf{Exponential model}: The simplest form, $G_{i} = r_i X_i$, assumes that cells grow at a constant rate, often used to describe tumor growth where tumor size is assumed to increase proportionally to its current size over time. 
\item \textbf{Power law model}: This generalization of the exponential is given by $G_{i} = r_i X_i^{\alpha_i}$, where the growth rate is proportional to the current cell population raised to the power of $\alpha$. 
\item \textbf{Logistic model}: In this model, $G_{i} = r_i \left(1-\frac{X_i}{K_i}\right)X_i$, growth slows as the cell population approaches its carrying capacity, $K_i$. A variant based on evolutionary game theory, the \textbf{competitive logistic model}, $G_{i} = r_i \left(1-\frac{\sum_{j=1}^{n} a_{ij}X_j}{K_i}\right)X_i $, describes competition among different cell subtypes \cite{West.ClinCancerRes.2019}. The logistic model can be generalized further to $G_{i} = r_i\left(1-(\frac{X_i}{K_i})^{\alpha_i}\right) X_i$, providing more flexibility in describing growth dynamics.
\item \textbf{Hill model}: Here, growth is expressed as a Hill model $G_i = r_i \frac{1}{1 + (X_i/K_i)^{\alpha_i}} X_i$, where $K_i$ represents the half effective inhibitory concentration. The Hill model is often used to model growth regulated by cytokines in the microenvironment \cite{Bernard:2003aa,Lei.JTheorBiol.2020}. 
\item \textbf{Gompertzian model}: This model, $G_{i} = r_i \log(\frac{K_i}{X_i})X_i$, describes tumor growth with an exponentially decreasing rate, commonly applied to model tumor vascular growth \cite{Norton.Nature.1976,Hahnfeldt.CancerRes.1999}. 
\item \textbf{von Bertalanffy model}: A lesser-known model, $G_{i} = a_i {X_i}^{\alpha_i} - b_i X_i$, which describes tumor growth in a form known as ``type II growth'' \cite{West.PNAS.2019}. 
\end{itemize}
These models provide various mathematical formulations for cell growth and are used in literature to describe different cell types within the tumor-immune interaction framework.

The mathematical form of the remaining terms in equation \eqref{eq:ODE} varies depending on the specific biological mechanisms and modeling objectives, offering flexibility to capture the complexity of tumor-immune interactions. 

\begin{figure}[htb!]
	\centering
	\includegraphics[width=13cm]{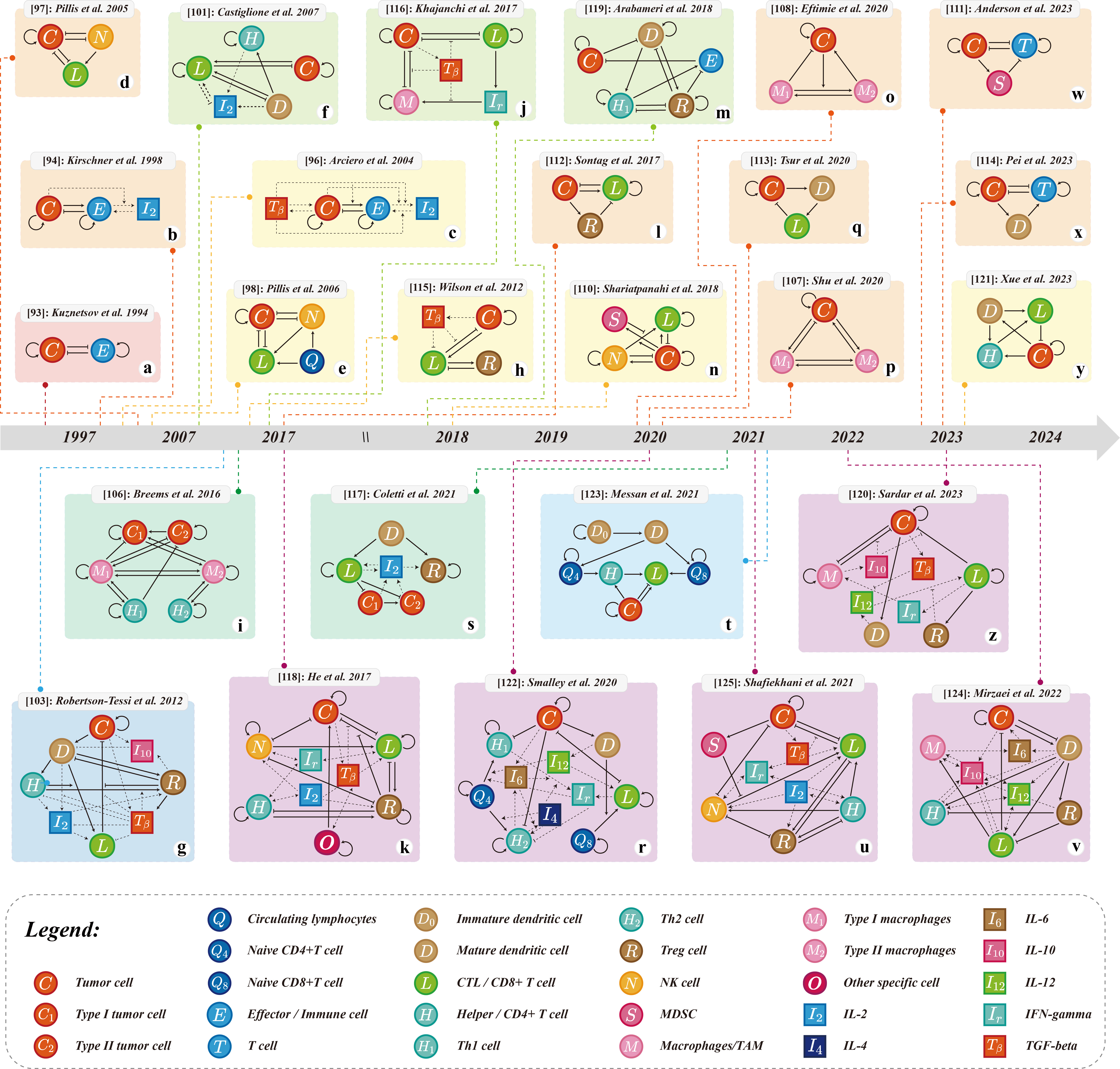}
	\caption{Application of ODE models in the description of tumor-immune regulatory networks. The content in the grey box indicates article information, with serial numbers corresponding to references. Solid lines represent cellular-level mechanisms, while dashed lines represent cytokine-level interactions. Arrows indicate proliferation or activation, and blocking arrows represent killing or inhibitions. }
	\label{3-2}
\end{figure}

A growing number of mathematical models have been developed to explain the complex regulatory mechanisms between tumors and the immune system, based on the principles of ODE model construction \cite{Eftimie.BullMathBiol.2011,Arabameri.MathBiosci.2018,Mahlbacher.JTheorBiol.2019}. This review focuses on summarizing the applications of these mathematical models in describing tumor-immune regulatory networks, as well as providing an overview of the development of ODE models in tumor-immune interactions modeling over the past three decades (Figure \ref{3-2}).

The Lotka-Volterra model, traditionally used to describe predator-prey dynamics in ecological systems, has been adapted to many mathematical models. In the 1990s, Kuznetsov and Makalkin \cite{Kuznetsov.BullMathBiol.1994} applied the Lotka-Volterra model principles to tumor-immune interactions (Figure \ref{3-2}a), highlighting how tumor growth stimulates immune responses and the phenomenon of tumor dormancy. Later, Kirschner and Panetta \cite{Kirschner.JMathBiol.1998} expanded this research by incorporating the role of IL-2, a cytokine that enhances T cell proliferation and function, in tumor-immune interactions (Figure \ref{3-2}b). This model has been instrumental in exploring adoptive cellular immunotherapy and analyzing behaviors such as short-term oscillations and long-term tumor recurrence. Wei et al. \cite{Wei.IntJBifurcatChaos.2013} further performed bifurcation analyses of the key parameters in \cite{Kirschner.JMathBiol.1998}, providing insights into their biological significance. Arciero et al. \cite{Arciero.DCDSB.2004}, building on Kirschner's model, incorporated the immunosuppressive and growth-promoting effects of TGF-$\beta$ in tumor immunology (Figure \ref{3-2}c). Their model predicted that increasing the production rate of TGF-$\beta$ could enhance tumor growth and its ability to evade immune surveillance.

Pillis et al. \cite{Pillis.CancerRes.2005} introduced an analytical framework to investigate the roles of NK cells and CD8+ T cells in tumor-immune surveillance (Figure \ref{3-2}d), introducing a new functional form for tumor cell killing by CD8+ T cells, which emphasized the different dynamics between NK and CD8+ T cells in tumor immunity. However, this model did not account for immune suppression. Subsequently, Pillis et al. \cite{Pillis.JTheorBiol.2006,Pillis.MathBiosci.2007} extended their model to include circulating lymphocytes, further exploring the effects of chemotherapy and immunotherapy on tumor evolution (Figure \ref{3-2}e), marking one of the early efforts to study optimal control in drug treatment. Similarly, Castiglione et al. \cite{Castiglione.BullMathBiol.2006,Castiglione.JTheorBiol.2007} established a population dynamics model of tumor-immune competition (Figure \ref{3-2}f) and used optimal control theory to identify the optimal dosing strategies for immunotherapy.

Tumor-immune interactions are exceedingly complex. While no single model can encompass all cell types and signaling molecules, overly simplified models fail to capture the intricate dynamics observed in experiments and clinical settings. Building on models involving IL-2 \cite{Kirschner.JMathBiol.1998}, TGF-$\beta$ \cite{Arciero.DCDSB.2004}, effector cells \cite{Pillis.CancerRes.2005}, and Tregs \cite{Leon.JTheorBiol.2007}, Robertson-Tessi et al. \cite{Robertson-Tessi.JTheorBiol.2012} developed a comprehensive mathematical model of tumor-immune interactions (Figure \ref{3-2}g). This model introduced an immune suppression mechanism, incorporating a negative feedback loop in the activation of the immune system. It suggested that tumors not only activate immune responses but also regulate immune suppression, weakening T cell function. Robertson-Tessi et al. \cite{Robertson-Tessi.JTheorBiol.2015} later extended this model to capture the interactions between tumors, the immune system, and chemotherapy. Soto-Ortiz et al. \cite{Soto-Ortiz.JTheorBiol.2016} built on these models, developing one that couples anti-angiogenic therapy targeting the tumor vasculature with immunotherapy targeting the tumor.

Macrophage polarization and transformation are typical biological phenomena where cancer cells remodel the TME. Breems et al. \cite{Breems.JTheorBiol.2016} developed a model of macrophage polarization (Figure \ref{3-2}i), which integrated interactions between two types of tumor cells, two subsets of Th cells, and two types of macrophages. Their results showed that tumor growth is strongly correlated with the Type II immune response characterized by Th2 and M2. Similarly, Shu et al. \cite{Shu.ApplMathModel.2020} proposed a model describing the interactions between tumor cells, M1 and M2 macrophages (Figure \ref{3-2}p), demonstrating that cancer evolution depends not only on tumor-induced activation of M1 and M2 macrophages but also on transitions between these macrophage states. Eftimie \cite{Eftimie.MathBiosci.2020} explored the impact of M1-to-M2 transformation driven by tumor cells (Figure \ref{3-2}o), analyzing how macrophage phenotype conversion influences tumor growth, control, and decay. Additionally, Eftimie et al. \cite{Eftimie.JTheorBiol.2021} investigated the role of transitional macrophages in tumor evolution. Analogous to M2 macrophages, MDSCs also exert potent immunosuppressive effects in the TME. Shariatpanahi et al. \cite{Shariatpanahi.JTheorBiol.2018} developed a model examining the interactions between tumors, CTLs, NK cells, and MDSCs (Figure \ref{3-2}n), assessing the impact of anti-MDSC drugs on tumor growth and immune system restoration. More recently, Anderson et al. \cite{Anderson.JMathBiol.2023} proposed an ODE model that provides insights into the tumor, T cell, and MDSC interactions  (Figure \ref{3-2}w), and suggested combining immune checkpoint inhibitors (ICIs) with MDSC inhibitors as a therapeutic strategy.

Sontag \cite{Sontag.CellSyst.2017} proposed an immune recognition model (Figure \ref{3-2}l) incorporating systems biology mechanisms such as negative feedback, incoherent feedforward loops, and bistability. This model captured the complex interactions among tumors, CTLs, and Tregs, using mathematical theory to elucidate key biological mechanisms. In recent years, significant research has focused on applying mathematical methods to tumor immunology models involving the regulation of three interacting elements. Tsur et al. \cite{Tsur.JTheorBiol.2020} developed a model (Figure \ref{3-2}q) incorporating tumors, CTLs, and DCs to predict the efficacy of ICIs in melanoma and analyze the system's local and global dynamics. Pei et al. \cite{Pei.JTheorBiol.2023} established a model (Figure \ref{3-2}x) incorporating tumors, T cells, and DCs to analyze the combined effects of RNA interference and ICIs in breast cancer, using machine learning methods to optimize treatment strategies.

Wilson et al. \cite{Wilson.BullMathBiol.2012} explored the synergistic effects of anti-TGF-$\beta$ and vaccine therapies by dividing the tumor immune response into four modules: tumor, CTLs, Tregs, and TGF-$\beta$ (Figure \ref{3-2}h). Building on this, Khajanchi et al. \cite{Khajanchi.MathBiosci.2017} integrated the interactions between tumors, macrophages, CTL, TGF-$\beta$, and IFN-$\gamma$ to examine tumor control through immunotherapy (Figure \ref{3-2}j). Coletti et al. \cite{Coletti.JTheorBiol.2021} developed a model (Figure \ref{3-2}s) incorporating two types of tumor cells, DCs, Tregs, CTLs, and IL-2, using bistability to explain the heterogeneity of tumor evolution. He et al. \cite{He.JBiolSyst.2017} proposed a model of the regulatory mechanisms within the TME (Figure \ref{3-2}k), demonstrating that combined therapies reduce Tregs and improve patient survival. Arabameri et al. \cite{Arabameri.JBiolSyst.2018} created a mathematical model of tumor-immune interactions, focusing on DC mechanisms (Figure \ref{3-2}m), which highlighted the role of DC vaccines in tumor progression. More recently, Sardar et al. \cite{Sardar.CommunNonlinearSci.2023} developed a nine-dimensional tumor immune dynamical system (Figure \ref{3-2}z) and employed a quasi-steady-state approximation to reduce it to a four-dimensional ODE model, capturing tumor immunity dynamics in response to various cytokines. Similarly, Xue et al. \cite{Xue.JTheorBiol.2023} established a four-dimensional ODE model of tumor immunity (Figure \ref{3-2}y), conducting Hopf bifurcation analysis and evaluating the combined therapeutic efficacy. This body of work, combining theoretical analysis with numerical simulations, provides a foundation for future studies in mathematical oncology.

T cell activation is crucial in the tumor-immune response, directly impacting the body's ability to mount an effective anti-tumor immune reaction. Smalley et al. \cite{Smalley.iScience.2020} constructed a tumor-immune interaction network (Figure \ref{3-2}r) to investigate the activation processes of CD4+ and CD8+ T cells, as well as their involvement in anti-tumor immune responses, using computer simulations to model dynamic responses to anti-PD-1 therapies. Messan et al. \cite{Rodriguez-Messan.PLoSComputBiol.2021} developed a mathematical model for cancer vaccine treatment (Figure \ref{3-2}t), focusing on DC activation, antigen presentation, and T cell-mediated immune attack on tumor cells. Similarly, Mirzaei et al. \cite{Mohammad-Mirzaei.PLoSComputBiol.2022} constructed a mathematical model (Figure \ref{3-2}v) that encompasses T cell activation and explores the intricate regulatory interactions between cells and cytokines. Shafiekhani et al. \cite{Shafiekhani.BMCCancer.2021} further examined the combined efficacy of chemotherapy and immunotherapy by developing a mathematical model driven by both cellular and cytokine interactions (Figure \ref{3-2}u).

\subsection{Delay differential equation (DDE) model}

\begin{figure}[htb!]
	\centering
	\includegraphics[width=13cm]{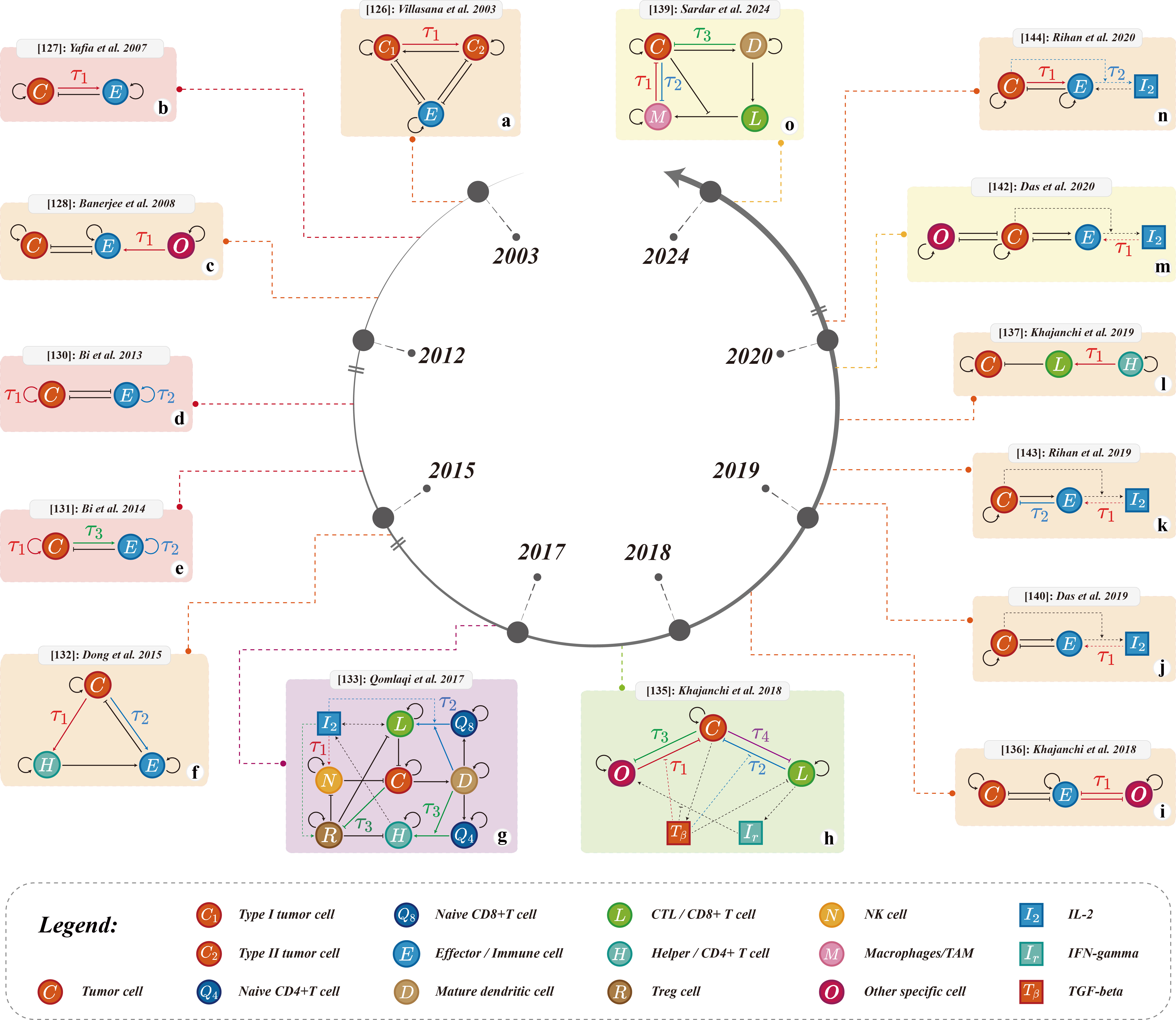}
	\caption{Application of DDE models in describing tumor-immune regulatory networks. The grey box contains article information, with serial numbers corresponding to the references. Solid lines represent cellular-level mechanisms, while dashed lines represent cytokine-level mechanisms. Sharp arrows indicate proliferation or activation, and blocked arrows indicate killing or inhibition. Red, blue, green, and purple lines correspond to the 1st, 2nd, 3rd, and 4th time delays, respectively. }
	\label{3-3}
\end{figure}

Time delays are an essential aspect of biological processes in the mathematical modeling of tumor-immune systems. These delays arise from various processes such as molecular production, cell proliferation and differentiation, tumor recognition and phagocytosis by the immune system, and the migration of cells between different tissues---each requiring a certain amount of time. Therefore, incorporating discrete time delays into mathematical oncology models helps improve the understanding of the dynamic interactions between tumors and the immune system. Based on equation \eqref{eq:ODE}, we can generalize the DDE model of the tumor-immune system in a unified form as:
\begin{equation}\label{eq:DDE}
	\frac{\mathrm{d} \vec{X}(t) }{\mathrm{d} t}  = \vec{F}(\vec{X}(t),\vec{X}(t-\tau_1),\vec{X}(t-\tau_2),\dots,\vec{X}(t-\tau_k);\Theta), 
\end{equation}
where $\tau_1$, $\tau_2$, $\dots$, $\tau_k$ represent the time delays. This review highlights several representative DDE models of tumor-immune interactions developed over the past two decades, with a particular focus on the biological mechanisms governed by discrete time delays. The regulatory networks and time-delay factors incorporated in these DDE models are visualized in Figure \ref{3-3}.

Villasana et al. \cite{Villasana.JMathBiol.2003} were pioneers in developing a DDE model (Figure \ref{3-3}a) that described interactions between tumor cell subpopulations in the interphase and mitotic phases with the immune system, examining the influence of cycle-specific drugs on tumor growth. Their theoretical and numerical analyses demonstrated that periodic solutions can arise through Hopf bifurcations. Additionally, Yafia \cite{Yafia.SIAMJApplMath.2007} expanded the work of Kuznetsov et al. \cite{Kuznetsov.BullMathBiol.1994} by introducing a two-dimensional DDE model (Figure \ref{3-3}b) of tumor-immune interactions, with a time delay representing the immune system's response time following tumor cell recognition. This model revealed that system dynamics are largely governed by the delay parameter, with Hopf bifurcations in this parameter predicting the emergence of limit cycles from non-trivial steady states. Similarly, Banerjee et al. \cite{Banerjee.Biosystems.2008} extended the model by Sarkar et al. \cite{Sarkar.MathBiosci.2005} to a three-dimensional DDE model framework (Figure \ref{3-3}c) by incorporating biologically relevant mechanisms and delays related to the conversion of resting cells to effector cells.

Bi and Ruan \cite{Bi.SIAMJApplDynSyst.2013} developed a two-dimensional tumor-immune model with two delays (Figure \ref{3-3}d), deriving general formulas to assess the direction, period, and stability of both codimension-one and codimension-two bifurcation periodic solutions. Building on this, Bi et al. \cite{Bi.Chaos.2014} advanced a similar two-dimensional model with three delays (Figure \ref{3-3}e), with each delay representing tumor proliferation, tumor-stimulated effector cell growth, and effector cell differentiation, respectively. Concurrently, Dong et al. \cite{Dong.ApplMathComput.2015} introduced a three-dimensional DDE model with two delays (Figure \ref{3-3}f), focusing on the immune activation delay of effector cells and the activation delay of Th cells. In computational modeling, Qomlaqi et al. \cite{Qomlaqi.MathBiosci.2017} developed a comprehensive nine-dimensional DDE model with three delays (Figure \ref{3-3}g), effectively illustrating the dynamic evolution of the tumor-immune interactions.

Khajanchi et al. \cite{Khajanchi.ApplMathComput.2014,Khajanchi.MathBiosci.2018,Khajanchi.Chaos.2018,Khajanchi.ApplMathComput.2019,Khajanchi.IntJBiomath.2020} proposed a series of influential DDE models for tumor-immune systems. Initially, Khajanchi et al. \cite{Khajanchi.ApplMathComput.2014} incorporated a discrete delay into the recruitment term for effector cells based on the model by Kuznetsov \cite{Kuznetsov.BullMathBiol.1994}, deriving explicit expressions for the direction of Hopf bifurcation and periodic solution stability using normal form theory and the center manifold theorem. Subsequently, Khajanchi et al. \cite{Khajanchi.MathBiosci.2018} introduced a five-dimensional DDE model with four nonlinear delay terms (Figure \ref{3-3}h), demonstrating the influence of multiple delays on tumor-immune interactions. In \cite{Khajanchi.Chaos.2018}, they also proposed a three-dimensional model depicting the interaction between tumors, effector cells, and healthy host cells (Figure \ref{3-3}i), which explores how tumor cells persist despite transient immune responses. Further models by \cite{Khajanchi.ApplMathComput.2019} and \cite{Khajanchi.IntJBiomath.2020} focused on interactions between tumors, CTLs, and Th cells (Figure \ref{3-3}l), incorporating delays associated with Th cell-mediated CTL activation. Recently, Sardar et al. \cite{Sardar.CSF.2024} developed an advanced tumor-immune interaction model with three discrete delays (Figure \ref{3-3}o), reducing a nine-dimensional ODE model to a four-dimensional DDE model through a quasi-steady-state approximation of cytokine levels. Their study extensively examined the model's foundational properties, including existence, uniqueness, positivity, boundedness, and uniform persistence.

Das et al. \cite{Das.ApplMathComput.2019,Das.Chaos.2020,Das.ChaosSolitonFract.2020} have made notable contributions to advancing DDE models in tumor-immune dynamics. In \cite{Das.ApplMathComput.2019}, they introduce a DDE model featuring Monod-Haldane response dynamics (Figure \ref{3-3}j), capturing the interactions among tumors, effector cells, and IL-2. Further expanding this framework, \cite{Das.Chaos.2020} and \cite{Das.ChaosSolitonFract.2020} developed a more comprehensive DDE model (Figure \ref{3-3}m) involving tumors, effector cells, Th cells, and IL-2, incorporating cytokine-mediated cell signaling with time delays to coordinate immune responses. Additionally, \cite{Das.ChaosSolitonFract.2020} explored an optimal control approach for combined immunotherapy and chemotherapy.

Rihan et al. \cite{Rihan.ApplMathComput.2019,Rihan.CSF.2020,Rihan.AlexEngJ.2022} have also contributed groundbreaking work to the field of DDE models of tumor-immune systems. Based on the foundational models in \cite{Kirschner.JMathBiol.1998} and \cite{Das.ApplMathComput.2019}, Rihan et al. \cite{Rihan.ApplMathComput.2019} introduced a model with two delay processes (Figure \ref{3-3}k), examining tumor-immune dynamics and optimal control under immunochemotherapy.  Building on this, \cite{Rihan.CSF.2020} introduced a fractional-order DDE model (Figure \ref{3-3}n) that analyzed conditions for stability and Hopf bifurcations with two distinct delays. More recently, Rihan et al. \cite{Rihan.AlexEngJ.2022} developed a DDE model incorporating stochastic noise, demonstrating that stochastic fluctuations can suppress tumor cell growth and that white noise can potentially lead to tumor dormancy or eradication. 

Recently, more biologically detailed DDE models have been formulated. Among them, Dickman and Kuang \cite{Dickman.Chaos.2020,Dickman.SIAMJApplMath.2020} presented a two-compartment DDE model that distinguishes the peripheral blood from the TME and integrates key mechanisms, including DC maturation and CTL cell activation. This work marks a substantial evolution from single-compartment to multi-compartment DDE models. Additionally, Wang et al. \cite{Wang.CSF.2022} introduced a DDE model featuring two specific delays to represent the dynamics between tumors and the lymphatic system, characterizing tumor proliferation and the maturation process of T lymphocytes.

\subsection{Stochastic differential equation (SDE) model}

Stochastic perturbations accompany nearly all living processes, encompassing intrinsic noise arising from molecular-level fluctuations and external noise stemming from environmental changes \cite{Tsimring:2014aa,Zechner:2020aa}. Integrating stochastic terms into models to capture these influences---such as intercellular communication and protein perturbations---on tumor-immune interactions is essential. Stochastic models can be constructed by introducing stochastic processes or parameters, providing a robust framework to study how randomness affects tumor-immune dynamics. SDE models allow for analysis of tumor-immune system behavior under stochastic perturbations, including asymptotic and stability analyses, periodic solutions, and tumor heterogeneity evolution. This review describes a general SDE model of the tumor-immune system as: 
\begin{equation}\label{eq:SDE}
	\mathrm{d} \vec{X}_t = \vec{\mu}(t,\mathbf{X}_t;\Theta) \mathrm{d} t + \vec{\sigma}(t,\vec{X}_t;\Theta) \mathrm{d} \vec{W}_t, 
\end{equation}
where $\vec{X}_t$ represents the stochastic state variable, $\vec{\mu}(t,\vec{X}_t;\Theta)$ is the drift term modeling the trend of changing, $\vec{\sigma}(t,\vec{X}_t;\Theta)$ is the diffusion term reflecting stochastic fluctuations, and $\vec{W}_t$ is a Wiener process capturing stochastic disturbances.

Mukhopadhyay et al. \cite{Mukhopadhyay.StochAnalAppl.2009} developed an SDE model for tumor-immune interactions, simulating white noise perturbations around system boundaries and equilibrium points---a standard method for adding stochastic perturbations to deterministic models. Caravagna et al. \cite{Caravagna.JTheorBiol.2010} extended the \cite{Kirschner.JMathBiol.1998} model to a hybrid stochastic framework, combining stochastic processes to capture cellular dynamics and differential equations for interleukin dynamics. Xu et al. \cite{Xu.PhysicaA.2013} investigated stochastic bifurcations in the tumor-immune system under symmetric non-Gaussian L\'{e}vy noise, linking bifurcation patterns with noise intensity and stability. Li et al. \cite{Li.CNSNS.2017} adapted a simplified tumor-immune ODE model to an SDE framework with Gaussian white noise, providing insights into the stochastic dynamics of tumor growth, immune response, and immunoediting. 

Subsequently, Caravagna et al. \cite{Caravagna.JTheorBiol.2010} examined the effects of stochastic shocks on tumor suppression, while Deng et al. \cite{Deng.ApplMathModel.2020} developed a pulsed stochastic tumor-immune model with mode transitions, emphasizing the link between stochastic and pulsed perturbations on system behavior. Liu et al. \cite{Liu.PhysicaA.2018} constructed a continuous time Markov chain model based on the branching processes theory to characterize the dynamics of tumor-immune interactions. Li et al. \cite{Li.SIAMJApplMath.2019,Chen.JMathAnalAppl.2023} extended the classical two-dimensional tumor-immune ODE model \cite{Kuznetsov.BullMathBiol.1994} to an SDE framework, utilizing stochastic Lyapunov analysis, comparison theorem, and strong ergodicity theorem to explore the system's asymptotic properties. Yang et al. \cite{Yang.CNSNS.2019} introduced a stochastic model for pulsatile therapy, examining the impact of fluctuations and combined immunotherapy and chemotherapy on treatment outcomes. Han and Hao et al. \cite{Han.ApplMathModel.2022,Hao.CSF.2022} studied the most probable trajectories of the proposed stochastic tumor-immunity model.

Recently, several three-dimensional SDE models have emerged to model tumor-immune interactions more accurately \cite{Bose.PRE.2011,Das.PhysicaA.2020,Phan.MathBiosciEng.2020,Yang.MathComputSimulat.2021,Alsakaji.MathBiosciEng.2023,Huang.ActaMathSci.2022}. For example, Bose et al. \cite{Bose.PRE.2011} investigated an SDE model involving tumors, effector cells, and tumor-detecting cells, showing that noise correlation parameters strongly influence tumor-immune dynamics. Phan et al. \cite{Phan.MathBiosciEng.2020} developed an SDE model to simulate viral therapy, while Yang et al. \cite{Yang.MathComputSimulat.2021} introduced a pulsed SDE model to describe interactions between the tumor, Th cells, and CTLs. Alsakaji et al. \cite{Alsakaji.MathBiosciEng.2023} proposed a stochastic delay differential model to simulate the tumor-immune system under white noise and treatment.

More recently, Lai et al. \cite{Lai.NPJSystBiolAppl.2024} developed an SDE model that characterizes the clinical course of chronic myeloid leukemia (CML) patients achieving treatment-free remission post-therapy. By modeling feedback interactions between leukemic stem cells and the bone marrow microenvironment, they identify early relapse and long-term remission as typical clinical manifestations following treatment cessation. This model suggests that the leukemic cell proportion in blood and the TME index could be important for TFR outcomes, representing a recent clinical application of SDE models in oncology.

\subsection{Partial differential equation (PDE) model}

PDE models effectively describe the spatiotemporal dynamics of tumors and immune cells, capturing changes in tumor-immune interactions. Recent research has highlighted the complex interactions among immune cells in the TME during tumor progression. In this review, we summarize mathematical models using reaction-diffusion equations to characterize tumor-immune interactions. The following unified framework describes the spatiotemporal dynamics of the tumor-immune system:
\begin{equation}
\label{eq:PDE}
	\frac{\partial X_i}{\partial t} + \nabla \cdot\left(\vec{u}_i\ X_i\right)-\delta_{i} \nabla^2 X_i=f_{i}\left(X_1, \ldots, X_n\right), \quad \vec{X} = (X_1, \cdots, X_n) \in \Omega(t)
\end{equation}
where $\nabla = \left ( \frac{\partial}{\partial x}, \frac{\partial}{\partial y}, \frac{\partial}{\partial z} \right)$, and $\nabla^2 = \frac{\partial^2}{\partial x^2}+\frac{\partial^2}{\partial y^2}+\frac{\partial^2}{\partial z^2}$. Here, $\vec{u}_i$ denotes the advective velocities , $\delta_ {i}>0$ are diffusion coefficients. The components $X_i$ can denote different cells or molecules, each exhibiting unique advective velocities and diffusion rates. Notably, in modeling molecular-scale dynamics, the convection term $\nabla \cdot\left(\vec{u}_i X_i\right)$ can be set to zero (i.e., $\vec{u}_i = \vec{0}$) to reflect the negligible effect of intercellular pressures on smaller molecules, distinguishing it from cellular-scale dynamics. The tumor is represented by $\Omega(t) \subset \mathbb{R}^3$ and is subject to a moving boundary condition. 

To simplify the model, it is often assumed that the tumor is spherical, with all variables radially symmetric. Consequently, the variables depend only on time $t$ and radial distance $r$, where $r = \vert \vec{x} \vert$. The velocity and bounded region are expressed as $\vec{u} = u(r,t)\frac{\vec{x}}{\vert \vec{x} \vert}$ and $\Omega(t)=\{r<R(t)\}$, respectively. In spherical coordinates, equation \eqref {eq:PDE} becomes:
\begin{equation}
\label{eq:PDE-r}
	\frac{\partial X_i}{\partial t}+\frac{1}{r^2} \frac{\partial}{\partial r}\left(r^2 u X_i \right)-\delta_{X_i} \frac{1}{r^2} \frac{\partial}{\partial r}\left(r^2 \frac{\partial X_i}{\partial r}\right)=f_{i}\left(X_1, \ldots, X_n\right).
\end{equation}
The free boundary $r=R(t)$ moves with the speed of the cellular population, hence  
\cite{Lai.PNAS.2018}
\begin{equation}
\label{eq:BVP}
\dfrac{\mathrm{d} R(t)}{\mathrm{d} t} = u(R(t), t),
\end{equation}
where the velocity $u$ is derived from pressure exerted by proliferating cancer cells. If we assume a constant total cell density, such that $\sum_{i=1}^{m}X_i(r,t) = X_0$. Integrating the cell dynamic equations allows for the derivation of $u(r, t)$, satisfying 
\begin{equation}
\label{eq:B1}
\frac{1}{r^2} \frac{\partial}{\partial r}\left(r^2 u\right)=\sum_{i=1}^{m} f_{i}.
\end{equation}

One notable study on a PDE model for combination therapy in breast cancer is by Lai et al. \cite{Lai.PNAS.2018}, which integrates eight cellular-level dynamic behaviors and fourteen molecular-level elements to assess therapeutic efficacy using evaluation indices. Their results demonstrate a positive correlation between BET inhibitors and CTLA-4 inhibitors in breast cancer, showing that tumor volume decreases as dosages increase for each drug. In a subsequent model, Lai et al. \cite{Lai.JTheorBiol.2019} explored breast cancer treatment by combining anti-angiogenic agents with chemotherapy. Given the antagonistic interaction observed between bevacizumab and docetaxel, the model examines various dosing strategies, suggesting that non-overlapping regimens may yield superior outcomes.

The BRAF mutation is one of the most commonly prevalent in melanoma patients. Lai et al. \cite{Lai.BMCSystBiol.2017} developed a PDE model for combined targeted therapy using BRAF inhibitors and ICIs in melanoma. This study reveals that the drugs have a synergistic effect at low doses, whereas high doses lead to antagonism. Thus, identifying these antagonistic regions early through animal studies or initial clinical trials is crucial to optimizing dosing in clinical applications. Similarly, Friedman et al. \cite{Friedman.BullMathBiol.2020} established a PDE model to investigate the efficacy of combining BRAF inhibitors with anti-CCL2, anti-PD-1, and anti-CTLA-4 antibodies, aiming to identify strategies that mitigate resistance induced by BRAF inhibition. Additionally, Liao et al. \cite{Liao.MathBiosci.2022} introduced a PDE model that incorporates both proinflammatory and anti-inflammatory effects of IFN-$\gamma$ for melanoma treatment using ICIs.

PDE models are frequently employed to explore the biological mechanisms underlying tumor evolution. Szomolay et al. \cite{Szomolay.JTheorBiol.2012} constructed a model to examine the role of GM-CSF in promoting vascular endothelial growth factor (VEGF) inactivation, which in turn slows tumor growth. Lee et al. \cite{Lee.PLoSComputBiol.2021} used a chemotaxis-reaction-diffusion model to analyze the interactions between tumor cells and neutrophils that drive tumor invasion. Kim et al. \cite{Kim.JMathBiol.2022} coupled this model with receptor dynamics to elucidate the dual role of cellular senescence in cancer progression. In other studies, Friedman et al. \cite{Friedman.BullMathBiol.2018,Siewe.BullMathBiol.2023} developed a PDE model to study tumor-immune interactions, focusing on the role of exosomes---extracellular vesicles containing mRNA, microRNA, and proteins---as predictive biomarkers for tumors. Jacobsen et al. \cite{Jacobsen.BullMathBiol.2015} created a PDE model to investigate the impact of CNN1, an extracellular matrix protein, on oncolytic virus therapy in gliomas, finding that CCN1 limits therapeutic efficacy by enhancing the activation and migration of pro-inflammatory macrophages.

In recent years, numerous PDE models have emerged to study the dynamic evolution of cancer under combination therapies. Lai et al. \cite{Lai.SciChinaMath.2020} developed a model explaining the effects of combined radiotherapy and anti-PD-L1 treatment. Their findings indicate that patients receiving concurrent therapy benefited more than those on weekly alternating schedules. Siewe et al. \cite{Siewe.JTheorBiol.2023} presented a PDE model for dual immunotherapy combining anti-PD-1 and anti-CSF-1. Kim et al. \cite{Kim.PNAS.2018} also contributed a model analyzing the role of NK cells in treating primary glioblastoma with oncolytic viruses (OV) and protease inhibitors, finding that NK cells exhibit significant anti-tumor effects, which increase when exogenous NK cells are injected into the tumor.

PDE models have also been used to quantify cancer immunoediting. Li et al. \cite{Li.ComputMathAppl.2022} developed a PDE model that encapsulates the interactions among tumor cells, immune cells, cancer-associated fibroblasts, and angiogenic cells, describing the phases of cancer evolution: Elimination, Equilibrium, and Escape. The model demonstrates how immune cells and cancer-associated fibroblasts facilitate transitions between these states, offering new insights into how changes in the TME influence cancer immunoediting.

\subsection{Integral differential equation (IDE) model}

\begin{figure}[htb!]
	\centering
	\includegraphics[width=13cm]{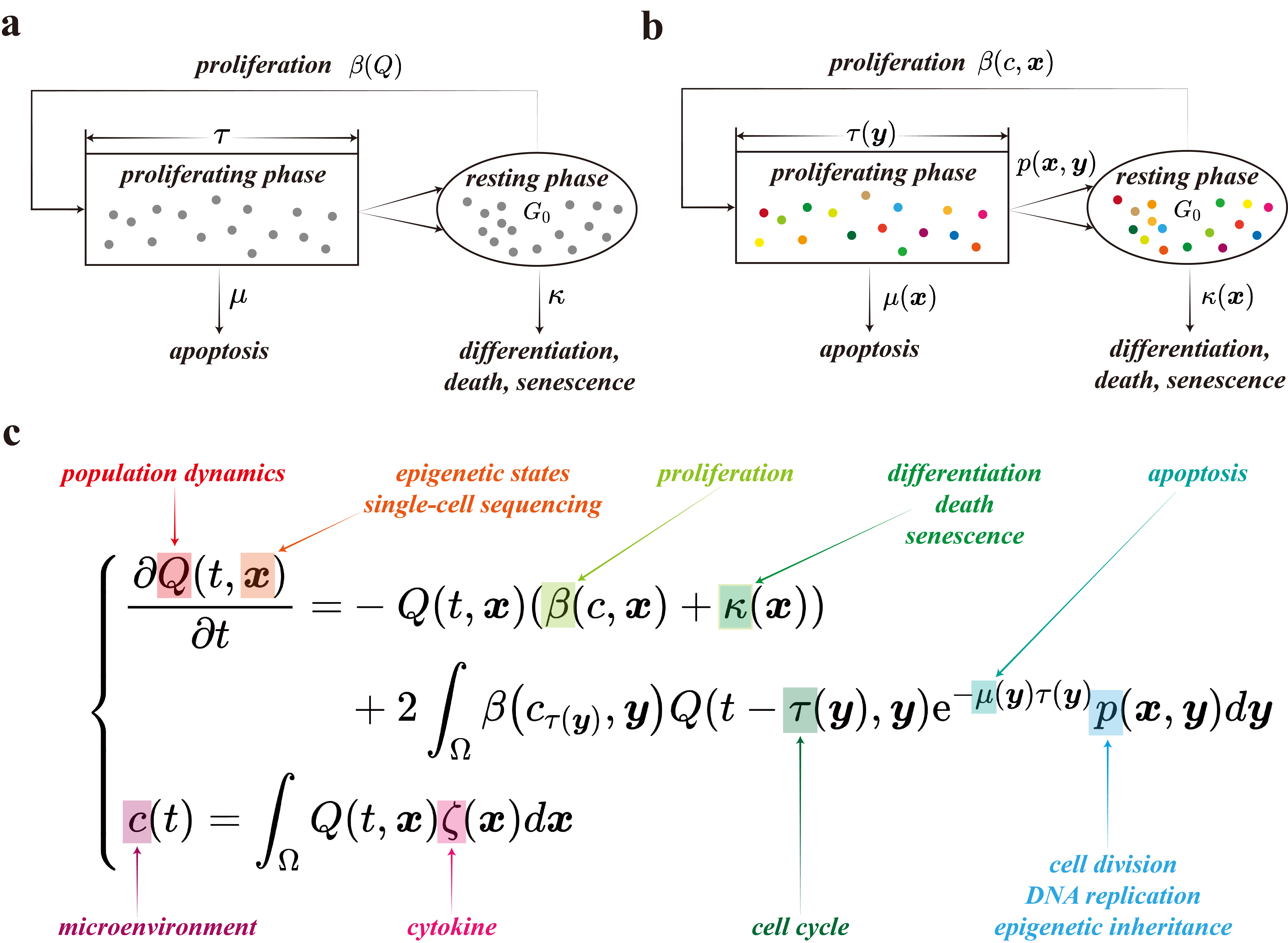}
	\caption{Mechanism illustration and general mathematical framework for stem cell regeneration dynamics. \textbf{a}. Mechanistic diagram of homogeneous stem cell regeneration dynamics. \textbf{b}. Mechanism diagram of heterogeneous stem cell regeneration dynamics. \textbf{c}. The framework of the mathematical model for heterogeneous stem cell regeneration. Here, $Q$ denotes the number of stem cells in the resting phase; $\vec{x}$ represents epigenetic status; $\Omega$ is the space encompassing all possible epigenetic states; $\beta$ represents the rate at which resting-phase cells return to the proliferating phase; $\kappa$ is the clearance rate (including differentiation, death, and senescence) of cells in the resting phase ($G_0$); $\tau$ denotes the cell cycle duration; $\mu$ is the apoptosis rate; $p(\vec{x}, \vec{y})$ represents the probability that a daughter cell in state $\vec{x}$ originates from a mother cell in state $\vec{y}$ after division; $c$ is the effective concentration of growth-inhibitory cytokines; and $\zeta(\vec{x})$ is the rate of cytokine secretion by a cell in state $\vec{x}$.}
	\label{3-4}
\end{figure}

Tumor cells are generally viewed as cells with malignant proliferative potential, resulting from genetic mutations that arise during the prolonged self-renewal processes of stem cells. In the 1970s, the G0 cell cycle model was introduced to describe the regenerative dynamics of homogeneous stem cells (Figure \ref{3-4}a) \cite{Burns.CellTissueKinet.1970,Mackey.Blood.1978}. Lei et al. \cite{Lei.PNAS.2014} were the first to incorporate epigenetic factors into models of stem cell regeneration, with the aim of exploring how genetic and epigenetic regulation interact in stem cell renewal. Building on this, to further characterize the regeneration dynamics of heterogeneous tumor stem cells (Figure \ref{3-4}b), Lei \cite{Lei.JTheorBiol.2020,Lei.SciChinaMath.2020} proposed an IDE model framework. This framework provides a general mathematical description of tumor stem cell regeneration dynamics with epigenetic transitions: 
\begin{equation}
\label{eq:IDE}
	\left\{
	\begin{aligned}
		\frac{\partial Q(t, \vec{x})}{\partial t}= & -Q(t, \vec{x})(\beta(c, \vec{x})+\kappa(\vec{x})) \\
		& +2 \int_{\Omega} \beta(c_{\tau(\vec{y})}, \vec{y}) Q(t-\tau(\vec{y}), \vec{y}) \mathrm{e}^{-\mu(\vec{y}) \tau(\vec{y})} p(\vec{x}, \vec{y}) \mathrm{d} \vec{y},\\
		c(t)=\int_{\Omega}Q&(t,\vec{x})\zeta(\vec{x})\mathrm{d}\vec{x}.
	\end{aligned}
	\right. 
\end{equation}
This equation extends the G0 cell cycle model to include stem cell heterogeneity and plasticity and can be applied to describe biological processes associated with stem cell regeneration, including development, aging, and tumor evolution \cite{Lei.JTheorBiol.2020,Lei.SciChinaMath.2020,Zhang.ComputSystOncol.2021,Liang.CommunApplMathComput.2023}. 

In modeling the dynamic mechanisms underlying tumor evolution, equation \eqref{eq:IDE} connects various components: epigenetic states $\vec{x}$, tumor dynamics ($\beta(c,\vec{x})$, $\kappa(\vec{x})$, $\mu(\vec{x})$), cell cycle duration $\tau(\vec{x})$, cytokine secretion ($\zeta(\vec{x})$), and the inheritance probability of epigenetic states $p(\vec{x}, \vec{y})$ (Figure \ref{3-4}c). The functions $\beta(c, \vec{x})$, $\kappa(\vec{x})$, $\mu(\vec{x})$, and $\tau(\vec{x})$, which describe cell cycle kinetics, are collectively termed the cell's \textit{kinetotype} as proposed in \cite{Lei.JTheorBiol.2020}. 

The inheritance function $p(\vec{x}, \vec{y})$ in equation \eqref{eq:IDE} is essential for capturing cell plasticity during the cell cycle. Although determining the exact form of $p(\vec{x}, \vec{y})$ biologically is challenging due to the complexity of biochemical processes involved in cell division, it can be considered as a conditional probability density:
$$
p(\vec{x}, \vec{y}) = P(\mbox{state of daughter cell}\ =\ \vec{x}\ \vert\ \mbox{state of mother cell}\ = \vec{y}).
$$
This allows us to focus on the epigenetic states before and after cell division, bypassing the intermediate processes. Huang et al. \cite{Huang.IntJModPhysB.2017,Huang.DCDSB.2019,Huang.JTheorBiol.2024} developed a computational model based on epigenetic mechanisms, specifically, histone modifications, showing that inheritance probabilities can be described using a conditions Beta distribution. 

The framework provided in equation \eqref{eq:IDE} establishes a foundational model encompassing the key elements of stem cell regeneration, including cell cycling, heterogeneity, and plasticity. This model can be extended to account for gene mutations and cell lineage evolution \cite{Lei.JTheorBiol.2020}. However, stem cell systems in biological processes may need to be incorporated, such as gene networks within the cell cycle, cell-to-cell interactions within specific niches, and interactions between cells and the microenvironment. For further discussion, please refer to \cite{Lei:arXiv2024}.

Utilizing the modeling mechanisms outlined by Lei et al. \cite{Lei.PNAS.2014}, Guo et al. \cite{Guo.CancerRes.2017} developed a multi-scale computational model to simulate the progression from inflammation to tumorigenesis. This model effectively reproduces the pathway of transformation from inflammation to cancer, comprising two primary stages: the transition from normal tissue to precancerous lesions and the progression from these lesions to malignant tumors. Computational results suggest that long-term, mild inflammation can initiate the development of precancerous lesions from a normal state, though it appears insufficient to drive full malignancy. In contrast, moderate and severe inflammation markedly enhances the progression from a precancerous state to tumor development.

Liang et al. \cite{Liang.CommunApplMathComput.2023} applied a generalized framework for heterogeneous stem cell regeneration to investigate the dynamics of epigenetic states in a one-dimensional context. The biological background of this study is to understand dynamic blood disorders, specifically fluctuations in blood cell counts. The model elucidates the influence of changes in cellular heterogeneity and plasticity on population dynamics, particularly cyclic and oscillatory behaviors. Results suggest that alterations in cellular heterogeneity and plasticity can affect conditions that give rise to oscillatory phenomena in stem cell regeneration systems.

In recent years, chimeric antigen receptor T (CAR-T) cell therapy has shown promising clinical benefits in treating B-cell acute lymphoblastic leukemia (B-ALL). Zhang et al. \cite{Zhang.ComputSystOncol.2021} combined biological experiments with a mathematical model to explore CAR-T-induced cellular plasticity leading to tumor recurrence. This study successfully replicates tumor evolution dynamics observed in biological models, predicting that CAR-T-induced cellular plasticity following CD19 CAR-T therapy could drive B-ALL recurrence. Both the model and experiments suggest that a combined CAR-T therapy targeting CD19 and CD123 at specific ratios may prevent disease relapse. 

Ma et al. \cite{Ma.JMathBiol.2023} recently developed a mathematical model based on equation \eqref{eq:IDE} to evaluate how heterogeneous PD-L1 expression affects disease progression in cancer patients. This model attributes tumor cell heterogeneity to stemness and PD-L1 expression levels, while T-cell heterogeneity is influenced by PD-1 expression. Results show that during the early stages of anti-PD-L1 therapy, response rate and efficacy correlate with PD-L1 expression levels in virtual patients. For patients with high PD-L1 expression, anti-PD-L1 treatment more effectively controls tumor growth. The model also reveals that a maximum-tolerated dose strategy offers superior survival benefits for PD-L1-positive esophageal cancer patients. 

In addition, Su et al. \cite{Su.JMPA.2023} conducted theoretical research on equation \eqref{eq:IDE}, focusing on the eigenvalue problems and asymptotic behaviors of both monogenotypic and polygenotypic stem cell regeneration models with epigenetic transitions. They examined the long-term dynamical and steady-state solutions associated with the three classes of quasilinear nonlocal diffusion evolution equations derived from these models, providing explicit formulas for thresholds pertinent to tissue development, degeneration, and abnormal growth.

\subsection{Quantitative systems pharmacology (QSP) model}

\begin{figure}[htb!]
	\centering
	\includegraphics[width=14cm]{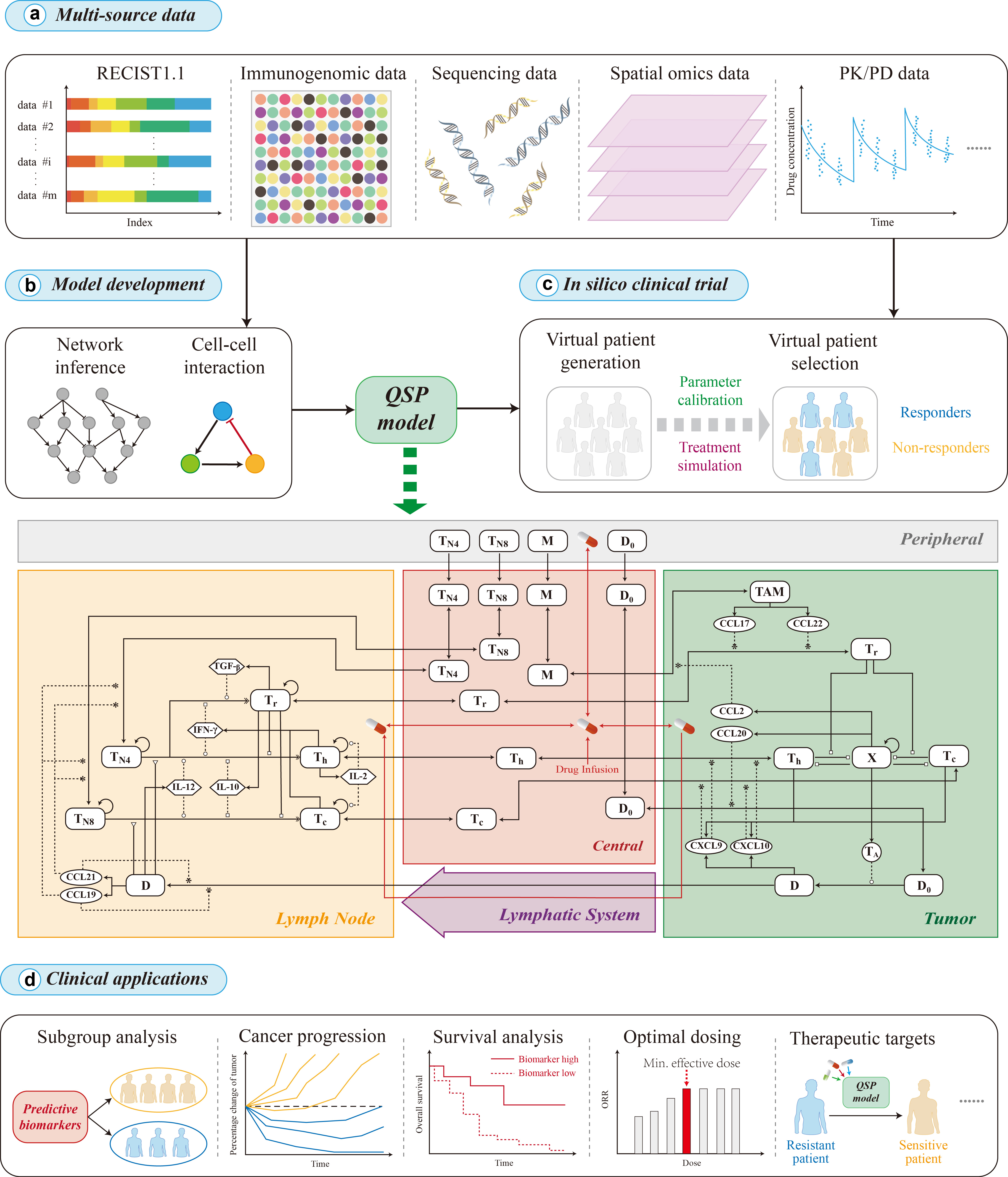}
	\caption{Schematic illustration of research methods integrating multi-source data with QSP models. \textbf{a}. Multi-source data help infer QSP model mechanisms and networks or guide the generation of effective virtual patients. \textbf{b}. A multi-compartmental QSP model is constructed based on tumor immunology's mechanisms and interaction networks. \textbf{c}. Calibration and simulation lead to selecting valid virtual patients for in silico clinical trials. \textbf{d}. QSP models can then be applied clinically to identify predictive biomarkers, project cancer progression, analyze survival, and optimize doses to enhance treatment sensitivity, especially in non-responders.}
	\label{3-5}
\end{figure}

QSP is a methodology that leverages traditional pharmacokinetics, pharmacology, and systems biology to quantitatively describe interactions between drugs and patients (Figure \ref{3-5}). QSP models focus on population characteristics, variability in drug response markers, and disease progression in drug analysis. In tumor-immune modeling, they highlight the mechanisms underlying tumor-immune interactions and the dynamic migration of immune cells across different compartments. The objective of QSP models is to provide quantitative descriptions of drug efficacy and predictive models for disease progression. In this review, we explore QSP models grounded in tumor-immune interactions and present recent advancements in the field.

Milberg et al. \cite{Milberg.SciRep.2019} developed a QSP model to predict the response of immune checkpoint blockade in melanoma treatment. This model examined the response of monotherapy, combination therapy, and sequential therapy with anti-PD-1, anti-PD-L1, and anti-CTLA-4, revealing the therapeutic variations among patients. Such models provide powerful tools for assessing the efficacy of immunotherapy and guiding clinical decisions. Similarly, Wang et al. \cite{Wang.RSocOpenSci.2019} developed a QSP model to investigate the pharmacokinetics and pharmacodynamics of anti-PD-1, anti-PD-L1, and anti-CTLA-4 therapies individually and in combination. Ma et al. \cite{Ma.JImmunotherCancer.2020,Ma.AAPSJ.2020} used QSP models to evaluate the efficacy of T-cell engager (TCE) monotherapy, anti-PD-L1 monotherapy, and combination therapy in colorectal cancer patients.

Wang et al. \cite{Wang.FrontBioengBiotechnol.2020} created a QSP model to conduct virtual clinical trials and identify predictive biomarkers. Their model, designed to evaluate immune checkpoint blockade therapy combined with epigenetic modulators in HER2-negative breast cancer, explored immune cell migration across four compartments: central, peripheral, tumor, and lymph node. The study confirmed that epigenetic modulators enhance ICIs' effects, proposing that tumor mutational burden, tumor-infiltrating effector T cell density, and the effector-to-regulatory T cell ratio in the TME as potential biomarkers for clinical trials.

In another application, Wang et al. \cite{Wang.JImmunotherCancer.2021} used a QSP model to predict the effectiveness of ICIs and chemotherapy in triple-negative breast cancer, optimizing drug dosages and treatment regimens. Recognizing the importance of TAMs as critical immunosuppressive cells, Wang et al. \cite{Wang.iScience.2022} expanded the QSP model to include TAM heterogeneity, examining their impact on tumor evolution within the TME.

Sov\'{e} et al. \cite{Sove.CPTPharmacometricsSystPharmacol.2020} developed a modular QSP platform for immuno-oncology (IO) research, which integrates essential tumor-immune interaction mechanisms. This modular approach allows for the creation of IO-QSP models with specific mechanisms to address targeted research questions. This work has facilitated and advanced the progress of QSP modeling research. Sov\'{e} et al. \cite{Sove.JImmunotherCancer.2022} also used this framework to examine ICIs in hepatocellular carcinoma, predicting clinical trial outcomes using a random forest model. Ippolito et al. \cite{Ippolito.CPTPharmacometricsSystPharmacol.2024} leveraged an IO-QSP model to explore the potential of conditionally activated molecules, which can enhance anti-tumor responses while reducing systemic toxicity, for breast cancer immunotherapy. Recently, Wang et al. \cite{Wang.ClinTranslSci.2024} focused on designing pharmacokinetic and pharmacodynamic modules within a QSP model to simulate the effects of targeted therapy combined with PD-L1 inhibitors in advanced non-small cell lung cancer. 

With advancements in imaging technologies and spatial transcriptomics, tumor spatial data is increasingly critical in guiding QSP models for improved predictive accuracy. Gong and Nikfar et al. \cite{Gong.CancersBasel.2021,Nikfar.CancersBasel.2023} developed a hybrid computational modeling platform, spQSP-IO, to simulate non-small cell lung cancer growth and immunotherapeutic responses based on spatial data, accounting for tumor heterogeneity and patient variability. Zhang et al. \cite{Zhang.ImmunoinformaticsAmst.2021} used single-cell sequencing and the spQSP platform to predict immunotherapy outcomes in triple-negative breast cancer. Ruiz-Martinez et al. \cite{Ruiz-Martinez.PLoSComputBiol.2022} extended the spQSP platform to analyze tumor growth dynamics across spatial and temporal scales.

Arulraj et al. \cite{Arulraj.SciAdv.2023} recently developed a transcriptome-informed QSP model to investigate metastasis in triple-negative breast cancer and predict PD-1 inhibitor efficacy. This model identified 30 key biomarkers, with Treg density variation within lymph nodes emerging as a promising indicator. Wang et al. \cite{Wang.NPJPrecisOncol.2023} further developed an immunogenomic-driven QSP model to forecast PD-L1 inhibitor response in non-small cell lung cancer patients. By adjusting model parameters, this study generated virtual patient cohorts to predict clinical responses and identify potential biomarkers, examining the pharmacokinetics of PD-L1 inhibitors and using compressed latent parameterization to account for individual variations in drug response.

\subsection{Agent-based model}

Tumor growth and development is a complex, multi-scale biological process encompassing molecular, cellular, microenvironmental, and tissue-level interactions \cite{Anderson.NatRevCancer.2008}. ABM is a computational approach that simulates complex systems by representing the behaviors of individual agents \cite{Abar.ComputSciRev.2017,West.TrendsCellBiol.2023}. ABM's capacity to model biological processes at the computational element level makes it an effective tool for simulating the multiscale nature of tumor development. Within ABMs, agents are entities with specific behaviors and functions, representing biological components like genes, proteins, blood vessels, or cells. In this review, we highlight ABM operational rules, available software packages, and primary applications in modeling tumor-immune system interactions. 

\begin{figure}[htb!]
	\centering
	\includegraphics[width=14cm]{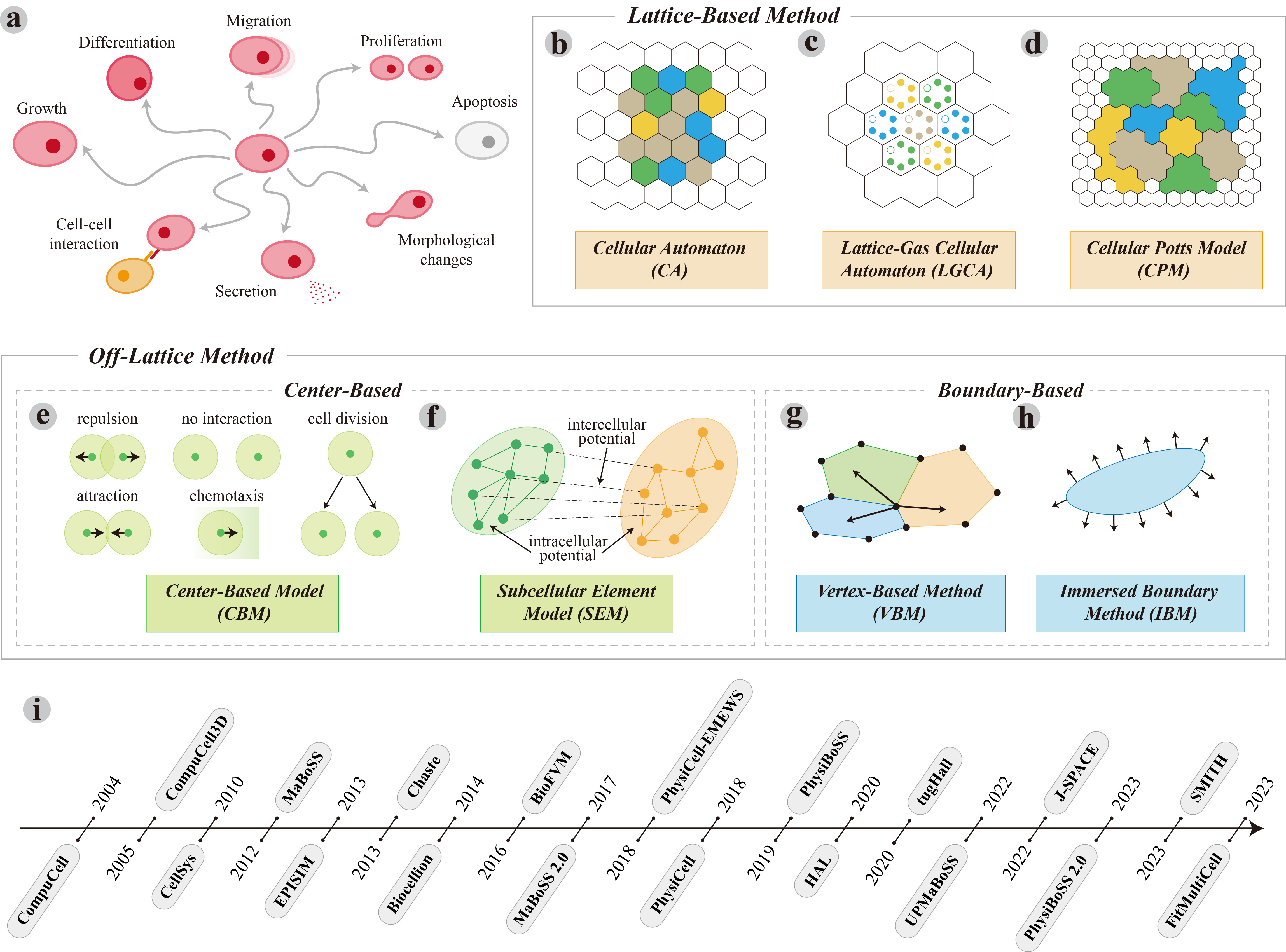}
	\caption{The biological mechanisms, modeling methods, and toolkits of ABMs for tumor-immune interactions. \textbf{a}. Biological mechanisms in ABMs. \textbf{b}. Cellular automaton method. \textbf{c}. Lattice-gas cellular automaton method. \textbf{d}. Cellular Potts method. \textbf{e}. Center-based method. \textbf{f}. Subcellular element method. \textbf{g}. Vertex-based method. \textbf{h}. Immersed boundary method. \textbf{i}. Toolkits of ABMs.}
	\label{3-6}
\end{figure}

Typically, cellular behaviors modeled in ABMs include migration, proliferation, differentiation, apoptosis, growth, morphological changes, secretion, and cell-cell interactions (Figure \ref{3-6}a). ABM frameworks are generally divided into two main paradigms: lattice-based and off-lattice methods. Lattice-based models use either structured or unstructured meshes. Structured meshes are easier to implement programmatically but have limitations in visualizing data and representing complex biological mechanisms. Unstructured meshes, like the hexagonal grids often used for tumor-immune models, help overcome these limitations. Off-lattice methods, meanwhile, include center-based and boundary-based models. 

Cellular Automata (CA) is one of the most foundational lattice-based approaches \cite{Valentim.ComputBiolMed.2023}, where each grid cell can accommodate at most one biological cell (Figure \ref{3-6}b). Operating within a discrete space-time framework, CA models update cell states based on predefined rules encompassing rest, movement (to adjacent sites), death (vacating a site), and division (placing daughter cells in adjacent grids). Another popular lattice-based approach, the Lattice-Gas Cellular Automaton (LGCA), allows multiple cells to occupy the same grid space (Figure \ref{3-6}c) \cite{Wolf.Springer.2004}. LGCA models follow simple particle movement and collision rules based on physical principles, ensuring the conservation of mass, momentum, and energy. LGCA has proven useful in simulating the spread of tumor cells, including their infiltration into surrounding tissues and distant metastasis. These models capture cell population dynamics effectively without requiring detailed descriptions of individual cell morphologies. In contrast, the Cellular Potts Model (CPM) uses multiple lattice sites to represent a single cell, enabling detailed modeling of cell morphology and mechanical properties (Figure \ref{3-6}d) \cite{Scianna.MultiscaleModelSim.2012}. Although CPM offers more detailed representations of cellular shape and behavior, it requires higher computational resources.

Center-based Model (CBM) is an off-lattice approach that characterizes cell behaviors and interactions within a system \cite{Mathias.BMCBioinformatics.2022}. In CBM, cells exert forces dependent on the distance from their neighbors, including repulsive forces when in close proximity, attractive forces when farther apart, and the chemokine-induced pulling force (Figure \ref{3-6}e). Another off-lattice model, the Subcellular Element Model (SEM), focuses on the dynamics and interactions of subcellular structures within cells (Figure \ref{3-6}f) \cite{Sandersius.PhysBiol.2008}. SEM can simulate processes such as the binding of signaling molecules---like hormones, antigens, and neurotransmitters---to cell membrane receptors, triggering biochemical cascades within the cell. In drug discovery, SEM models simulate drug-target binding, aiding in predictions of mechanisms of action and potential side effects. Beyond CBM and SEM, boundary-based models, which simulate dynamic changes in complex systems, have gained prominence. The Vertex-based Method (VBM) represents cells as polygons or polyhedra and calculates forces on vertices to depict cell morphological changes (Figure \ref{3-6}g) \cite{Fletcher.BiophysJ.2014}. VBM is crucial for processing vertex data in mesh models and identifying key points in imaging. The Immersed Boundary Method (IBM), a biomechanical approach, models tissues as clusters of heterogeneous cells (Figure \ref{3-6}h) \cite{Rejniak.JTheorBiol.2007}, emphasizing biomechanical properties and cell-microenvironment interactions.

A range of open-source ABM software packages has emerged based on the principles of ABM construction \cite{Abar.ComputSciRev.2017,West.TrendsCellBiol.2023}. Here, we highlight toolkits valuable for studying tumor evolution and tumor-immune interactions (Figure \ref{3-6}i). Initial studies focused on intracellular signaling pathways and gene networks in tumor growth, resulting in software packages like CompuCell \cite{Izaguirre.Bioinformatics.2004}, CompuCell3D \cite{Cickovski.IEEE/ACMTransComputBiolBioinform.2005}, MaBoSS \cite{Stoll.BMCSystBiol.2012}, MaBoSS 2.0 \cite{Stoll.Bioinformatics.2017}, tugHall \cite{Nagornov.Bioinformatics.2020}, and UPMaBoSS \cite{Stoll.FrontMolBiosci.2022}. As the importance of the TME was recognized, new toolkits emerged to analyze the TME and multicellular interactions, including CellSys \cite{Hoehme.Bioinformatics.2010}, EPISIM \cite{Sutterlin.Bioinformatics.2013}, Chaste \cite{Mirams.PLoSComputBiol.2013}, Biocellion \cite{Kang.Bioinformatics.2014}, PhysiCell \cite{Ghaffarizadeh.PLoSComputBiol.2018}, PhysiBoSS \cite{Letort.Bioinformatics.2019}, PhysiBoSS 2.0 \cite{Ponce-de-Leon.NPJSystBiolAppl.2023}, and FitMultiCell \cite{Alamoudi.Bioinformatics.2023}. These tools bridge molecular-level cellular signaling and gene networks with the TME, facilitating multi-scale integration. Concurrently, tools for mathematical oncology models in spatially complex systems, such as BioFVM \cite{Ghaffarizadeh.Bioinformatics.2016}, HAL \cite{Bravo.PLoSComputBiol.2020}, and PhysiCell-EMEWS \cite{Ozik.BMCBioinformatics.2018}, have been developed. Emulating Darwinian evolution, cancer is seen as an evolving system with competing subpopulations. Consequently, toolkits like J-SPACE \cite{Angaroni.BMCBioinformatics.2022} and SMITH \cite{Streck.Bioinformatics.2023} focus on tumor branching evolution and heterogeneity.

Numerous multi-scale models have been developed to explore intricate tumor-immune interactions. Anderson et al. \cite{Anderson.Cell.2006} pioneered a multi-scale cancer invasion model, enabling studies of how the microenvironment affects solid tumor growth and therapeutic responses. Building on this, Sun et al. \cite{Sun.BMCBioinformatics.2012} developed a multi-scale ABM to evaluate tyrosine kinase inhibitor (TKI) efficacy in brain tumors, incorporating biological and physical features such as blood flow and pressure from tumor growth. This model showed that tumor growth is influenced by the EGFR signaling pathway and cell cycle. Additionally, Liang et al. \cite{Liang.BMCBioinformatics.2019} employed multi-scale modeling to predict the synergistic effects of targeting both EGFR and VEGFR pathways in brain tumor treatment.

Recently, ABMs have advanced the study of tumor heterogeneity and drug resistance. Gong et al. \cite{Gong.JRSocInterface.2017} developed an ABM to model tumor-immune interactions, focusing on the effects of ICIs on tumor progression. This study categorized tumors as PD-L1$^+$ and PD-L1$^-$ and demonstrated decreasing T cell distribution over time in tumor sites, alongside spatial and temporal variations in cell type distributions. Jenner et al. \cite{Jenner.PLoSComputBiol.2023} used ABM to assess locoregional gemcitabine treatment efficacy in pancreatic cancer, accounting for cancer cell sensitivity, drug resistance, and drug distribution. Genderen et al. \cite{Genderen.NPJSystBiolAppl.2024}  studied prostate TME with ABM, revealing spatial constraints on tumor growth and immune regulation.

ABMs have increasingly integrated machine learning, statistical techniques, and multi-modal imaging to enhance quantitative analyses of tumor-immune interactions. Cess et al. \cite{Cess.PLoSComputBiol.2020} combined ABM with neural networks to create a multi-scale model examining how macrophage-based immunotherapies may alter immune responses. Bull et al. \cite{Bull.PLoSComputBiol.2023}  employed spatial autocorrelation and clustering methods to analyze ABM-generated data, quantifying spatial and phenotypic heterogeneity in simulated tumors, offering novel perspectives and approaches for comprehending the complexity and dynamics of tumor progression. Hickey et al. \cite{Hickey.CellSyst.2024} integrated multi-modal imaging with multi-scale modeling, capturing intricate biological processes in tumors. This approach provides valuable tools for understanding tumor dynamics and enhancing cancer therapy development. 

The fusion of hybrid modeling, multi-scale modeling, and machine learning in mathematical oncology has introduced innovative approaches to tumor research \cite{West.ClinCancerRes.2019,Issa:2021aa,Stephan:2024aa}. These interdisciplinary studies have advanced tumor immunology and offer theoretical and practical foundations for developing effective immunotherapies. With ongoing research and technological advancements, tumor immunotherapy continues to evolve, promising improved treatments and hope for cancer patients. 

\section{Mathematical models of cancer therapy approaches}
\label{sec:4}

\begin{figure}[htb!]
	\centering
	\includegraphics[width=10.5cm]{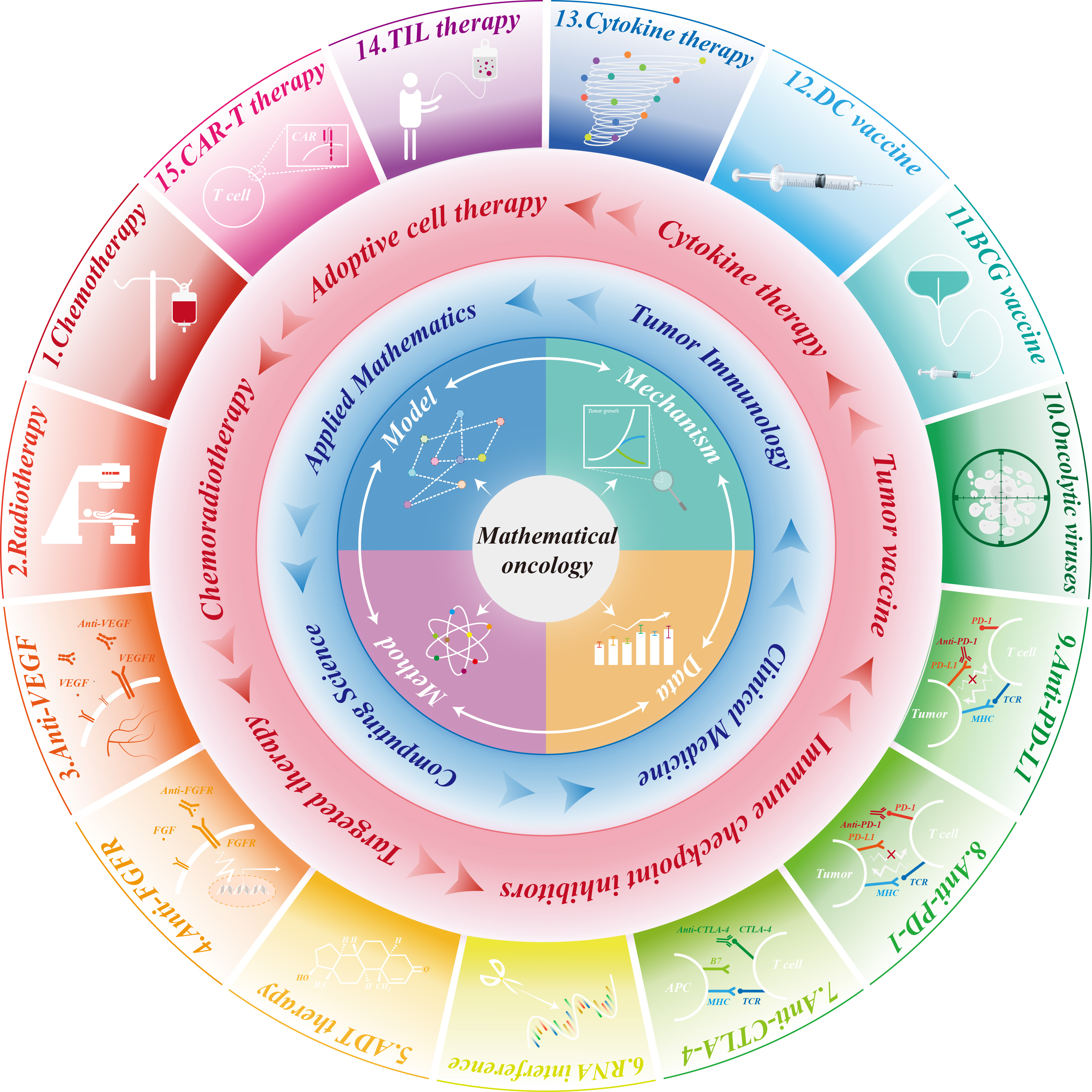}
	\caption{Mathematical models and mechanisms of cancer therapy methods.}
	\label{4}
\end{figure}

Mathematical models are invaluable in cancer research, offering theoretical frameworks to decipher cancer's complexity, forecast disease progression, and assess treatment strategies. The immune microenvironment, biological characteristics, and treatment approaches vary significantly across cancer types. Table \ref{Tab1} presents the most common cancer types and the corresponding mathematical modeling methods. Figure \ref{4} illustrates six primary categories encompassing 15 prominent cancer treatment modalities. The following section provides a concise overview of the biological mechanisms and mathematical models underlying various cancer therapies. 

\begin{tiny}
	\renewcommand{\arraystretch}{1.6}
	\begin{longtable}{c|c|c|c}
		\caption{Mathematical oncology models of various cancer types} \\
		\hline
		\textbf{Type} & \textbf{Research} & \textbf{Model}  & \textbf{Treatment method} \\ 
		\hline
		\endfirsthead
		
		\multicolumn{0}{c}{continued table} \\
		\hline
		\textbf{Type} & \textbf{Research} & \textbf{Model}  & \textbf{Treatment method}  \\
		\hline
		\endhead
		
		\endfoot
		
		\hline
		\endlastfoot
		
		\multirow{3}*{Leukemia} & Moore et al. \cite{Moore.JTheorBiol.2004}  & ODE & --  \\ \cline{2-4}
		
		& Lai et al. \cite{Lai.NPJSystBiolAppl.2024}  & SDE & Tyrosine kinase inhibitor  \\ \cline{2-4}
		
		& Zhang et al. \cite{Zhang.ComputSystOncol.2021} & IDE & CAR-T therapy  \\ \cline{1-4}

		\multirow{6}*{Brain cancer} & Kogan et al. \cite{Kogan.SIAMJApplMath.2010}  & ODE & T cell infusion therapy  \\ \cline{2-4}
		
		& Sun et al. \cite{Sun.BMCBioinformatics.2012} & ABM& Tyrosine kinase inhibitors   \\ \cline{2-4}
		
		& Khajanchi et al. \cite{Khajanchi.MathBiosci.2017}  & ODE & Immunotherapy  \\ \cline{2-4}
		
		& Khajanchi et al. \cite{Khajanchi.MathBiosci.2018}  & DDE &  Immunotherapy   \\ \cline{2-4}
		
		& Liang et al. \cite{Liang.BMCBioinformatics.2019} & ABM &  anti-EGFR + anti+VEGFR  \\ \cline{2-4}
		
		& Anderson et al. \cite{Anderson.JMathBiol.2023}  &ODE & --  \\ \cline{1-4}
			
		\multirow{4}*{Bladder cancer} & Bunimovich-Mendrazitsky et al. \cite{Bunimovich-Mendrazitsky.BullMathBiol.2007}  & ODE & BCG vaccine  \\ \cline{2-4}
		
		& Bunimovich-Mendrazitsky et al. \cite{Bunimovich-Mendrazitsky.JTheorBiol.2011}  & ODE & BCG vaccine + IL-2 treatment   \\ \cline{2-4}
		
		& Okuneye et al. \cite{Okuneye.ComputSystOncol.2021}  & ODE & Anti-FGFR + Immune checkpoint inhibitor   \\   \cline{2-4}
		
		& Li et al. \cite{Li.BullMathBiol.2024}  & ODE & Anti-FGFR + Immune checkpoint inhibitor   \\ \cline{1-4}
		
		\multirow{8}*{Melanoma} & Lai et al. \cite{Lai.BMCSystBiol.2017}  & PDE & BRAF inhibitor + Immune checkpoint inhibitor  \\ \cline{2-4}
		
		& Tsur et al. \cite{Tsur.JTheorBiol.2020}  & ODE & Immune checkpoint inhibitor  \\ \cline{2-4}
		
		& Friedman et al. \cite{Friedman.BullMathBiol.2020}  & PDE & BRAF inhibitor + Immune checkpoint inhibitor  \\ \cline{2-4}
		
		& Dickman et al. \cite{Dickman.SIAMJApplMath.2020}  & DDE & DC vaccine  \\ \cline{2-4}
		
		& Liao et al. \cite{Liao.MathBiosci.2022}  & PDE & Immune checkpoint inhibitor + IFN-$\gamma$ treatment  \\ \cline{2-4}
		
		& Milberg et al. \cite{Milberg.SciRep.2019} & QSP & Immune checkpoint inhibitor  \\ \cline{2-4}
		
		& Xue et al. \cite{Xue.JTheorBiol.2023}  & ODE & DC vaccine + Immune checkpoint inhibitor  \\ \cline{2-4} 
		
		& Ramaj et al. \cite{Ramaj.MathBiosci.2023}  & ODE & Oncolytic virotherapy  \\ \cline{1-4}

		\multirow{5}*{Prostatic cancer} & Valle et al. \cite{Valle.ApplMathModel.2021}   & ODE & Cancer vaccine + Chemotherapy \\ \cline{2-4}
		
		& Kogan et al. \cite{Kogan.CancerRes.2012}  & ODE & Immunotherapy  \\ \cline{2-4}
		
		& Ji et al. \cite{Ji.PLoSComputBiol.2019}  & ODE & --  \\ \cline{2-4}
		
		& Coletti et al. \cite{Coletti.JTheorBiol.2021}  & ODE & DC vaccine + Anti-CTLA-4  \\ \cline{2-4}
		
		& Genderen et al. \cite{Genderen.NPJSystBiolAppl.2024} & ABM & Androgen deprivation therapy  \\  \cline{1-4}
		
		\multirow{9}*{Breast cancer} & Lai et al. \cite{Lai.PNAS.2018}    & PDE & BET inhibitor + Immune checkpoint inhibitor  \\ \cline{2-4}
		
		& Szomolay et al. \cite{Szomolay.JTheorBiol.2012}  & PDE & GM-CSF treatment  \\ \cline{2-4}
		
		& Lai et al. \cite{Lai.JTheorBiol.2019}  & PDE & VEGF inhibitor + Chemotherapy  \\ \cline{2-4}
		
		& Wang et al. \cite{Wang.FrontBioengBiotechnol.2020} & QSP & Immune checkpoint inhibitor + epigenetic modulator   \\ \cline{2-4}
		
		& Wang et al. \cite{Wang.JImmunotherCancer.2021} & QSP & Chemotherapy + Immune checkpoint inhibitor  \\ \cline{2-4}
		
		& Pei et al. \cite{Pei.JTheorBiol.2023}  & ODE & RNA interference + Immune checkpoint inhibitor \\ \cline{2-4}
		
		& Mirzaei et al. \cite{Mohammad-Mirzaei.PLoSComputBiol.2022}  & ODE & --  \\ \cline{2-4}
		
		& Bitsouni et al. \cite{Bitsouni.JTheorBiol.2022}  & ODE & Anti-CD20\ ( Rituximab )  \\ \cline{2-4}
		
		& Siewe et al. \cite{Siewe.BullMathBiol.2023}  & ODE &  Anti-CD20\ ( Rituximab ) \\  \cline{1-4}
		
		\multirow{3}*{Head and neck cancer} & Smalley et al. \cite{Smalley.iScience.2020}    & ODE & Immune checkpoint inhibitor  \\ \cline{2-4}
		
		& Nazari et al.  \cite{Nazari.PLoSComputBiol.2018}  & ODE & Anti-IL-6  \\ \cline{2-4}
		
		& Pang et al. \cite{Pang.ApplMathModel.2021}  & ODE & Radiotherapy  \\  \cline{1-4}
		
		\multirow{3}*{Pancreatic cancer} & Shafiekhani et al. \cite{Shafiekhani.BMCCancer.2021}    & ODE & Anti-CD25 + Chemotherapy   \\ \cline{2-4}
		
		& Louzoun et al. \cite{Louzoun.JTheorBiol.2014}  & ODE & EGFR silencing + TGF-$\beta$ silencing  \\ \cline{2-4}
		
		& Jenner et al. \cite{Jenner.PLoSComputBiol.2023}  & ABM & Chemotherapy  \\ \cline{1-4}
		
		\multirow{4}*{Lung cancer} & Eftimie et al. \cite{Eftimie.JTheorBiol.2021}  & ODE & -- \\ \cline{2-4}
		
		& Lourenco et al. \cite{Lourenco.JBiolSyst.2023}  & ODE & -- \\ \cline{2-4}
		
		& Wang et al. \cite{Wang.NPJPrecisOncol.2023} & QSP & Immune checkpoint inhibitor  \\  \cline{2-4}
		
		& Wang et al. \cite{Wang.ClinTranslSci.2024} & QSP & Macrophage-targeted therapy + Immune checkpoint inhibitor  \\   \cline{1-4}
		
		\multirow{5}*{Colorectal cancer} & Fletcher et al. \cite{Fletcher.JTheorBiol.2012}  & ABM & -- \\ \cline{2-4}
		
		& Sameen et al. \cite{Sameen.JTheorBiol.2016}  & ODE & EGFR inhibitor + Chemotherapy  \\ \cline{2-4}
		
		& Lo et al. \cite{Lo.JTheorBiol.2013}  & ODE & --  \\ \cline{2-4}
		
		& Ma et al. \cite{Ma.JImmunotherCancer.2020} & QSP & TCE therapy + Immune checkpoint inhibitor   \\ \cline{2-4}
		
		& Mohammad-Mirzaei et al. \cite{Mohammad-Mirzaei.iScience.2023}  & PDE & --  \\  \cline{1-4}
		
		\multirow{3}*{Myeloma} & Koenders et al. \cite{Koenders.JTheorBiol.2016}  & ODE & -- \\ \cline{2-4}
		
		& Gallaher et al. \cite{Gallaher.JTheorBiol.2018}  & ODE & -- \\ \cline{2-4}
		
		& Bouchnita et al. \cite{Bouchnita.JTheorBiol.2024}  & PDE & -- \\  \cline{1-4}
		
		\multirow{1}*{Thyroid cancer} & Da et al. \cite{Da.JBiolSyst.2020}  & ODE & Radiotherapy  \\  \cline{1-4}
		
		\multirow{2}*{Liver cancer} & Delitala et al. \cite{Delitala.JTheorBiol.2012}  & ODE & Radiotherapy  \\ \cline{2-4}
		
		& Sov$\acute{e}$ et al. \cite{Sove.JImmunotherCancer.2022} & QSP & Immune checkpoint inhibitor
		
		\label{Tab1}
	\end{longtable}
\end{tiny}

\subsection{Chemotherapy and radiotherapy}

Chemotherapy, a longstanding cancer treatment, utilizes chemical agents to kill or inhibit the proliferation of cancer cells. While it is crucial in preventing cancer spread and metastasis, chemotherapy can also damage normal tissues and immune cells within the tumor environment. Recently, researchers have developed mathematical models to examine metronomic chemotherapy approaches, which involve continuous low-dose regimens \cite{Valle.ApplMathModel.2021}, as well as pulse chemotherapy, characterized by intermittent high-dose treatments \cite{Yang.MathComputSimulat.2021,Huang.ActaMathSci.2022}. With the increasing success of combination therapies, mathematical models have also explored chemotherapy in conjunction with radiotherapy \cite{Barazzuol.JTheorBiol.2010}, immunotherapy \cite{Pillis.JTheorBiol.2006,Pillis.MathBiosci.2007,Shafiekhani.BMCCancer.2021,Das.ChaosSolitonFract.2020}, or antiangiogenic therapy \cite{Lai.JTheorBiol.2019}.

Radiotherapy remains one of the most widely used cancer treatments, benefiting nearly half of all cancer patients. This approach employs high-energy radiation to damage the DNA within tumor cells, thereby inhibiting their growth and replication to achieve therapeutic goals. While few mathematical models focused on radiotherapy in the past, recent research has led to models addressing standalone radiotherapy \cite{Pang.ApplMathModel.2021}, chemoradiotherapy combinations \cite{Barazzuol.JTheorBiol.2010}, and radiotherapy paired with immunotherapy \cite{Serre.CancerRes.2016,Lai.SciChinaMath.2020}. These developments underscore the growing role of mathematical modeling as an effective tool for studying and optimizing cancer treatments. 

\subsection{Targeted therapy}

Targeted therapy is a precision-based approach in cancer treatment that disrupts specific molecular pathways essential for tumor growth and survival, contrasting with traditional chemotherapy that broadly affects both healthy and cancerous cells \cite{Labrie:2022aa,Waarts:2022aa}. This targeted inhibition of oncogenic pathways leverages unique or dysregulated proteins and genes within cancer cells, thereby improving treatment specificity. 

Angiogenesis, essential for tumor nutrient supply, is a primary target in solid tumors, driven by factors such as VEGF. Anti-VEGF therapies inhibit blood vessel formation by blocking VEGF signaling, effectively starving the tumor. Mathematical models have been instrumental in understanding VEGF dynamics, evaluating anti-VEGF efficacy, and predicting resistance patterns. For example, Liang et al. \cite{Liang.BMCBioinformatics.2019} and Hutchinson et al. \cite{Hutchinson.JTheorBiol.2016} developed multiscale models that capture VEGF signaling within the TME, simulating tumor growth inhibition through VEGF targeting. Additionally, pharmacokinetics/pharmacodynamics (PK/PD) models by He et al. \cite{He.BullMathBiol.2018} and Zheng et al. \cite{Zheng.JMathBiol.2018} predict optimal dosing and timing of anti-angiogenic therapies to improve clinical outcomes. 

Combination therapies are frequently pursued to counteract resistance associated with monotherapy. Hybrid models incorporating anti-VEGF with immunotherapies, such as checkpoint inhibitors, reveal enhanced immune cell infiltration and reduced immune evasion within the tumor \cite{Lai.BMCSystBiol.2017}. Similarly, models combining anti-VEGF with chemotherapy illustrate the modulation of tumor sensitivity to chemotherapeutics, supporting strategies that maximize synergistic effects while minimizing toxicity \cite{Lai.JTheorBiol.2019}.

Fibroblast growth factor receptor (FGFR) targeting is another avenue, especially in cancers where FGFR contributes to tumor progression and resistance. Mathematical models by Okuneye et al. \cite{Okuneye.ComputSystOncol.2021} and Li et al. \cite{Li.BullMathBiol.2024} explored the co-targeting of FGFR and VEGF pathways in bladder cancer, revealing that FGFR inhibition can mitigate resistance mechanisms against anti-VEGF therapy. Additionally, RNA interference (RNAi) therapies show promise in silencing oncogenes and resistance genes, with models developed by Arcieto et al. \cite{Arciero.DCDSB.2004} and Pei et al. \cite{Pei.JTheorBiol.2023} helping predict gene silencing impacts on tumor progression.

In hormone-dependent cancers like prostate cancer, androgen deprivation therapy (ADT) plays a crucial role. Models developed by Coletti et al. \cite{Coletti.JTheorBiol.2021} and West et al. \cite{West.ClinCancerRes.2019} have elucidated androgen receptor dynamics, illustrating feedback mechanisms leading to resistance. Such models guide adaptive ADT strategies, aiming to sustain tumor sensitivity over prolonged treatment periods. 

Beyond microenvironmental and hormonal targets, direct approaches to disrupt oncogenic drivers in cancer cells include TKIs, such as imatinib, which selectively targets the BCR-ABL fusion protein in CML \cite{Druker:2001aa,Breccia:2010aa}, and EGFR-targeting TKIs in non-small cell lung cancer, like gefitinib and erlotinib, which significantly enhance outcomes by inhibiting tumor growth pathways \cite{Johnson:2022aa}. PK/PD models for these TKIs help optimize dosing regimens to balance efficacy and minimize resistance and toxicity \cite{Lai.NPJSystBiolAppl.2024,Sun.BMCBioinformatics.2012,Rodriguez:2023aa,Bouchnita:2020aa}.

Some therapies aim to directly activate apoptotic pathways in cancer cells. BH3 mimetics, such as Venetoclax, inhibit the anti-apoptotic protein BCL-2, reactivating apoptosis in cancers like chronic lymphocytic leukemia (CLL) \cite{Perini:2018aa}. Models incorporating cell-death kinetics and pathway dynamics are used in predicting resistance and optimizing combination strategies with BH-3 mimetics \cite{Lee:2021aa,Ballesta:2013aa}.

Emerging multi-omics and patient-specific data, combined with machine learning, enhance the predictive power of mathematical models in targeted therapy \cite{Mukherjee:2024aa}. Future research is expected to integrate real-time patient data, enabling adaptive dosing and personalized treatment adjustments, with the potential to further refine therapeutic responses and combat resistance effectively.

\subsection{Immune checkpoint inhibitors}

ICIs target key molecules that regulate immune responses by inhibiting T cell activity, primarily through pathways involving CTLA-4, PD-1, and PD-L1. CTLA-4 reduces T cell activation by binding to B7 molecules on APCs, while PD-1 on T cells and PD-L1 on tumor cells interact to enable immune evasion by tumors. Blocking these immune checkpoints with ICIs enables a robust anti-tumor immune response, making a major breakthrough for patients with advanced cancers. 

Mathematical models have evolved alongside the clinical use of ICIs, enhancing understanding of T cell dynamics, tumor progression, and therapy optimization. Models for anti-CTLA-4 \cite{Wang.RSocOpenSci.2019,Lai.PNAS.2018,Milberg.SciRep.2019,Coletti.JTheorBiol.2021,Sove.JImmunotherCancer.2022,Xue.JTheorBiol.2023}, anti-PD-1 \cite{Lai.BMCSystBiol.2017,Milberg.SciRep.2019,Ma.JImmunotherCancer.2020,Smalley.iScience.2020,Liao.MathBiosci.2022,Sove.JImmunotherCancer.2022,Pei.JTheorBiol.2023,Xue.JTheorBiol.2023,Arulraj.SciAdv.2023}, and anti-PD-L1 \cite{Wang.RSocOpenSci.2019,Milberg.SciRep.2019,Lai.SciChinaMath.2020,Okuneye.ComputSystOncol.2021,Wang.NPJPrecisOncol.2023,Wang.ClinTranslSci.2024} have been widely developed, aiming to simulate the effects of ICIs on T cell proliferation and tumor rejection in solid tumors. These models assist in identifying optimal dosing schedules, assessing TME variations, and exploring resistance mechanisms, thereby providing actionable insights for improved treatment protocols. 

Recent studies have integrated ICIs with combination therapies to reflect current clinical strategies, pairing ICIs with chemotherapy \cite{Galluzzi:2020aa,Zouein:2022aa}, radiotherapy \cite{Liao:2024aa}, and anti-angiogenic agents \cite{Lee:2020aa}. Quantitative approaches using mathematical models to analyze such combinations must account for synergistic and antagonistic interactions among drugs to reflect real-world dynamics. Incorporating multiple layers of immune interactions, tumor heterogeneity, and drug effects, mathematical models of ICIs serve as a theoretical foundation for optimizing personalized ICI therapies and could significantly inform precision treatment strategies \cite{Li.BullMathBiol.2024,Li2024.04.30.591845}. 

\subsection{Adoptive cell therapy}

Adoptive cell therapy (ACT) is a promising strategy in cancer immunotherapy that uses patients' own immune cells to target and eliminate tumor cells. The primary ACT methods currently utilized in clinical settings include tumor-infiltrating lymphocyte (TIL) therapy and CAR-T therapy. In TIL therapy, lymphocytes are extracted from a patient's tumor tissue, expanded in vitro, and reintroduced to the patient. These TILs are highly specific to the tumor, allowing them to recognize and effectively eliminate cancer cells within the TME. Kogan et al. \cite{Kogan.SIAMJApplMath.2010} developed a mathematical model assessing the therapeutic impact of T-cell infusion, providing theoretical insights into outcomes for high-grade malignant gliomas. Similarly, Yang et al. \cite{Yang.MathComputSimulat.2015} explored the therapeutic potential of pulsed IL-2 administration alongside ACT, demonstrating the cytokine's ability to enhance therapeutic efficacy. 

CAR-T cell therapy, a transformative form of ACT, involves genetically modifying T cells to express chimeric antigen receptors that specifically target antigens on tumor cells. This approach has shown remarkable success in treating hematological cancers, with ongoing research expanding its potential applications to solid tumors \cite{June:2018aa}. Mathematical modeling has become instrumental in optimizing CAR-T therapy, providing insights into T cell proliferation dynamics, tumor-cell interactions, cytokine release, and patient-specific treatment protocols. Both deterministic and ABMs allow researchers to simulate CAR-T cell expansion and immune response and predict optimal dosing schedules, which also consider side effects like cytokine release syndrome \cite{Zhang.ComputSystOncol.2021,Owens.BullMathBiol.2021,Adhikarla:2024aa}. 

To address the specific challenges CAR-T cells face in solid tumors, spatial and multi-scale models have been employed to explore barriers to CAR-T cell infiltration and interactions within immunosuppressive TME. The models offer insights into immune evasion mechanisms and T cell exhaustion, both crucial for enhancing CAR-T efficacy in solid tumors \cite{Kara:2024aa,Giorgadze:2022aa,Sahoo:2020aa,Prybutok:2022aa}. In addition, advanced machine learning and neural network methods have recently been applied to analyze CAR-T cell therapy, examining correlations between CAR-T cell subtype dynamics in vivo and therapeutic outcomes \cite{Kirouac.NatBiotechnol.2023}. 

Mathematical modeling of ACT therapies has allowed researchers to simulate complex tumor-immune interactions and optimize various therapeutic parameters. These models play a critical role in identifying variables affecting treatment efficacy, such as cell dosage, cytokine support, and immune-tumor interactions. Consequently, mathematical frameworks provide a foundational basis for refining ACT strategies, enhancing their effectiveness, and broadening their applicability to diverse cancer types. 

\subsection{Tumor vaccine}

Cancer vaccines, a form of active immunotherapy, aim to activate or amplify the body's immune defenses to slow tumor progression or eradicate cancer cells. Common types include OV, Bacillus Calmette-Gu\'{e}rin (BCG) vaccines, and DC vaccines. 

OV are genetically modified to effectively infect and destroy cancer cells. Jacobsen et al. \cite{Jacobsen.BullMathBiol.2015} developed a mathematical model to explore how extracellular matrix protein CCN1 impacts OV efficacy in glioma treatment. Additionally, Kim et al. \cite{Kim.PNAS.2018} proposed a framework evaluating NK cell activity in OV and Bortezomib therapy for glioblastoma, while Ramaj and Zhou \cite{Ramaj.MathBiosci.2023} studied hypoxia's effect on OV outcomes, showing environmental factors can influence treatment success.

The BCG vaccine, derived from attenuated \textit{Mycobacterium bovis}, is widely used to prevent tuberculosis and has applications in treating non-muscle-invasive bladder cancer. Bunimovich-Mendrazitsky et al. \cite{Bunimovich-Mendrazitsky.BullMathBiol.2007,Bunimovich-Mendrazitsky.JTheorBiol.2011} modeled BCG therapy, both alone and combined with IL-2, concluding that IL-2 does not enhance BCG's anti-tumor effects in bladder cancer, highlighting the need for precise treatment combinations. 

As a promising approach in immunotherapy, DC vaccines present new avenues for cancer treatment with a potential for personalized medicine. Sardar et al. \cite{Sardar.CommunNonlinearSci.2023} examined the effects of pulsed DC vaccine therapy on immune response and tumor control, while Dickman et al. \cite{Dickman.SIAMJApplMath.2020} used a compartmental model to analyze tumor elimination, control, and escape during DC therapy for melanoma. Importantly, DC vaccines reinforce personalized treatment by targeting specific tumor antigens, increasing therapeutic accuracy. The value of combination therapies has also been explored; Coletti et al. \cite{Coletti.JTheorBiol.2021} and Xue et al. \cite{Xue.JTheorBiol.2023} investigated dual therapy with DC vaccines and immune checkpoint inhibitors, providing a theoretical basis for future preclinical trials in dual immunotherapy. 

\subsection{Cytokine inhibitor}

Cytokine inhibitors serve a vital function in cancer therapy by regulating cytokine activity to affect tumor cell growth, metastasis, and invasion. Mathematical models enable the simulation of therapeutic effects for various doses, administration times, and delivery methods, providing a scientific foundation for refining clinical treatment approaches. Wilson et al. \cite{Wilson.BullMathBiol.2012} investigated the synergy between anti-TGF-$\beta$ therapy and vaccine treatment, shedding light on combination therapies' impact on immune modulation. Yang et al. \cite{Yang.MathComputSimulat.2015} further explored the efficacy of pulsed dosing of adoptive cell therapy with IL-2 in cancer treatment, while Ratajczyk et al. \cite{Ratajczyk.MathBiosciEng.2017} developed a model combining TNF-$\alpha$ inhibitors with virotherapy, demonstrating the potential benefits of integrated approaches. Although monotherapy with cytokine inhibitors can have limited efficacy, combining them with other immunotherapies has shown synergistic effects. This integration highlights the importance of mathematical modeling in elucidating underlying biological mechanisms and optimizing treatment strategies.

\section{Discussions}

Mathematical models describing tumor-immune interactions are increasingly recognized as vital tools in understanding the complex dynamics between tumor evolution and immune response \cite{Eftimie.BullMathBiol.2011,Arabameri.MathBiosci.2018,Mahlbacher.JTheorBiol.2019}. These models provide a quantitative framework for investigating tumor-immune interactions, predicting treatment outcomes, and optimizing therapeutic strategies, paving the way for individualized precision medicine. Despite notable progress, significant challenges remain.

\textbf{Uncertainty in model parameters.} Mathematical models of tumor-immune interaction require numerous biological variables and parameters to accurately represent complex system dynamics. This complexity, however, creates challenges in parameter estimation. Experimental limitations and the scarcity of precise biological data often hinder the direct measurement of these parameters. Moreover, many of these parameters are not static; they shift dynamically with changes in the tumor and immune environment, further complicating estimation. Addressing this issue will require a stronger emphasis on experimental data collection and analysis alongside the development of more sophisticated methods for parameter analysis and estimation \cite{Marino.JTheorBiol.2008,Lillacci.PLoSComputBiol.2010,Mitra.NatCommun.2018}. Recently, specialized and efficient parameter estimation methods and tools have been proposed in computational systems biology. For example, Bayesian parameter estimation methods \cite{Linden.PLoSComputBiol.2022,Liepe.NatProtoc.2014}, Monte Carlo methods \cite{Kramer.BMCBioinformatics.2014}, optimization methods \cite{Schmiester.JMathBiol.2020}, neural networks \cite{Giampiccolo.NPJSystBiolAppl.2024}, SensSB \cite{Rodriguez.Bioinformatics.2010}, BioModels \cite{Glont.Bioinformatics.2020}, and pyPESTO \cite{Schalte.Bioinformatics.2023}.

\textbf{Discrepancies between simplified models and tumor-immune system complexity.} To reduce complexity and enhance mathematical tractability, existing models often simplify the biological landscape of tumor-immune interactions. However, excessive simplification can overlook critical complexities inherent to real-world biology. Many models, for example, focus only on tumor-T cell interactions, often neglecting the roles of other immune cells and non-cellular components (e.g., oxygen, cytokines, or chemokines) that significantly influence tumor progression. Additionally, essential biological mechanisms, such as genetic mutations, tumor heterogeneity, and plasticity, are often excluded. Advancing model accuracy will require the integration of these crucial factors to better reflect tumor-immune dynamics \cite{Masoudi-Nejad.SeminCancerBiol.2015,Meyer.CellSyst.2021,Yue.NPJSystBiolAppl.2022}. Recently, mathematical models of the interaction between tumors, immunity, and microorganisms have been proposed to explore the role of microorganisms in tumor evolution dynamics \cite{Chen.MathBiosci.2022}. Meanwhile, multi-scale models integrating molecules, cells, microenvironments, and tissues have also been developed \cite{Liang.BMCBioinformatics.2019}.

\textbf{Computational challenges in multiscale modeling.} Modeling tumor-immune interactions requires capturing processes across multiple scales, from molecular to tissue levels, necessitating complex interscale connections and imposing high computational demands. This data exchange between scales consumes substantial resources, with specific regions often requiring high-precision models or algorithms to improve accuracy. However, higher precision amplifies computational complexity and strains power resources. Additionally, solving extensive multiscale models often entails prolonged simulation times, especially for interactive or long-term scenarios, which can increase costs and reduce modeling efficiency. Addressing these challenges will require innovative modeling approaches, optimized algorithms, and advancements in data processing and storage capabilities \cite{Heath.ComputSciRev.2009,Walpole.AnnuRevBiomedEng.2013,Fletcher.WIREsMechDis.2022}. FitMultiCell has recently been developed for modeling, simulating, and parameterizing multi-scale multicellular processes \cite{Alamoudi.Bioinformatics.2023}. PhysiBoSS offers simulations for complex events across various spatial and temporal scales \cite{Letort.Bioinformatics.2019,Ponce-de-Leon.NPJSystBiolAppl.2023}. These methods aid in modeling multi-scale tumor immune systems and enhance computational performance.

Mathematical oncology integrates tumor immunology, clinical medicine, applied mathematics, and computational science, forming a powerful approach to tackling complex challenges in tumor research \cite{Gatenby.Nature.2003,Anderson.NatRevCancer.2008,Byrne.NatRevCancer.2010,Altrock.NatRevCancer.2015,Rockne.PhysBiol.2019}. As mathematical models advance, they promise greater precision, personalization, and integration with intelligent technologies. Future progression in this field will depend on multidisciplinary collaborations, allowing for the continuous evolution of mathematical approaches. Modeling tumor-immune interactions, a crucial core of this field, elucidates the immune system's role in tumor progression, dormancy, and immune evasion, informing broader models of cancer growth and treatment response. Building on recent developments, we highlight key research directions to guide future studies in the mathematical modeling of tumor-immune interactions. 

\textbf{Systematic modeling and quantitative analysis of the TME.} Systematic modeling and quantitative analysis are essential for investigating the dynamic changes within the TME \cite{Altrock.NatRevCancer.2015,Rockne.PhysBiol.2019,Cappuccio.BriefBioinform.2016,Yue.NPJSystBiolAppl.2022,Kazerouni.iScience.2020}. By developing detailed tumor-immune regulatory network models, researchers can quantitatively represent interactions between tumors and the immune system, which aids in identifying potential immune biomarkers predictive of tumor behavior. Quantitative metrics derived from these models provide theoretical foundations for understanding cancer immunoediting and classifying cancer immune phenotypes. Such metrics not only shed light on tumor-immune system evolution but also facilitate cancer subtyping. Furthermore, mathematical oncology models help reveal mechanisms behind TME-mediated drug resistance and recurrence. Ultimately, systematic modeling and quantitative analysis offer novel perspectives for cancer therapy, significantly supporting individualized treatment plans for patients.  

\textbf{Development of multiscale and multiphysics mathematical models.} Multiscale modeling enables the integration of biological processes occurring across molecular, cellular, and tissue scales, allowing mathematical models to more accurately capture the complex dynamics of tumor progression \cite{Cappuccio.BriefBioinform.2016,Rockne.PhysBiol.2019,Bull.PIEEE.2022,Masoudi-Nejad.SeminCancerBiol.2015,West.TrendsCellBiol.2023}. This capability supports the exploration of drug diffusion and distribution by simulating anticancer drug mechanisms across different biological levels, thereby improving the prediction of therapeutic outcomes. Multiphysics models further enhance this by combining different physical fields to simulate tumor behavior in various environments. For example, mechanical fields can represent pressure gradients in the TME and model cell migration, while chemical fields can depict drug distribution and metabolic processes. The integration of multiscale and multiphysics modeling in oncology provides a powerful tool for understanding and predicting tumor growth, metastasis, and response to treatments.    

\textbf{Development of mathematical models integrating multisource data.} Integrating multisource data into mathematical models offers an enriched understanding of tumor biology, immune responses, disease progression, and optimized treatment approaches \cite{Rockne.PhysBiol.2019,Wang.NPJPrecisOncol.2023,Zhao.InformationFusion.2024,Boehm.NatRevCancer.2022,Butner.NatComputSci.2022,Lorenzo.AnnuRevBiomedEng.2024}. With the establishment of extensive public cancer databases---such as SEER, TCGA, and NCDB---researchers have access to detailed clinical, biomarker, genomic, transcriptomic, and proteomic data. Translating this diverse data into formats compatible with mathematical models bridges a critical gap, enhancing model validation and addressing biases in predictive accuracy. Additionally, data-model integration enables the development of early warning systems for cancer progression. As data and models become more interoperable, this integration stands to be a major focus in advancing tumor research and predictive oncology. 
 
\textbf{Exploring the application of machine learning in mathematical oncology.} Machine learning (ML) introduces new capabilities to model optimization, parameter estimation, and cancer classification \cite{Rockne.PhysBiol.2019,Alber.NPJDigitMed.2019,Kozowska.PLoSComputBiol.2020,Metzcar.FrontImmunol.2024,Perez-Lopez.NatRevCancer.2024}. Techniques like neural networks, support vector machines, and Gaussian mixture models enhance the predictive power of mathematical models and facilitate the creation of virtual cancer cohorts. Machine learning optimization techniques also facilitate model parameter adjustments, reducing prediction errors and boosting overall model performance. Recent research has also highlighted the promise of neural networks in solving differential equations, especially physics-informed neural networks (PINNs) and neural ODE approaches, which improve both solution accuracy and model generalization. Thus, machine learning integration into mathematical oncology not only enhances model precision and efficiency but also opens up new avenues for individualized cancer research and treatment planning. 

In summary, mathematical models of tumor-immune interactions offer a robust framework for exploring tumor dynamics and informing clinical treatment strategies. While challenges remain, advancements in technology and interdisciplinary collaboration promise to elevate the role of mathematical models in tumor immunology research, promoting closer cooperation between mathematicians and immunologists to drive cross-disciplinary breakthroughs. 

%%%% Acknowledgments %%%%%%%%
\section*{Acknowledgments}

This work was funded by the National Natural Science Foundation of China (NSFC 12331018).

%%%% Bibliography  %%%%%%%%%%

%\bibliographystyle{unsrt} 
%{\footnotesize \bibliography{sn-bibliography}}

\begin{thebibliography}{100}

\bibitem{Chen.CancerCommun.2022}
Xueman Chen and Erwei Song.
\newblock The theory of tumor ecosystem.
\newblock {\em Cancer Commun (Lond)}, 42(7):587--608, Jul 2022.

\bibitem{Anderson.CurrBiol.2020}
Nicole~M Anderson and M~Celeste Simon.
\newblock The tumor microenvironment.
\newblock {\em Curr Biol}, 30(16):R921--R925, Aug 2020.

\bibitem{Visser.CancerCell.2023}
Karin~E de~Visser and Johanna~A Joyce.
\newblock {The evolving tumor microenvironment: From cancer initiation to
  metastatic outgrowth.}
\newblock {\em Cancer Cell}, 41(3):374--403, Mar 2023.

\bibitem{Gajewski.NatImmunol.2013}
Thomas~F Gajewski, Hans Schreiber, and Yang-Xin Fu.
\newblock Innate and adaptive immune cells in the tumor microenvironment.
\newblock {\em Nat Immunol}, 14(10):1014--1022, Oct 2013.

\bibitem{Hinshaw.CancerRes.2019}
Dominique~C Hinshaw and Lalita~A Shevde.
\newblock The tumor microenvironment innately modulates cancer progression.
\newblock {\em Cancer Res}, 79(18):4557--4566, Sep 2019.

\bibitem{Meads.NatRevCancer.2009}
Mark~B Meads, Robert~A Gatenby, and William~S Dalton.
\newblock {Environment-mediated drug resistance: A major contributor to minimal
  residual disease.}
\newblock {\em Nat Rev Cancer}, 9(9):665--674, Sep 2009.

\bibitem{Quail.NatMed.2013}
Daniela~F Quail and Johanna~A Joyce.
\newblock Microenvironmental regulation of tumor progression and metastasis.
\newblock {\em Nat Med}, 19(11):1423--1437, Nov 2013.

\bibitem{Rabinovich.AnnuRevImmunol.2007}
Gabriel~A Rabinovich, Dmitry Gabrilovich, and Eduardo~M Sotomayor.
\newblock Immunosuppressive strategies that are mediated by tumor cells.
\newblock {\em Annu Rev Immunol}, 25:267--296, Apr 2007.

\bibitem{Joyce.Science.2015}
Johanna~A Joyce and Douglas~T Fearon.
\newblock T cell exclusion, immune privilege, and the tumor microenvironment.
\newblock {\em Science}, 348(6230):74--80, Apr 2015.

\bibitem{Ozga.Immunity.2021}
Aleksandra~J Ozga, Melvyn~T Chow, and Andrew~D Luster.
\newblock Chemokines and the immune response to cancer.
\newblock {\em Immunity}, 54(5):859--874, May 2021.

\bibitem{Nagarsheth.NatRevImmunol.2017}
Nisha Nagarsheth, Max~S Wicha, and Weiping Zou.
\newblock Chemokines in the cancer microenvironment and their relevance in
  cancer immunotherapy.
\newblock {\em Nat Rev Immunol}, 17(9):559--572, Sep 2017.

\bibitem{Prendergast.CancerRes.2007}
George~C Prendergast and Elizabeth~M Jaffee.
\newblock {Cancer immunologists and cancer biologists: Why we didn't talk then
  but need to now.}
\newblock {\em Cancer Res}, 67(8):3500--3504, Apr 2007.

\bibitem{Galon.Immunity.2020}
J{\'e}r{\^o}me Galon and Daniela Bruni.
\newblock {Tumor immunology and tumor evolution: Intertwined histories.}
\newblock {\em Immunity}, 52(1):55--81, Jan 2020.

\bibitem{Dunn.NatImmunol.2002}
Gavin~P Dunn, Allen~T Bruce, Hiroaki Ikeda, Lloyd~J Old, and Robert~D
  Schreiber.
\newblock {Cancer immunoediting: From immunosurveillance to tumor escape.}
\newblock {\em Nat Immunol}, 3(11):991--998, Nov 2002.

\bibitem{Dunn.AnnuRevImmunol.2004}
Gavin~P Dunn, Lloyd~J Old, and Robert~D Schreiber.
\newblock {The three Es of cancer immunoediting.}
\newblock {\em Annu Rev Immunol}, 22:329--360, Apr 2004.

\bibitem{Dunn.Immunity.2004}
Gavin~P Dunn, Lloyd~J Old, and Robert~D Schreiber.
\newblock The immunobiology of cancer immunosurveillance and immunoediting.
\newblock {\em Immunity}, 21(2):137--148, Aug 2004.

\bibitem{Schreiber.Science.2011}
Robert~D Schreiber, Lloyd~J Old, and Mark~J Smyth.
\newblock {Cancer immunoediting: Integrating immunity's roles in cancer
  suppression and promotion.}
\newblock {\em Science}, 331(6024):1565--1570, Mar 2011.

\bibitem{Ren.AnnuRevImmunol.2021}
Xianwen Ren, Lei Zhang, Yuanyuan Zhang, Ziyi Li, Nathan Siemers, and Zemin
  Zhang.
\newblock Insights gained from single-cell analysis of immune cells in the
  tumor microenvironment.
\newblock {\em Annu Rev Immunol}, 39:583--609, Apr 2021.

\bibitem{Xu.SignalTransductTargetTher.2021}
Ying Xu, Guan-Hua Su, Ding Ma, Yi~Xiao, Zhi-Ming Shao, and Yi-Zhou Jiang.
\newblock {Technological advances in cancer immunity: From immunogenomics to
  single-cell analysis and artificial intelligence.}
\newblock {\em Signal Transduct Target Ther}, 6(1):312, Aug 2021.

\bibitem{Kashyap.TrendsBiotechnol.2022}
Aditya Kashyap, Maria~Anna Rapsomaniki, Vesna Barros, Anna
  Fomitcheva-Khartchenko, Adriano~Luca Martinelli, Antonio~Foncubierta
  Rodriguez, Maria Gabrani, Michal Rosen-Zvi, and Govind Kaigala.
\newblock {Quantification of tumor heterogeneity: From data acquisition to
  metric generation.}
\newblock {\em Trends Biotechnol}, 40(6):647--676, Jun 2022.

\bibitem{Gatenby.Nature.2003}
Robert~A Gatenby and Philip~K Maini.
\newblock {Mathematical oncology: Cancer summed up.}
\newblock {\em Nature}, 421(6921):321, Jan 2003.

\bibitem{Anderson.NatRevCancer.2008}
Alexander R~A Anderson and Vito Quaranta.
\newblock Integrative mathematical oncology.
\newblock {\em Nat Rev Cancer}, 8(3):227--234, Mar 2008.

\bibitem{Byrne.NatRevCancer.2010}
Helen~M Byrne.
\newblock {Dissecting cancer through mathematics: From the cell to the animal
  model.}
\newblock {\em Nat Rev Cancer}, 10(3):221--230, Mar 2010.

\bibitem{Altrock.NatRevCancer.2015}
Philipp~M Altrock, Lin~L Liu, and Franziska Michor.
\newblock {The mathematics of cancer: Integrating quantitative models.}
\newblock {\em Nat Rev Cancer}, 15(12):730--745, Dec 2015.

\bibitem{Rockne.PhysBiol.2019}
Russell~C Rockne, Andrea Hawkins-Daarud, Kristin~R Swanson, James~P Sluka,
  James~A Glazier, Paul Macklin, David~A Hormuth, Angela~M Jarrett, Ernesto A
  B~F Lima, J~Tinsley~Oden, George Biros, Thomas~E Yankeelov, Kit Curtius,
  Ibrahim Al~Bakir, Dominik Wodarz, Natalia Komarova, Luis Aparicio, Mykola
  Bordyuh, Raul Rabadan, Stacey~D Finley, Heiko Enderling, Jimmy Caudell,
  Eduardo~G Moros, Alexander R~A Anderson, Robert~A Gatenby, Artem Kaznatcheev,
  Peter Jeavons, Nikhil Krishnan, Julia Pelesko, Raoul~R Wadhwa, Nara Yoon,
  Daniel Nichol, Andriy Marusyk, Michael Hinczewski, and Jacob~G Scott.
\newblock The 2019 mathematical oncology roadmap.
\newblock {\em Phys Biol}, 16(4):041005, Jun 2019.

\bibitem{Michor.Cell.2015}
Franziska Michor and Kathryn Beal.
\newblock {Improving cancer treatment via mathematical modeling: Surmounting
  the challenges is worth the effort.}
\newblock {\em Cell}, 163(5):1059--1063, Nov 2015.

\bibitem{Konstorum.JRSocInterface.2017}
Anna Konstorum, Anthony~T Vella, Adam~J Adler, and Reinhard~C Laubenbacher.
\newblock {Addressing current challenges in cancer immunotherapy with
  mathematical and computational modelling.}
\newblock {\em J R Soc Interface}, 14(131), Jun 2017.

\bibitem{Clarke.NatRevCancer.2020}
Matthew~A Clarke and Jasmin Fisher.
\newblock {Executable cancer models: Successes and challenges.}
\newblock {\em Nat Rev Cancer}, 20(6):343--354, Jun 2020.

\bibitem{Butner.NatComputSci.2022}
Joseph~D Butner, Prashant Dogra, Caroline Chung, Renata Pasqualini, Wadih Arap,
  John Lowengrub, Vittorio Cristini, and Zhihui Wang.
\newblock Mathematical modeling of cancer immunotherapy for personalized
  clinical translation.
\newblock {\em Nat Comput Sci}, 2(12):785--796, Dec 2022.

\bibitem{Hanahan.Cell.2000}
D~Hanahan and R~A Weinberg.
\newblock The hallmarks of cancer.
\newblock {\em Cell}, 100(1):57--70, Jan 2000.

\bibitem{Hanahan.Cell.2011}
Douglas Hanahan and Robert~A Weinberg.
\newblock {Hallmarks of cancer: The next generation.}
\newblock {\em Cell}, 144(5):646--674, Mar 2011.

\bibitem{Hanahan.CancerDiscov.2022}
Douglas Hanahan.
\newblock {Hallmarks of cancer: New dimensions.}
\newblock {\em Cancer Discov}, 12(1):31--46, Jan 2022.

\bibitem{Bull.PIEEE.2022}
Joshua~Adam Bull and Helen~Mary Byrne.
\newblock The hallmarks of mathematical oncology.
\newblock {\em Proceedings of the IEEE}, 110(5):523--540, Feb 2022.

\bibitem{Grivennikov.Cell.2010}
Sergei~I Grivennikov, Florian~R Greten, and Michael Karin.
\newblock Immunity, inflammation, and cancer.
\newblock {\em Cell}, 140(6):883--899, Mar 2010.

\bibitem{Yang.SignalTransductTargetTher.2023}
Li~Yang, Aitian Li, Ying Wang, and Yi~Zhang.
\newblock {Intratumoral microbiota: Roles in cancer initiation, development and
  therapeutic efficacy.}
\newblock {\em Signal Transduct Target Ther}, 8(1):35, Jan 2023.

\bibitem{Parkin.Lancet.2001}
J~Parkin and B~Cohen.
\newblock An overview of the immune system.
\newblock {\em Lancet}, 357(9270):1777--1789, Jun 2001.

\bibitem{Delves.NEnglJMed.2000}
P~J Delves and I~M Roitt.
\newblock {The immune system. First of two parts.}
\newblock {\em N Engl J Med}, 343(1):37--49, Jul 2000.

\bibitem{Wculek.NatRevImmunol.2020}
Stefanie~K Wculek, Francisco~J Cueto, Adriana~M Mujal, Ignacio Melero,
  Matthew~F Krummel, and David Sancho.
\newblock Dendritic cells in cancer immunology and immunotherapy.
\newblock {\em Nat Rev Immunol}, 20(1):7--24, Jan 2020.

\bibitem{Zhu.Blood.2008}
Jinfang Zhu and William~E Paul.
\newblock {CD4 T cells: Fates, functions, and faults.}
\newblock {\em Blood}, 112(5):1557--1569, Sep 2008.

\bibitem{Zhou.Immunity.2009}
Liang Zhou, Mark M~W Chong, and Dan~R Littman.
\newblock {Plasticity of CD4+ T cell lineage differentiation.}
\newblock {\em Immunity}, 30(5):646--655, May 2009.

\bibitem{OShea.Science.2010}
John~J O'Shea and William~E Paul.
\newblock {Mechanisms underlying lineage commitment and plasticity of helper
  CD4+ T cells.}
\newblock {\em Science}, 327(5969):1098--1102, Feb 2010.

\bibitem{Liew.NatRevImmunol.2002}
Foo~Y Liew.
\newblock {T(H)1 and T(H)2 cells: A historical perspective.}
\newblock {\em Nat Rev Immunol}, 2(1):55--60, Jan 2002.

\bibitem{Stadhouders.JAutoimmun.2018}
Ralph Stadhouders, Erik Lubberts, and Rudi~W Hendriks.
\newblock {A cellular and molecular view of T helper 17 cell plasticity in
  autoimmunity.}
\newblock {\em J Autoimmun}, 87:1--15, Feb 2018.

\bibitem{Crotty.AnnuRevImmunol.2011}
Shane Crotty.
\newblock {Follicular helper CD4 T cells (TFH).}
\newblock {\em Annu Rev Immunol}, 29:621--663, Apr 2011.

\bibitem{Beyer.Blood.2006}
Marc Beyer and Joachim~L Schultze.
\newblock Regulatory {T} cells in cancer.
\newblock {\em Blood}, 108(3):804--811, Aug 2006.

\bibitem{Zou.NatRevImmunol.2006}
Weiping Zou.
\newblock Regulatory {T} cells, tumour immunity and immunotherapy.
\newblock {\em Nat Rev Immunol}, 6(4):295--307, Apr 2006.

\bibitem{Barry.NatRevImmunol.2002}
Michele Barry and R~Chris Bleackley.
\newblock {Cytotoxic T lymphocytes: All roads lead to death.}
\newblock {\em Nat Rev Immunol}, 2(6):401--409, Jun 2002.

\bibitem{Wherry.NatImmunol.2011}
E~John Wherry.
\newblock T cell exhaustion.
\newblock {\em Nat Immunol}, 12(6):492--499, Jun 2011.

\bibitem{McLane.AnnuRevImmunol.2019}
Laura~M McLane, Mohamed~S Abdel-Hakeem, and E~John Wherry.
\newblock {CD8 T cell exhaustion during chronic viral Infection and cancer.}
\newblock {\em Annu Rev Immunol}, 37:457--495, Apr 2019.

\bibitem{Sharonov.NatRevImmunol.2020}
George~V Sharonov, Ekaterina~O Serebrovskaya, Diana~V Yuzhakova, Olga~V
  Britanova, and Dmitriy~M Chudakov.
\newblock B cells, plasma cells and antibody repertoires in the tumour
  microenvironment.
\newblock {\em Nat Rev Immunol}, 20(5):294--307, May 2020.

\bibitem{Schumacher.Science.2022}
Ton~N Schumacher and Daniela~S Thommen.
\newblock Tertiary lymphoid structures in cancer.
\newblock {\em Science}, 375(6576):eabf9419, Jan 2022.

\bibitem{Fridman.Immunity.2023}
Wolf~H Fridman, Maxime Meylan, Guilhem Pupier, Anne Calvez, Isa{\"\i}as
  Hernandez, and Catherine Saut{\`e}s-Fridman.
\newblock {Tertiary lymphoid structures and B cells: An intratumoral immunity
  cycle.}
\newblock {\em Immunity}, 56(10):2254--2269, Oct 2023.

\bibitem{Chiossone.NatRevImmunol.2018}
Laura Chiossone, Pierre-Yves Dumas, Margaux Vienne, and Eric Vivier.
\newblock Natural killer cells and other innate lymphoid cells in cancer.
\newblock {\em Nat Rev Immunol}, 18(11):671--688, Nov 2018.

\bibitem{Eisenbarth.NatRevImmunol.2019}
S~C Eisenbarth.
\newblock {Dendritic cell subsets in T cell programming: Location dictates
  function.}
\newblock {\em Nat Rev Immunol}, 19(2):89--103, Feb 2019.

\bibitem{Biswas.NatImmunol.2010}
Subhra~K Biswas and Alberto Mantovani.
\newblock {Macrophage plasticity and interaction with lymphocyte subsets:
  Cancer as a paradigm.}
\newblock {\em Nat Immunol}, 11(10):889--896, Oct 2010.

\bibitem{Galli.NatImmunol.2011}
Stephen~J Galli, Niels Borregaard, and Thomas~A Wynn.
\newblock {Phenotypic and functional plasticity of cells of innate immunity:
  Macrophages, mast cells and neutrophils.}
\newblock {\em Nat Immunol}, 12(11):1035--1044, Oct 2011.

\bibitem{Noy.Immunity.2014}
Roy Noy and Jeffrey~W Pollard.
\newblock {Tumor-associated macrophages: From mechanisms to therapy.}
\newblock {\em Immunity}, 41(1):49--61, Jul 2014.

\bibitem{Basak.FrontImmunol.2023}
Udit Basak, Tania Sarkar, Sumon Mukherjee, Sourio Chakraborty, Apratim Dutta,
  Saikat Dutta, Debadatta Nayak, Subhash Kaushik, Tanya Das, and Gaurisankar
  Sa.
\newblock {Tumor-associated macrophages: An effective player of the tumor
  microenvironment.}
\newblock {\em Front Immunol}, 14:1295257, Nov 2023.

\bibitem{Giese.Blood.2019}
Morgan~A Giese, Laurel~E Hind, and Anna Huttenlocher.
\newblock Neutrophil plasticity in the tumor microenvironment.
\newblock {\em Blood}, 133(20):2159--2167, May 2019.

\bibitem{Shaul.NatRevClinOncol.2019}
Merav~E Shaul and Zvi~G Fridlender.
\newblock Tumour-associated neutrophils in patients with cancer.
\newblock {\em Nat Rev Clin Oncol}, 16(10):601--620, Oct 2019.

\bibitem{Sansores-Espana.IntJMolSci.2022}
Luis~Daniel Sansores-Espa{\~n}a, Samanta Melgar-Rodr{\'\i}guez, Rolando Vernal,
  Bertha~Arelly Carrillo-{\'A}vila, V{\'\i}ctor~Manuel Mart{\'\i}nez-Aguilar,
  and Jaime D{\'\i}az-Z{\'u}{\~n}iga.
\newblock {Neutrophil N1 and N2 subsets and their possible association with
  periodontitis: A scoping review.}
\newblock {\em Int J Mol Sci}, 23(20), Oct 2022.

\bibitem{Hegde.Immunity.2021}
Samarth Hegde, Andrew~M Leader, and Miriam Merad.
\newblock {MDSC: Markers, development, states, and unaddressed complexity.}
\newblock {\em Immunity}, 54(5):875--884, May 2021.

\bibitem{Groth.BrJCancer.2019}
Christopher Groth, Xiaoying Hu, Rebekka Weber, Viktor Fleming, Peter Altevogt,
  Jochen Utikal, and Viktor Umansky.
\newblock Immunosuppression mediated by myeloid-derived suppressor cells
  ({MDSCs}) during tumour progression.
\newblock {\em Br J Cancer}, 120(1):16--25, Jan 2019.

\bibitem{Li.SignalTransductTargetTher.2021}
Kai Li, Houhui Shi, Benxia Zhang, Xuejin Ou, Qizhi Ma, Yue Chen, Pei Shu, Dan
  Li, and Yongsheng Wang.
\newblock Myeloid-derived suppressor cells as immunosuppressive regulators and
  therapeutic targets in cancer.
\newblock {\em Signal Transduct Target Ther}, 6(1):362, Oct 2021.

\bibitem{Zitvogel.NatRevImmunol.2006}
Laurence Zitvogel, Antoine Tesniere, and Guido Kroemer.
\newblock {Cancer despite immunosurveillance: Immunoselection and
  immunosubversion.}
\newblock {\em Nat Rev Immunol}, 6(10):715--727, Oct 2006.

\bibitem{Kareva.FrontImmunol.2021}
Irina Kareva, Kimberly~A Luddy, Cliona O'Farrelly, Robert~A Gatenby, and Joel~S
  Brown.
\newblock {Predator-Prey in tumor-immune interactions: A wrong model or just an
  incomplete one?}
\newblock {\em Front Immunol}, 12:668221, Aug 2021.

\bibitem{Gubin.ClinCancerRes.2022}
Matthew~M Gubin and Matthew~D Vesely.
\newblock Cancer immunoediting in the era of immuno-oncology.
\newblock {\em Clin Cancer Res}, 28(18):3917--3928, Sep 2022.

\bibitem{Aguirre-Ghiso.NatRevCancer.2007}
Julio~A Aguirre-Ghiso.
\newblock Models, mechanisms and clinical evidence for cancer dormancy.
\newblock {\em Nat Rev Cancer}, 7(11):834--846, Nov 2007.

\bibitem{Manjili.CancerRes.2017}
Masoud~H Manjili.
\newblock {Tumor dormancy and relapse: From a natural byproduct of evolution to
  a disease state.}
\newblock {\em Cancer Res}, 77(10):2564--2569, May 2017.

\bibitem{Santos-de-Frutos.CommunBiol.2021}
Karla Santos-de Frutos and Nabil Djouder.
\newblock When dormancy fuels tumour relapse.
\newblock {\em Commun Biol}, 4(1):747, Jun 2021.

\bibitem{Chen.Immunity.2013}
Daniel~S Chen and Ira Mellman.
\newblock {Oncology meets immunology: The cancer-immunity cycle.}
\newblock {\em Immunity}, 39(1):1--10, Jul 2013.

\bibitem{Mellman.Immunity.2023}
Ira Mellman, Daniel~S Chen, Thomas Powles, and Shannon~J Turley.
\newblock {The cancer-immunity cycle: Indication, genotype, and immunotype.}
\newblock {\em Immunity}, 56(10):2188--2205, Oct 2023.

\bibitem{Lewis.IntRevImmunol.2003}
Jennifer~D Lewis, Brian~D Reilly, and Robert~K Bright.
\newblock {Tumor-associated antigens: From discovery to immunity.}
\newblock {\em Int Rev Immunol}, 22(2):81--112, Mar-Apr 2003.

\bibitem{Duan.TrendsCancer.2020}
Qianqian Duan, Hualing Zhang, Junnian Zheng, and Lianjun Zhang.
\newblock {Turning cold into hot: Firing up the tumor microenvironment.}
\newblock {\em Trends Cancer}, 6(7):605--618, Jul 2020.

\bibitem{Liu.Theranostics.2021}
Yuan-Tong Liu and Zhi-Jun Sun.
\newblock Turning cold tumors into hot tumors by improving {T}-cell
  infiltration.
\newblock {\em Theranostics}, 11(11):5365--5386, Mar 2021.

\bibitem{Zhang.TrendsImmunol.2022}
Jiahui Zhang, Di~Huang, Phei~Er Saw, and Erwei Song.
\newblock {Turning cold tumors hot: From molecular mechanisms to clinical
  applications.}
\newblock {\em Trends Immunol}, 43(7):523--545, Jul 2022.

\bibitem{Chen.Nature.2017}
Daniel~S Chen and Ira Mellman.
\newblock Elements of cancer immunity and the cancer-immune set point.
\newblock {\em Nature}, 541(7637):321--330, Jan 2017.

\bibitem{Anandappa.CancerDiscov.2020}
Annabelle~J Anandappa, Catherine~J Wu, and Patrick~A Ott.
\newblock {Directing traffic: How to effectively drive T cells into tumors.}
\newblock {\em Cancer Discov}, 10(2):185--197, Feb 2020.

\bibitem{Khosravi.CancerCommunLond.2024}
Gholam-Reza Khosravi, Samaneh Mostafavi, Sanaz Bastan, Narges Ebrahimi,
  Roya~Safari Gharibvand, and Nahid Eskandari.
\newblock Immunologic tumor microenvironment modulators for turning cold tumors
  hot.
\newblock {\em Cancer Commun (Lond)}, 44(5):521--553, May 2024.

\bibitem{Eftimie.BullMathBiol.2011}
Raluca Eftimie, Jonathan~L Bramson, and David J~D Earn.
\newblock Interactions between the immune system and cancer: {A} brief review
  of non-spatial mathematical models.
\newblock {\em Bull Math Biol}, 73(1):2--32, Jan 2011.

\bibitem{Arabameri.MathBiosci.2018}
Abazar Arabameri, Davud Asemani, and Jamshid Hadjati.
\newblock A structural methodology for modeling immune-tumor interactions
  including pro- and anti-tumor factors for clinical applications.
\newblock {\em Math Biosci}, 304:48--61, Oct 2018.

\bibitem{Mahlbacher.JTheorBiol.2019}
Grace~E Mahlbacher, Kara~C Reihmer, and Hermann~B Frieboes.
\newblock Mathematical modeling of tumor-immune cell interactions.
\newblock {\em J Theor Biol}, 469:47--60, May 2019.

\bibitem{Gerlee.CancerRes.2013}
Philip Gerlee.
\newblock {The model muddle: In search of tumor growth laws.}
\newblock {\em Cancer Res}, 73(8):2407--2411, Apr 2013.

\bibitem{Sarapata.BullMathBiol.2014}
E~A Sarapata and L~G de~Pillis.
\newblock A comparison and catalog of intrinsic tumor growth models.
\newblock {\em Bull Math Biol}, 76(8):2010--2024, Aug 2014.

\bibitem{Benzekry.PLoSComputBiol.2014}
S{\'e}bastien Benzekry, Clare Lamont, Afshin Beheshti, Amanda Tracz, John M~L
  Ebos, Lynn Hlatky, and Philip Hahnfeldt.
\newblock Classical mathematical models for description and prediction of
  experimental tumor growth.
\newblock {\em PLoS Comput Biol}, 10(8):e1003800, Aug 2014.

\bibitem{Talkington.BullMathBiol.2015}
Anne Talkington and Rick Durrett.
\newblock Estimating tumor growth rates in vivo.
\newblock {\em Bull Math Biol}, 77(10):1934--1954, Oct 2015.

\bibitem{Lei.JTheorBiol.2020}
Jinzhi Lei.
\newblock A general mathematical framework for understanding the behavior of
  heterogeneous stem cell regeneration.
\newblock {\em J Theor Biol}, 492:110196, May 2020.

\bibitem{West.ClinCancerRes.2019}
Jeffrey~B West, Mina~N Dinh, Joel~S Brown, Jingsong Zhang, Alexander~R
  Anderson, and Robert~A Gatenby.
\newblock {Multidrug Cancer Therapy in Metastatic Castrate-Resistant Prostate
  Cancer: {An} Evolution-Based Strategy.}
\newblock {\em Clin Cancer Res}, 25(14):4413--4421, Jul 2019.

\bibitem{Bernard:2003aa}
Samuel Bernard, Jacques B{\'e}lair, and Michael~C Mackey.
\newblock Oscillations in cyclical neutropenia: new evidence based on
  mathematical modeling.
\newblock {\em J Theor Biol}, 223(3):283--298, Aug 2003.

\bibitem{Norton.Nature.1976}
L~Norton, R~Simon, H~D Brereton, and A~E Bogden.
\newblock Predicting the course of gompertzian growth.
\newblock {\em Nature}, 264(5586):542--545, Dec 1976.

\bibitem{Hahnfeldt.CancerRes.1999}
P~Hahnfeldt, D~Panigrahy, J~Folkman, and L~Hlatky.
\newblock {Tumor development under angiogenic signaling: A dynamical theory of
  tumor growth, treatment response, and postvascular dormancy.}
\newblock {\em Cancer Res}, 59(19):4770--4775, Oct 1999.

\bibitem{West.PNAS.2019}
Jeffrey West and Paul~K Newton.
\newblock Cellular interactions constrain tumor growth.
\newblock {\em Proc Natl Acad Sci USA}, 116(6):1918--1923, Feb 2019.

\bibitem{Kuznetsov.BullMathBiol.1994}
V~A Kuznetsov, I~A Makalkin, M~A Taylor, and A~S Perelson.
\newblock {Nonlinear dynamics of immunogenic tumors: Parameter estimation and
  global bifurcation analysis.}
\newblock {\em Bull Math Biol}, 56(2):295--321, Mar 1994.

\bibitem{Kirschner.JMathBiol.1998}
D~Kirschner and J~C Panetta.
\newblock Modeling immunotherapy of the tumor-immune interaction.
\newblock {\em J Math Biol}, 37(3):235--252, Sep 1998.

\bibitem{Wei.IntJBifurcatChaos.2013}
Hsiu-Chuan Wei and Jenn-Tsann Lin.
\newblock Periodically pulsed immunotherapy in a mathematical model of
  tumor-immune interaction.
\newblock {\em Int J Bifurcat Chaos}, 23(04):1350068, 2013.

\bibitem{Arciero.DCDSB.2004}
JC~Arciero, TL~Jackson, and DE~Kirschner.
\newblock A mathematical model of tumor-immune evasion and {siRNA} treatment.
\newblock {\em Discrete and Continuous Dynamical Systems-B}, 4(1):39--58, Feb
  2004.

\bibitem{Pillis.CancerRes.2005}
Lisette~G de~Pillis, Ami~E Radunskaya, and Charles~L Wiseman.
\newblock A validated mathematical model of cell-mediated immune response to
  tumor growth.
\newblock {\em Cancer Res}, 65(17):7950--7958, Sep 2005.

\bibitem{Pillis.JTheorBiol.2006}
L~G de~Pillis, W~Gu, and A~E Radunskaya.
\newblock {Mixed immunotherapy and chemotherapy of tumors: Modeling,
  applications and biological interpretations.}
\newblock {\em J Theor Biol}, 238(4):841--862, Feb 2006.

\bibitem{Pillis.MathBiosci.2007}
L~G de~Pillis, W~Gu, K~R Fister, T~Head, K~Maples, A~Murugan, T~Neal, and
  K~Yoshida.
\newblock {Chemotherapy for tumors: An analysis of the dynamics and a study of
  quadratic and linear optimal controls.}
\newblock {\em Math Biosci}, 209(1):292--315, Sep 2007.

\bibitem{Castiglione.BullMathBiol.2006}
Filippo Castiglione and Benedetto Piccoli.
\newblock Optimal control in a model of dendritic cell transfection cancer
  immunotherapy.
\newblock {\em Bull Math Biol}, 68(2):255--274, Feb 2006.

\bibitem{Castiglione.JTheorBiol.2007}
F~Castiglione and B~Piccoli.
\newblock Cancer immunotherapy, mathematical modeling and optimal control.
\newblock {\em J Theor Biol}, 247(4):723--732, Aug 2007.

\bibitem{Leon.JTheorBiol.2007}
Kalet Leon, Karina Garcia, Jorge Carneiro, and Agustin Lage.
\newblock How regulatory {CD25(+)CD4(+) T} cells impinge on tumor
  immunobiology? {On} the existence of two alternative dynamical classes of
  tumors.
\newblock {\em J Theor Biol}, 247(1):122--137, Jul 2007.

\bibitem{Robertson-Tessi.JTheorBiol.2012}
Mark Robertson-Tessi, Ardith El-Kareh, and Alain Goriely.
\newblock A mathematical model of tumor-immune interactions.
\newblock {\em J Theor Biol}, 294:56--73, Feb 2012.

\bibitem{Robertson-Tessi.JTheorBiol.2015}
Mark Robertson-Tessi, Ardith El-Kareh, and Alain Goriely.
\newblock A model for effects of adaptive immunity on tumor response to
  chemotherapy and chemoimmunotherapy.
\newblock {\em J Theor Biol}, 380:569--584, Sep 2015.

\bibitem{Soto-Ortiz.JTheorBiol.2016}
Luis Soto-Ortiz and Stacey~D Finley.
\newblock A cancer treatment based on synergy between anti-angiogenic and
  immune cell therapies.
\newblock {\em J Theor Biol}, 394:197--211, Apr 2016.

\bibitem{Breems.JTheorBiol.2016}
Nicoline~Y den Breems and Raluca Eftimie.
\newblock The re-polarisation of {M2 and M1} macrophages and its role on cancer
  outcomes.
\newblock {\em J Theor Biol}, 390:23--39, Feb 2016.

\bibitem{Shu.ApplMathModel.2020}
Yaqin Shu, Jicai Huang, Yueping Dong, and Yasuhiro Takeuchi.
\newblock Mathematical modeling and bifurcation analysis of pro-and anti-tumor
  macrophages.
\newblock {\em Appl Math Model}, 88:758--773, Jul 2020.

\bibitem{Eftimie.MathBiosci.2020}
R~Eftimie.
\newblock Investigation into the role of macrophages heterogeneity on solid
  tumour aggregations.
\newblock {\em Math Biosci}, 322:108325, Apr 2020.

\bibitem{Eftimie.JTheorBiol.2021}
Raluca Eftimie and Charlotte Barelle.
\newblock {Mathematical investigation of innate immune responses to lung
  cancer: The role of macrophages with mixed phenotypes.}
\newblock {\em J Theor Biol}, 524:110739, Sep 2021.

\bibitem{Shariatpanahi.JTheorBiol.2018}
Seyed~Peyman Shariatpanahi, Seyed~Pooya Shariatpanahi, Keivan Madjidzadeh,
  Moustapha Hassan, and Manuchehr Abedi-Valugerdi.
\newblock Mathematical modeling of tumor-induced immunosuppression by
  myeloid-derived suppressor cells: Implications for therapeutic targeting
  strategies.
\newblock {\em J Theor Biol}, 442:1--10, Apr 2018.

\bibitem{Anderson.JMathBiol.2023}
Hannah~G Anderson, Gregory~P Takacs, Duane~C Harris, Yang Kuang, Jeffrey~K
  Harrison, and Tracy~L Stepien.
\newblock {Global stability and parameter analysis reinforce therapeutic
  targets of PD-L1-PD-1 and MDSCs for glioblastoma.}
\newblock {\em J Math Biol}, 88(1):10, Dec 2023.

\bibitem{Sontag.CellSyst.2017}
Eduardo~D Sontag.
\newblock A dynamic model of immune responses to antigen presentation predicts
  different regions of tumor or pathogen elimination.
\newblock {\em Cell Syst}, 4(2):231--241, Feb 2017.

\bibitem{Tsur.JTheorBiol.2020}
N~Tsur, Y~Kogan, M~Rehm, and Z~Agur.
\newblock Response of patients with melanoma to immune checkpoint blockade -
  insights gleaned from analysis of a new mathematical mechanistic model.
\newblock {\em J Theor Biol}, 485:110033, Jan 2020.

\bibitem{Pei.JTheorBiol.2023}
Yongzhen Pei, Siqi Han, Changguo Li, Jinzhi Lei, and Fengxi Wen.
\newblock Data-based modeling of breast cancer and optimal therapy.
\newblock {\em J Theor Biol}, 573:111593, Sep 2023.

\bibitem{Wilson.BullMathBiol.2012}
Shelby Wilson and Doron Levy.
\newblock A mathematical model of the enhancement of tumor vaccine efficacy by
  immunotherapy.
\newblock {\em Bull Math Biol}, 74(7):1485--1500, Jul 2012.

\bibitem{Khajanchi.MathBiosci.2017}
Subhas Khajanchi and Sandip Banerjee.
\newblock Quantifying the role of immunotherapeutic drug {T11} target structure
  in progression of malignant gliomas: {Mathematical} modeling and dynamical
  perspective.
\newblock {\em Math Biosci}, 289:69--77, Jul 2017.

\bibitem{Coletti.JTheorBiol.2021}
Roberta Coletti, Andrea Pugliese, and Luca Marchetti.
\newblock Modeling the effect of immunotherapies on human castration-resistant
  prostate cancer.
\newblock {\em J Theor Biol}, 509:110500, Jan 2021.

\bibitem{He.JBiolSyst.2017}
Dan-Hua He and Jian-Xin Xu.
\newblock A mathematical model of pancreatic cancer with two kinds of
  treatments.
\newblock {\em J Biol Syst}, 25(01):83--104, Jan 2017.

\bibitem{Arabameri.JBiolSyst.2018}
Abazar Arabameri, Davud Asemani, and Jamshid Hajati.
\newblock Mathematical modeling of in-vivo tumor-immune interactions for the
  cancer immunotherapy using matured dendritic cells.
\newblock {\em J Biol Syst}, 26(01):167--188, Mar 2018.

\bibitem{Sardar.CommunNonlinearSci.2023}
Mrinmoy Sardar, Subhas Khajanchi, and Bashir Ahmad.
\newblock A tumor-immune interaction model with the effect of impulse therapy.
\newblock {\em Commun Nonlinear Sci}, 126:107430, Jul 2023.

\bibitem{Xue.JTheorBiol.2023}
Ling Xue, Hongyu Zhang, Xiaoming Zheng, Wei Sun, and Jinzhi Lei.
\newblock Treatment of melanoma with dendritic cell vaccines and immune
  checkpoint inhibitors: {A} mathematical modeling study.
\newblock {\em J Theor Biol}, 568:111489, Jul 2023.

\bibitem{Smalley.iScience.2020}
Munisha Smalley, Michelle Przedborski, Saravanan Thiyagarajan, Moriah Pellowe,
  Amit Verma, Nilesh Brijwani, Debika Datta, Misti Jain, Basavaraja~U
  Shanthappa, Vidushi Kapoor, Kodaganur~S Gopinath, D~C Doval, K~S Sabitha,
  Gaspar Taroncher-Oldenburg, Biswanath Majumder, Pradip Majumder, Mohammad
  Kohandel, and Aaron Goldman.
\newblock Integrating systems biology and an ex vivo human tumor model
  elucidates {PD-1} blockade response dynamics.
\newblock {\em iScience}, 23(6):101229, Jun 2020.

\bibitem{Rodriguez-Messan.PLoSComputBiol.2021}
Marisabel Rodriguez~Messan, Osman~N Yogurtcu, Joseph~R McGill, Ujwani Nukala,
  Zuben~E Sauna, and Hong Yang.
\newblock Mathematical model of a personalized neoantigen cancer vaccine and
  the human immune system.
\newblock {\em PLoS Comput Biol}, 17(9):e1009318, Sep 2021.

\bibitem{Mohammad-Mirzaei.PLoSComputBiol.2022}
Navid Mohammad~Mirzaei, Navid Changizi, Alireza Asadpoure, Sumeyye Su, Dilruba
  Sofia, Zuzana Tatarova, Ioannis~K Zervantonakis, Young~Hwan Chang, and Leili
  Shahriyari.
\newblock Investigating key cell types and molecules dynamics in {PyMT} mice
  model of breast cancer through a mathematical model.
\newblock {\em PLoS Comput Biol}, 18(3):e1009953, Mar 2022.

\bibitem{Shafiekhani.BMCCancer.2021}
Sajad Shafiekhani, Hojat Dehghanbanadaki, Azam~Sadat Fatemi, Sara Rahbar,
  Jamshid Hadjati, and Amir~Homayoun Jafari.
\newblock Prediction of {anti-CD25 and 5-FU} treatments efficacy for pancreatic
  cancer using a mathematical model.
\newblock {\em BMC Cancer}, 21(1):1226, Nov 2021.

\bibitem{Villasana.JMathBiol.2003}
Minaya Villasana and Ami Radunskaya.
\newblock A delay differential equation model for tumor growth.
\newblock {\em J Math Biol}, 47(3):270--294, Sep 2003.

\bibitem{Yafia.SIAMJApplMath.2007}
Radouane Yafia.
\newblock Hopf bifurcation in differential equations with delay for
  tumor--immune system competition model.
\newblock {\em SIAM J Appl Math}, 67(6):1693--1703, 2007.

\bibitem{Banerjee.Biosystems.2008}
Sandip Banerjee and Ram~Rup Sarkar.
\newblock Delay-induced model for tumor-immune interaction and control of
  malignant tumor growth.
\newblock {\em Biosystems}, 91(1):268--288, Jan 2008.

\bibitem{Sarkar.MathBiosci.2005}
Ram~Rup Sarkar and Sandip Banerjee.
\newblock Cancer self remission and tumor stability-- a stochastic approach.
\newblock {\em Math Biosci}, 196(1):65--81, Jul 2005.

\bibitem{Bi.SIAMJApplDynSyst.2013}
Ping Bi and Shigui Ruan.
\newblock Bifurcations in delay differential equations and applications to
  tumor and immune system interaction models.
\newblock {\em SIAM J Appl Dyn Syst}, 12(4):1847--1888, 2013.

\bibitem{Bi.Chaos.2014}
Ping Bi, Shigui Ruan, and Xinan Zhang.
\newblock Periodic and chaotic oscillations in a tumor and immune system
  interaction model with three delays.
\newblock {\em Chaos}, 24(2), Jun 2014.

\bibitem{Dong.ApplMathComput.2015}
Yueping Dong, Gang Huang, Rinko Miyazaki, and Yasuhiro Takeuchi.
\newblock Dynamics in a tumor immune system with time delays.
\newblock {\em Appl Math Comput}, 252:99--113, Feb 2015.

\bibitem{Qomlaqi.MathBiosci.2017}
Milad Qomlaqi, Fariba Bahrami, Maryam Ajami, and Jamshid Hajati.
\newblock An extended mathematical model of tumor growth and its interaction
  with the immune system, to be used for developing an optimized immunotherapy
  treatment protocol.
\newblock {\em Math Biosci}, 292:1--9, Oct 2017.

\bibitem{Khajanchi.ApplMathComput.2014}
Subhas Khajanchi and Sandip Banerjee.
\newblock Stability and bifurcation analysis of delay induced tumor immune
  interaction model.
\newblock {\em Appl Math Comput}, 248:652--671, Dec 2014.

\bibitem{Khajanchi.MathBiosci.2018}
Subhas Khajanchi and Sandip Banerjee.
\newblock Influence of multiple delays in brain tumor and immune system
  interaction with {T11} target structure as a potent stimulator.
\newblock {\em Math Biosci}, 302:116--130, Aug 2018.

\bibitem{Khajanchi.Chaos.2018}
Subhas Khajanchi, Matja{\v{z}} Perc, and Dibakar Ghosh.
\newblock The influence of time delay in a chaotic cancer model.
\newblock {\em Chaos}, 28(10), Oct 2018.

\bibitem{Khajanchi.ApplMathComput.2019}
Subhas Khajanchi and Juan~J Nieto.
\newblock Mathematical modeling of tumor-immune competitive system, considering
  the role of time delay.
\newblock {\em Appl Math Comput}, 340:180--205, Jan 2019.

\bibitem{Khajanchi.IntJBiomath.2020}
Subhas Khajanchi.
\newblock Chaotic dynamics of a delayed tumor--immune interaction model.
\newblock {\em Int J Biomath}, 13(02):2050009, Jan 2020.

\bibitem{Sardar.CSF.2024}
Mrinmoy Sardar, Subhas Khajanchi, Santosh Biswas, and Sumana Ghosh.
\newblock A mathematical model for tumor-immune competitive system with
  multiple time delays.
\newblock {\em Chaos Soliton Fract}, 179:114397, Feb 2024.

\bibitem{Das.ApplMathComput.2019}
Parthasakha Das, Pritha Das, and Samhita Das.
\newblock An investigation on monod--haldane immune response based
  tumor-effector--interleukin-2 interactions with treatments.
\newblock {\em Appl Math Comput}, 361:536--551, Jul 2019.

\bibitem{Das.Chaos.2020}
Parthasakha Das, Ranjit~Kumar Upadhyay, Pritha Das, and Dibakar Ghosh.
\newblock Exploring dynamical complexity in a time-delayed tumor-immune model.
\newblock {\em Chaos}, 30(12), Dec 2020.

\bibitem{Das.ChaosSolitonFract.2020}
Parthasakha Das, Samhita Das, Ranjit~Kumar Upadhyay, and Pritha Das.
\newblock Optimal treatment strategies for delayed cancer-immune system with
  multiple therapeutic approach.
\newblock {\em Chaos Soliton Fract}, 136:109806, Apr 2020.

\bibitem{Rihan.ApplMathComput.2019}
Fathalla~A Rihan, Shanmugam Lakshmanan, and H~Maurer.
\newblock Optimal control of tumour-immune model with time-delay and
  immuno-chemotherapy.
\newblock {\em Appl Math Comput}, 353:147--165, Jul 2019.

\bibitem{Rihan.CSF.2020}
FA~Rihan and G~Velmurugan.
\newblock Dynamics of fractional-order delay differential model for
  tumor-immune system.
\newblock {\em Chaos Soliton Fract}, 132:109592, Jan 2020.

\bibitem{Rihan.AlexEngJ.2022}
FA~Rihan, HJ~Alsakaji, S~Kundu, and O~Mohamed.
\newblock Dynamics of a time-delay differential model for tumour-immune
  interactions with random noise.
\newblock {\em Alex Eng J}, 61(12):11913--11923, Dec 2022.

\bibitem{Dickman.Chaos.2020}
Lauren~R Dickman and Yang Kuang.
\newblock Analysis of tumor-immune dynamics in a delayed dendritic cell therapy
  model.
\newblock {\em Chaos}, 30(11), Nov 2020.

\bibitem{Dickman.SIAMJApplMath.2020}
Lauren~R Dickman, Evan Milliken, and Yang Kuang.
\newblock Tumor control, elimination, and escape through a compartmental model
  of dendritic cell therapy for melanoma.
\newblock {\em SIAM J Appl Math}, 80(2):906--928, Apr 2020.

\bibitem{Wang.CSF.2022}
Jingnan Wang, Hongbin Shi, Li~Xu, and Lu~Zang.
\newblock Hopf bifurcation and chaos of tumor-lymphatic model with two time
  delays.
\newblock {\em Chaos Soliton Fract}, 157:111922, Apr 2022.

\bibitem{Tsimring:2014aa}
Lev~S Tsimring.
\newblock {Noise in biology.}
\newblock {\em Rep Prog Phys}, 77(2):026601, 2014.

\bibitem{Zechner:2020aa}
Christoph Zechner, Elisa Nerli, and Caren Norden.
\newblock {Stochasticity and determinism in cell fate decisions.}
\newblock {\em Development}, 147(14), Jul 2020.

\bibitem{Mukhopadhyay.StochAnalAppl.2009}
B~Mukhopadhyay and R~Bhattacharyya.
\newblock {A nonlinear mathematical model of virus-tumor-immune system
  interaction: Deterministic and stochastic analysis}.
\newblock {\em Stoch Anal Appl}, 27(2):409--429, Mar 2009.

\bibitem{Caravagna.JTheorBiol.2010}
Giulio Caravagna, Alberto d'Onofrio, Paolo Milazzo, and Roberto Barbuti.
\newblock Tumour suppression by immune system through stochastic oscillations.
\newblock {\em J Theor Biol}, 265(3):336--345, Aug 2010.

\bibitem{Xu.PhysicaA.2013}
Yong Xu, Jing Feng, JuanJuan Li, and Huiqing Zhang.
\newblock Stochastic bifurcation for a tumor--immune system with symmetric
  l{\'e}vy noise.
\newblock {\em Physica A}, 392(20):4739--4748, Oct 2013.

\bibitem{Li.CNSNS.2017}
Dongxi Li and Fangjuan Cheng.
\newblock Threshold for extinction and survival in stochastic tumor immune
  system.
\newblock {\em Commun Nonlinear Sci}, 51:1--12, Mar 2017.

\bibitem{Deng.ApplMathModel.2020}
Ying Deng and Meng Liu.
\newblock Analysis of a stochastic tumor-immune model with regime switching and
  impulsive perturbations.
\newblock {\em Appl Math Model}, 78:482--504, Oct 2020.

\bibitem{Liu.PhysicaA.2018}
Xiangdong Liu, Qingze Li, and Jianxin Pan.
\newblock A deterministic and stochastic model for the system dynamics of
  tumor--immune responses to chemotherapy.
\newblock {\em Physica A}, 500:162--176, Jun 2018.

\bibitem{Li.SIAMJApplMath.2019}
Xiaoyue Li, Guoting Song, Yang Xia, and Chenggui Yuan.
\newblock Dynamical behaviors of the tumor-immune system in a stochastic
  environment.
\newblock {\em SIAM J Appl Math}, 79(6):2193--2217, Feb 2019.

\bibitem{Chen.JMathAnalAppl.2023}
Xing Chen, Xiaoyue Li, Yuting Ma, and Chenggui Yuan.
\newblock The threshold of stochastic tumor-immune model with regime switching.
\newblock {\em J Math Anal Appl}, 522(1):126956, Jun 2023.

\bibitem{Yang.CNSNS.2019}
Jin Yang, Yuanshun Tan, and Robert~A Cheke.
\newblock Thresholds for extinction and proliferation in a stochastic
  tumour-immune model with pulsed comprehensive therapy.
\newblock {\em Commun Nonlinear Sci}, 73:363--378, Jul 2019.

\bibitem{Han.ApplMathModel.2022}
Ping Han, Wei Xu, Liang Wang, Hongxia Zhang, and Zhicong Ren.
\newblock Most probable trajectories in a two-dimensional tumor-immune system
  under stochastic perturbation.
\newblock {\em Appl Math Model}, 105:800--814, May 2022.

\bibitem{Hao.CSF.2022}
Mengli Hao, Wantao Jia, Liang Wang, and Fuxiao Li.
\newblock Most probable trajectory of a tumor model with immune response
  subjected to asymmetric l{\'e}vy noise.
\newblock {\em Chaos Soliton Fract}, 165:112765, Dec 2022.

\bibitem{Bose.PRE.2011}
Thomas Bose and Steffen Trimper.
\newblock Noise-assisted interactions of tumor and immune cells.
\newblock {\em Phys Rev E}, 84(2 Pt 1):021927, Aug 2011.

\bibitem{Das.PhysicaA.2020}
Parthasakha Das, Pritha Das, and Sayan Mukherjee.
\newblock Stochastic dynamics of michaelis--menten kinetics based tumor-immune
  interactions.
\newblock {\em Physica A}, 541:123603, Mar 2020.

\bibitem{Phan.MathBiosciEng.2020}
Tuan~Anh Phan and Jianjun~Paul Tian.
\newblock Basic stochastic model for tumor virotherapy.
\newblock {\em Math Biosci Eng}, 17(4):4271, Jun 2020.

\bibitem{Yang.MathComputSimulat.2021}
Huan Yang, Yuanshun Tan, Jin Yang, and Zijian Liu.
\newblock Extinction and persistence of a tumor-immune model with white noise
  and pulsed comprehensive therapy.
\newblock {\em Math Comput Simulat}, 182:456--470, Apr 2021.

\bibitem{Alsakaji.MathBiosciEng.2023}
H~J Alsakaji, F~A Rihan, K~Udhayakumar, and F~El Ktaibi.
\newblock {Stochastic tumor-immune interaction model with external treatments
  and time delays: An optimal control problem.}
\newblock {\em Math Biosci Eng}, 20(11):19270--19299, Oct 2023.

\bibitem{Huang.ActaMathSci.2022}
Mingzhan Huang, Shouzong Liu, Xinyu Song, and Xiufen Zou.
\newblock Control strategies for a tumor-immune system with impulsive drug
  delivery under a random environment.
\newblock {\em Acta Mathematica Scientia}, 42(3):1141--1159, Apr 2022.

\bibitem{Lai.NPJSystBiolAppl.2024}
Xiulan Lai, Xiaopei Jiao, Haojian Zhang, and Jinzhi Lei.
\newblock Computational modeling reveals key factors driving treatment-free
  remission in chronic myeloid leukemia patients.
\newblock {\em NPJ Syst Biol Appl}, 10(1):45, Apr 2024.

\bibitem{Lai.PNAS.2018}
Xiulan Lai, Andrew Stiff, Megan Duggan, Robert Wesolowski, William E~3rd
  Carson, and Avner Friedman.
\newblock {Modeling combination therapy for breast cancer with BET and immune
  checkpoint inhibitors.}
\newblock {\em Proc Natl Acad Sci USA}, 115(21):5534--5539, May 2018.

\bibitem{Lai.JTheorBiol.2019}
Xiulan Lai and Avner Friedman.
\newblock Mathematical modeling in scheduling cancer treatment with combination
  of {VEGF} inhibitor and chemotherapy drugs.
\newblock {\em J Theor Biol}, 462:490--498, Feb 2019.

\bibitem{Lai.BMCSystBiol.2017}
Xiulan Lai and Avner Friedman.
\newblock {Combination therapy for melanoma with BRAF/MEK inhibitor and immune
  checkpoint inhibitor: A mathematical model.}
\newblock {\em BMC Syst Biol}, 11(1):70, Jul 2017.

\bibitem{Friedman.BullMathBiol.2020}
Avner Friedman and Nourridine Siewe.
\newblock Overcoming drug resistance to {BRAF} inhibitor.
\newblock {\em Bull Math Biol}, 82(1):8, Jan 2020.

\bibitem{Liao.MathBiosci.2022}
Kang-Ling Liao and Kenton~D Watt.
\newblock {Mathematical modeling for the combination treatment of IFN-$\gamma$
  and anti-PD-1 in cancer immunotherapy.}
\newblock {\em Math Biosci}, 353:108911, Nov 2022.

\bibitem{Szomolay.JTheorBiol.2012}
Barbara Szomolay, Tim~D Eubank, Ryan~D Roberts, Clay~B Marsh, and Avner
  Friedman.
\newblock Modeling the inhibition of breast cancer growth by {GM-CSF}.
\newblock {\em J Theor Biol}, 303:141--151, Jun 2012.

\bibitem{Lee.PLoSComputBiol.2021}
Junho Lee, Donggu Lee, Sean Lawler, and Yangjin Kim.
\newblock {Role of neutrophil extracellular traps in regulation of lung cancer
  invasion and metastasis: Structural insights from a computational model.}
\newblock {\em PLoS Comput Biol}, 17(2):e1008257, Feb 2021.

\bibitem{Kim.JMathBiol.2022}
Yangjin Kim, Junho Lee, Chaeyoung Lee, and Sean Lawler.
\newblock {Role of senescent tumor cells in building a cytokine shield in the
  tumor microenvironment: Mathematical modeling.}
\newblock {\em J Math Biol}, 86(1):14, Dec 2022.

\bibitem{Friedman.BullMathBiol.2018}
Avner Friedman and Wenrui Hao.
\newblock The role of exosomes in pancreatic cancer microenvironment.
\newblock {\em Bull Math Biol}, 80(5):1111--1133, May 2018.

\bibitem{Siewe.BullMathBiol.2023}
Nourridine Siewe and Avner Friedman.
\newblock Breast cancer exosomal {microRNAs} facilitate pre-metastatic niche
  formation in the bone: {A} mathematical model.
\newblock {\em Bull Math Biol}, 85(2):12, Jan 2023.

\bibitem{Jacobsen.BullMathBiol.2015}
Karly Jacobsen, Luke Russell, Balveen Kaur, and Avner Friedman.
\newblock Effects of ccn1 and macrophage content on glioma virotherapy: {A}
  mathematical model.
\newblock {\em Bull Math Biol}, 77(6):984--1012, Jun 2015.

\bibitem{Lai.SciChinaMath.2020}
Xiulan Lai and Avner Friedman.
\newblock {Mathematical modeling of cancer treatment with radiation and PD-L1
  inhibitor.}
\newblock {\em Sci China Math}, 63:465--484, Feb 2020.

\bibitem{Siewe.JTheorBiol.2023}
Nourridine Siewe and Avner Friedman.
\newblock {Cancer therapy with immune checkpoint inhibitor and CSF-1 blockade:
  A mathematical model.}
\newblock {\em J Theor Biol}, 556:111297, Jan 2023.

\bibitem{Kim.PNAS.2018}
Yangjin Kim, Ji~Young Yoo, Tae~Jin Lee, Joseph Liu, Jianhua Yu, Michael~A
  Caligiuri, Balveen Kaur, and Avner Friedman.
\newblock Complex role of {NK} cells in regulation of oncolytic
  virus-bortezomib therapy.
\newblock {\em Proc Natl Acad Sci USA}, 115(19):4927--4932, May 2018.

\bibitem{Li.ComputMathAppl.2022}
Sicheng Li, Shun Wang, and Xiufen Zou.
\newblock Data-driven mathematical modeling and quantitative analysis of cell
  dynamics in the tumor microenvironment.
\newblock {\em Comput Math Appl}, 113:300--314, May 2022.

\bibitem{Burns.CellTissueKinet.1970}
Burns F~J and Tannock I~F.
\newblock On the existence of a {G0}-phase in the cell cycle.
\newblock {\em Cell Tissue Kinet}, 3:321--334, Oct 1970.

\bibitem{Mackey.Blood.1978}
M~C Mackey.
\newblock Unified hypothesis for the origin of aplastic anemia and periodic
  hematopoiesis.
\newblock {\em Blood}, 51(5):941--956, May 1978.

\bibitem{Lei.PNAS.2014}
Jinzhi Lei, Simon~A Levin, and Qing Nie.
\newblock Mathematical model of adult stem cell regeneration with cross-talk
  between genetic and epigenetic regulation.
\newblock {\em Proc Natl Acad Sci USA}, 111(10):E880--7, Mar 2014.

\bibitem{Lei.SciChinaMath.2020}
Jinzhi Lei.
\newblock {Evolutionary dynamics of cancer: From epigenetic regulation to cell
  population dynamics---mathematical model framework, applications, and open
  problems.}
\newblock {\em Sci China Math}, 63:411--424, Jan 2020.

\bibitem{Zhang.ComputSystOncol.2021}
Zhang Can, Shao Changrong, Jiao Xiaopei, Bai Yue, Li~Miao, Shi Hanping, Lei
  Jinzhi, and Zhong Xiaosong.
\newblock Individual cell-based modeling of tumor cell plasticity-induced
  immune escape after {CAR-T} therapy.
\newblock {\em Comput Syst Oncol}, 1(3):e21029, Sep 2021.

\bibitem{Liang.CommunApplMathComput.2023}
Liang Xiyin and Lei Jinzhi.
\newblock Oscillatory dynamics of heterogeneous stem cell regeneration.
\newblock {\em Commun Appl Math Comput}, 6(1):431--453, Apr 2023.

\bibitem{Huang.IntJModPhysB.2017}
Rongsheng Huang and Jinzhi Lei.
\newblock Dynamics of gene expression with positive feedback to histone
  modifications at bivalent domains.
\newblock {\em Int J Mod Phys B}, 32(7), Nov 2017.

\bibitem{Huang.DCDSB.2019}
Rongsheng Huang and Jinzhi Lei.
\newblock Cell-type switches induced by stochastic histone modification
  inheritance.
\newblock {\em Discrete and Continuous Dynamical Systems-B}, 24(10):5601--5619,
  Aug 2019.

\bibitem{Huang.JTheorBiol.2024}
Rongsheng Huang, Qiaojun Situ, and Jinzhi Lei.
\newblock Dynamics of cell-type transition mediated by epigenetic
  modifications.
\newblock {\em J Theor Biol}, 577:111664, Jan 2024.

\bibitem{Lei:arXiv2024}
Jinzhi Lei.
\newblock {Mathematical modeling of heterogeneous stem cell regeneration: From
  cell division to Waddington's epigenetic landscape}.
\newblock {\em arXiv}, page arXiv:2309.08064, 2024.

\bibitem{Guo.CancerRes.2017}
Yucheng Guo, Qing Nie, Adam~L MacLean, Yanda Li, Jinzhi Lei, and Shao Li.
\newblock Multiscale modeling of inflammation-induced tumorigenesis reveals
  competing oncogenic and oncoprotective roles for inflammation.
\newblock {\em Cancer Res}, 77(22):6429--6441, Nov 2017.

\bibitem{Ma.JMathBiol.2023}
Shizhao Ma, Jinzhi Lei, and Xiulan Lai.
\newblock Modeling tumour heterogeneity of {PD-L1} expression in tumour
  progression and adaptive therapy.
\newblock {\em J Math Biol}, 86(3):38, Jan 2023.

\bibitem{Su.JMPA.2023}
Yuan-Hang Su, Wan-Tong Li, Yuan Lou, and Xuefeng Wang.
\newblock Principal spectral theory for nonlocal systems and applications to
  stem cell regeneration models.
\newblock {\em Journal de Math{\'e}matiques Pures et Appliqu{\'e}es},
  176:226--281, 2023.

\bibitem{Milberg.SciRep.2019}
Oleg Milberg, Chang Gong, Mohammad Jafarnejad, Imke~H Bartelink, Bing Wang,
  Paolo Vicini, Rajesh Narwal, Lorin Roskos, and Aleksander~S Popel.
\newblock A {QSP} model for predicting clinical responses to monotherapy,
  combination and sequential therapy following {CTLA-4, PD-1, and PD-L1}
  checkpoint blockade.
\newblock {\em Sci Rep}, 9(1):11286, Aug 2019.

\bibitem{Wang.RSocOpenSci.2019}
Hanwen Wang, Oleg Milberg, Imke~H Bartelink, Paolo Vicini, Bing Wang, Rajesh
  Narwal, Lorin Roskos, Cesar~A Santa-Maria, and Aleksander~S Popel.
\newblock {In silico simulation of a clinical trial with anti-CTLA-4 and
  anti-PD-L1 immunotherapies in metastatic breast cancer using a systems
  pharmacology model.}
\newblock {\em R Soc Open Sci}, 6(5):190366, May 2019.

\bibitem{Ma.JImmunotherCancer.2020}
Huilin Ma, Hanwen Wang, Richard~J Sov{\'e}, Jun Wang, Craig Giragossian, and
  Aleksander~S Popel.
\newblock {Combination therapy with {T} cell engager and {PD-L1} blockade
  enhances the antitumor potency of {T} cells as predicted by a {QSP} model.}
\newblock {\em J Immunother Cancer}, 8(2), Aug 2020.

\bibitem{Ma.AAPSJ.2020}
Huilin Ma, Hanwen Wang, Richard~J Sove, Mohammad Jafarnejad, Chia-Hung Tsai,
  Jun Wang, Craig Giragossian, and Aleksander~S Popel.
\newblock A quantitative systems pharmacology model of {T} cell engager applied
  to solid tumor.
\newblock {\em AAPS J}, 22(4):85, Jun 2020.

\bibitem{Wang.FrontBioengBiotechnol.2020}
Hanwen Wang, Richard~J Sov{\'e}, Mohammad Jafarnejad, Sondra Rahmeh,
  Elizabeth~M Jaffee, Vered Stearns, Evanthia~T Roussos~Torres, Roisin~M
  Connolly, and Aleksander~S Popel.
\newblock Conducting a virtual clinical trial in {HER2}-negative breast cancer
  using a quantitative systems pharmacology model with an epigenetic modulator
  and immune checkpoint inhibitors.
\newblock {\em Front Bioeng Biotechnol}, 8:141, Feb 2020.

\bibitem{Wang.JImmunotherCancer.2021}
Hanwen Wang, Huilin Ma, Richard~J Sov{\'e}, Leisha~A Emens, and Aleksander~S
  Popel.
\newblock Quantitative systems pharmacology model predictions for efficacy of
  atezolizumab and nab-paclitaxel in triple-negative breast cancer.
\newblock {\em J Immunother Cancer}, 9(2), Feb 2021.

\bibitem{Wang.iScience.2022}
Hanwen Wang, Chen Zhao, Cesar~A Santa-Maria, Leisha~A Emens, and Aleksander~S
  Popel.
\newblock {Dynamics of tumor-associated macrophages in a quantitative systems
  pharmacology model of immunotherapy in triple-negative breast cancer.}
\newblock {\em iScience}, 25(8):104702, Aug 2022.

\bibitem{Sove.CPTPharmacometricsSystPharmacol.2020}
Richard~J Sov{\'e}, Mohammad Jafarnejad, Chen Zhao, Hanwen Wang, Huilin Ma, and
  Aleksander~S Popel.
\newblock {QSP-IO}: {A} quantitative systems pharmacology toolbox for
  mechanistic multiscale modeling for immuno-oncology applications.
\newblock {\em CPT Pharmacometrics Syst Pharmacol}, 9(9):484--497, Sep 2020.

\bibitem{Sove.JImmunotherCancer.2022}
Richard~J Sov{\'e}, Babita~K Verma, Hanwen Wang, Won~Jin Ho, Mark Yarchoan, and
  Aleksander~S Popel.
\newblock {Virtual clinical trials of anti-PD-1 and anti-CTLA-4 immunotherapy
  in advanced hepatocellular carcinoma using a quantitative systems
  pharmacology model.}
\newblock {\em J Immunother Cancer}, 10(11), Nov 2022.

\bibitem{Ippolito.CPTPharmacometricsSystPharmacol.2024}
Alberto Ippolito, Hanwen Wang, Yu~Zhang, Vahideh Vakil, Hojjat Bazzazi, and
  Aleksander~S Popel.
\newblock {Eliciting the antitumor immune response with a conditionally
  activated PD-L1 targeting antibody analyzed with a quantitative systems
  pharmacology model.}
\newblock {\em CPT Pharmacometrics Syst Pharmacol}, 13(1):93--105, Jan 2024.

\bibitem{Wang.ClinTranslSci.2024}
Hanwen Wang, Theinmozhi Arulraj, Samira Anbari, and Aleksander~S Popel.
\newblock {Quantitative systems pharmacology modeling of macrophage-targeted
  therapy combined with PD-L1 inhibition in advanced NSCLC.}
\newblock {\em Clin Transl Sci}, 17(6):e13811, Jun 2024.

\bibitem{Gong.CancersBasel.2021}
Chang Gong, Alvaro Ruiz-Martinez, Holly Kimko, and Aleksander~S Popel.
\newblock A spatial quantitative systems pharmacology platform {spQSP-IO} for
  simulations of tumor-immune interactions and effects of checkpoint inhibitor
  immunotherapy.
\newblock {\em Cancers (Basel)}, 13(15), Jul 2021.

\bibitem{Nikfar.CancersBasel.2023}
Mehdi Nikfar, Haoyang Mi, Chang Gong, Holly Kimko, and Aleksander~S Popel.
\newblock {Quantifying intratumoral heterogeneity and immunoarchitecture
  generated in-silico by a spatial quantitative systems pharmacology model.}
\newblock {\em Cancers (Basel)}, 15(10), May 2023.

\bibitem{Zhang.ImmunoinformaticsAmst.2021}
Shuming Zhang, Chang Gong, Alvaro Ruiz-Martinez, Hanwen Wang, Emily
  Davis-Marcisak, Atul Deshpande, Aleksander~S Popel, and Elana~J Fertig.
\newblock Integrating single cell sequencing with a spatial quantitative
  systems pharmacology model {spQSP} for personalized prediction of
  triple-negative breast cancer immunotherapy response.
\newblock {\em Immunoinformatics (Amst)}, 1-2, Oct 2021.

\bibitem{Ruiz-Martinez.PLoSComputBiol.2022}
Alvaro Ruiz-Martinez, Chang Gong, Hanwen Wang, Richard~J Sov{\'e}, Haoyang Mi,
  Holly Kimko, and Aleksander~S Popel.
\newblock Simulations of tumor growth and response to immunotherapy by coupling
  a spatial agent-based model with a whole-patient quantitative systems
  pharmacology model.
\newblock {\em PLoS Comput Biol}, 18(7):e1010254, Jul 2022.

\bibitem{Arulraj.SciAdv.2023}
Theinmozhi Arulraj, Hanwen Wang, Leisha~A Emens, Cesar~A Santa-Maria, and
  Aleksander~S Popel.
\newblock {A transcriptome-informed QSP model of metastatic triple-negative
  breast cancer identifies predictive biomarkers for PD-1 inhibition.}
\newblock {\em Sci Adv}, 9(26):eadg0289, Jun 2023.

\bibitem{Wang.NPJPrecisOncol.2023}
Hanwen Wang, Theinmozhi Arulraj, Holly Kimko, and Aleksander~S Popel.
\newblock {Generating immunogenomic data-guided virtual patients using a QSP
  model to predict response of advanced NSCLC to PD-L1 inhibition.}
\newblock {\em NPJ Precis Oncol}, 7(1):55, Jun 2023.

\bibitem{Abar.ComputSciRev.2017}
Sameera Abar, Georgios~K Theodoropoulos, Pierre Lemarinier, and Gregory~MP
  O'Hare.
\newblock {Agent based modelling and simulation tools: A review of the
  state-of-art software.}
\newblock {\em Comput Sci Rev}, 24:13--33, May 2017.

\bibitem{West.TrendsCellBiol.2023}
Jeffrey West, Mark Robertson-Tessi, and Alexander R~A Anderson.
\newblock Agent-based methods facilitate integrative science in cancer.
\newblock {\em Trends Cell Biol}, 33(4):300--311, Apr 2023.

\bibitem{Valentim.ComputBiolMed.2023}
Carlos~A Valentim, Jos{\'e}A Rabi, and Sergio~A David.
\newblock {Cellular-automaton model for tumor growth dynamics: Virtualization
  of different scenarios.}
\newblock {\em Comput Biol Med}, 153:106481, Feb 2023.

\bibitem{Wolf.Springer.2004}
Dieter~A Wolf-Gladrow.
\newblock {\em {Lattice-gas cellular automata and lattice boltzmann models: An
  introduction}}.
\newblock Springer, 2004.

\bibitem{Scianna.MultiscaleModelSim.2012}
Marco Scianna and Luigi Preziosi.
\newblock Multiscale developments of the cellular potts model.
\newblock {\em Multiscale Model Sim}, 10(2):342--382, Jan 2012.

\bibitem{Mathias.BMCBioinformatics.2022}
Sonja Mathias, Adrien Coulier, and Andreas Hellander.
\newblock {CBMOS: a GPU-enabled python framework for the numerical study of
  center-based models.}
\newblock {\em BMC Bioinformatics}, 23(1):55, Jan 2022.

\bibitem{Sandersius.PhysBiol.2008}
Sebastian~A Sandersius and Timothy~J Newman.
\newblock {Modeling cell rheology with the subcellular element model.}
\newblock {\em Phys Biol}, 5(1):015002, Apr 2008.

\bibitem{Fletcher.BiophysJ.2014}
Alexander~G Fletcher, Miriam Osterfield, Ruth~E Baker, and Stanislav~Y
  Shvartsman.
\newblock Vertex models of epithelial morphogenesis.
\newblock {\em Biophys J}, 106(11):2291--2304, Jun 2014.

\bibitem{Rejniak.JTheorBiol.2007}
Katarzyna~A Rejniak.
\newblock {An immersed boundary framework for modelling the growth of
  individual cells: An application to the early tumour development.}
\newblock {\em J Theor Biol}, 247(1):186--204, Jul 2007.

\bibitem{Izaguirre.Bioinformatics.2004}
J~A Izaguirre, R~Chaturvedi, C~Huang, T~Cickovski, J~Coffland, G~Thomas,
  G~Forgacs, M~Alber, G~Hentschel, S~A Newman, and J~A Glazier.
\newblock {CompuCell, a multi-model framework for simulation of morphogenesis.}
\newblock {\em Bioinformatics}, 20(7):1129--1137, May 2004.

\bibitem{Cickovski.IEEE/ACMTransComputBiolBioinform.2005}
Trevor~M Cickovski, Chenbang Huang, Rajiv Chaturvedi, Tilmann Glimm, H~George~E
  Hentschel, Mark~S Alber, James~A Glazier, Stuart~A Newman, and Jes{\'u}s~A
  Izaguirre.
\newblock A framework for three-dimensional simulation of morphogenesis.
\newblock {\em IEEE/ACM Trans Comput Biol Bioinform}, 2(4):273--288, Oct-Dec
  2005.

\bibitem{Stoll.BMCSystBiol.2012}
Gautier Stoll, Eric Viara, Emmanuel Barillot, and Laurence Calzone.
\newblock {Continuous time boolean modeling for biological signaling:
  Application of Gillespie algorithm.}
\newblock {\em BMC Syst Biol}, 6:116, Aug 2012.

\bibitem{Stoll.Bioinformatics.2017}
Gautier Stoll, Barth{\'e}l{\'e}my Caron, Eric Viara, Aur{\'e}lien Dugourd,
  Andrei Zinovyev, Aur{\'e}lien Naldi, Guido Kroemer, Emmanuel Barillot, and
  Laurence Calzone.
\newblock {MaBoSS 2.0: an environment for stochastic boolean modeling.}
\newblock {\em Bioinformatics}, 33(14):2226--2228, Jul 2017.

\bibitem{Nagornov.Bioinformatics.2020}
Iurii~S Nagornov and Mamoru Kato.
\newblock {tugHall: A simulator of cancer-cell evolution based on the hallmarks
  of cancer and tumor-related genes.}
\newblock {\em Bioinformatics}, 36(11):3597--3599, Jun 2020.

\bibitem{Stoll.FrontMolBiosci.2022}
Gautier Stoll, Aur{\'e}lien Naldi, Vincent No{\"e}l, Eric Viara, Emmanuel
  Barillot, Guido Kroemer, Denis Thieffry, and Laurence Calzone.
\newblock {UPMaBoSS: A novel framework for dynamic cell population modeling.}
\newblock {\em Front Mol Biosci}, 9:800152, Mar 2022.

\bibitem{Hoehme.Bioinformatics.2010}
Stefan Hoehme and Dirk Drasdo.
\newblock A cell-based simulation software for multi-cellular systems.
\newblock {\em Bioinformatics}, 26(20):2641--2642, Oct 2010.

\bibitem{Sutterlin.Bioinformatics.2013}
Thomas S{\"u}tterlin, Christoph Kolb, Hartmut Dickhaus, Dirk J{\"a}ger, and
  Niels Grabe.
\newblock {Bridging the scales: Semantic integration of quantitative SBML in
  graphical multi-cellular models and simulations with EPISIM and COPASI.}
\newblock {\em Bioinformatics}, 29(2):223--229, Jan 2013.

\bibitem{Mirams.PLoSComputBiol.2013}
Gary~R Mirams, Christopher~J Arthurs, Miguel~O Bernabeu, Rafel Bordas, Jonathan
  Cooper, Alberto Corrias, Yohan Davit, Sara-Jane Dunn, Alexander~G Fletcher,
  Daniel~G Harvey, Megan~E Marsh, James~M Osborne, Pras Pathmanathan, Joe
  Pitt-Francis, James Southern, Nejib Zemzemi, and David~J Gavaghan.
\newblock Chaste: an open source c++ library for computational physiology and
  biology.
\newblock {\em PLoS Comput Biol}, 9(3):e1002970, Mar 2013.

\bibitem{Kang.Bioinformatics.2014}
Seunghwa Kang, Simon Kahan, Jason McDermott, Nicholas Flann, and Ilya
  Shmulevich.
\newblock {Biocellion: Accelerating computer simulation of multicellular
  biological system models.}
\newblock {\em Bioinformatics}, 30(21):3101--3108, Nov 2014.

\bibitem{Ghaffarizadeh.PLoSComputBiol.2018}
Ahmadreza Ghaffarizadeh, Randy Heiland, Samuel~H Friedman, Shannon~M
  Mumenthaler, and Paul Macklin.
\newblock {PhysiCell: An open source physics-based cell simulator for 3-D
  multicellular systems.}
\newblock {\em PLoS Comput Biol}, 14(2):e1005991, Feb 2018.

\bibitem{Letort.Bioinformatics.2019}
Gaelle Letort, Arnau Montagud, Gautier Stoll, Randy Heiland, Emmanuel Barillot,
  Paul Macklin, Andrei Zinovyev, and Laurence Calzone.
\newblock {PhysiBoSS: a multi-scale agent-based modelling framework integrating
  physical dimension and cell signalling.}
\newblock {\em Bioinformatics}, 35(7):1188--1196, Apr 2019.

\bibitem{Ponce-de-Leon.NPJSystBiolAppl.2023}
Miguel Ponce-de Leon, Arnau Montagud, Vincent No{\"e}l, Annika Meert, Gerard
  Pradas, Emmanuel Barillot, Laurence Calzone, and Alfonso Valencia.
\newblock {PhysiBoSS 2.0: A sustainable integration of stochastic boolean and
  agent-based modelling frameworks.}
\newblock {\em NPJ Syst Biol Appl}, 9(1):54, Oct 2023.

\bibitem{Alamoudi.Bioinformatics.2023}
Emad Alamoudi, Yannik Sch{\"a}lte, Robert M{\"u}ller, J{\"o}rn Starru{\ss},
  Nils Bundgaard, Frederik Graw, Lutz Brusch, and Jan Hasenauer.
\newblock {FitMultiCell: simulating and parameterizing computational models of
  multi-scale and multi-cellular processes.}
\newblock {\em Bioinformatics}, 39(11), Nov 2023.

\bibitem{Ghaffarizadeh.Bioinformatics.2016}
Ahmadreza Ghaffarizadeh, Samuel~H Friedman, and Paul Macklin.
\newblock {BioFVM: An efficient, parallelized diffusive transport solver for
  3-D biological simulations.}
\newblock {\em Bioinformatics}, 32(8):1256--1258, Apr 2016.

\bibitem{Bravo.PLoSComputBiol.2020}
Rafael~R Bravo, Etienne Baratchart, Jeffrey West, Ryan~O Schenck, Anna~K
  Miller, Jill Gallaher, Chandler~D Gatenbee, David Basanta, Mark
  Robertson-Tessi, and Alexander R~A Anderson.
\newblock Hybrid automata library: A flexible platform for hybrid modeling with
  real-time visualization.
\newblock {\em PLoS Comput Biol}, 16(3):e1007635, Mar 2020.

\bibitem{Ozik.BMCBioinformatics.2018}
Jonathan Ozik, Nicholson Collier, Justin~M Wozniak, Charles Macal, Chase
  Cockrell, Samuel~H Friedman, Ahmadreza Ghaffarizadeh, Randy Heiland, Gary An,
  and Paul Macklin.
\newblock {High-throughput cancer hypothesis testing with an integrated
  PhysiCell-EMEWS workflow.}
\newblock {\em BMC Bioinformatics}, 19(Suppl 18):483, Dec 2018.

\bibitem{Angaroni.BMCBioinformatics.2022}
Fabrizio Angaroni, Alessandro Guidi, Gianluca Ascolani, Alberto d'Onofrio,
  Marco Antoniotti, and Alex Graudenzi.
\newblock {J-SPACE: A Julia package for the simulation of spatial models of
  cancer evolution and of sequencing experiments.}
\newblock {\em BMC Bioinformatics}, 23(1):269, Jul 2022.

\bibitem{Streck.Bioinformatics.2023}
Adam Streck, Tom~L Kaufmann, and Roland~F Schwarz.
\newblock {SMITH: Spatially constrained stochastic model for simulation of
  intra-tumour heterogeneity.}
\newblock {\em Bioinformatics}, 39(3), Mar 2023.

\bibitem{Anderson.Cell.2006}
Alexander R~A Anderson, Alissa~M Weaver, Peter~T Cummings, and Vito Quaranta.
\newblock Tumor morphology and phenotypic evolution driven by selective
  pressure from the microenvironment.
\newblock {\em Cell}, 127(5):905--915, Dec 2006.

\bibitem{Sun.BMCBioinformatics.2012}
Xiaoqiang Sun, Le~Zhang, Hua Tan, Jiguang Bao, Costas Strouthos, and Xiaobo
  Zhou.
\newblock {Multi-scale agent-based brain cancer modeling and prediction of TKI
  treatment response: Incorporating EGFR signaling pathway and angiogenesis.}
\newblock {\em BMC Bioinformatics}, 13:218, Aug 2012.

\bibitem{Liang.BMCBioinformatics.2019}
Weishan Liang, Yongjiang Zheng, Ji~Zhang, and Xiaoqiang Sun.
\newblock Multiscale modeling reveals angiogenesis-induced drug resistance in
  brain tumors and predicts a synergistic drug combination targeting {EGFR and
  VEGFR} pathways.
\newblock {\em BMC Bioinformatics}, 20(Suppl 7):203, May 2019.

\bibitem{Gong.JRSocInterface.2017}
Chang Gong, Oleg Milberg, Bing Wang, Paolo Vicini, Rajesh Narwal, Lorin Roskos,
  and Aleksander~S Popel.
\newblock {A computational multiscale agent-based model for simulating
  spatio-temporal tumour immune response to PD1 and PDL1 inhibition.}
\newblock {\em J R Soc Interface}, 14(134), Sep 2017.

\bibitem{Jenner.PLoSComputBiol.2023}
Adrianne~L Jenner, Wayne Kelly, Michael Dallaston, Robyn Araujo, Isobelle
  Parfitt, Dominic Steinitz, Pantea Pooladvand, Peter~S Kim, Samantha~J Wade,
  and Kara~L Vine.
\newblock Examining the efficacy of localised gemcitabine therapy for the
  treatment of pancreatic cancer using a hybrid agent-based model.
\newblock {\em PLoS Comput Biol}, 19(1):e1010104, Jan 2023.

\bibitem{Genderen.NPJSystBiolAppl.2024}
Maisa N~G van Genderen, Jeroen Kneppers, Anniek Zaalberg, Elise~M Bekers,
  Andries~M Bergman, Wilbert Zwart, and Federica Eduati.
\newblock Agent-based modeling of the prostate tumor microenvironment uncovers
  spatial tumor growth constraints and immunomodulatory properties.
\newblock {\em NPJ Syst Biol Appl}, 10(1):20, Feb 2024.

\bibitem{Cess.PLoSComputBiol.2020}
Colin~G Cess and Stacey~D Finley.
\newblock Multi-scale modeling of macrophage-{T} cell interactions within the
  tumor microenvironment.
\newblock {\em PLoS Comput Biol}, 16(12):e1008519, Dec 2020.

\bibitem{Bull.PLoSComputBiol.2023}
Joshua~A Bull and Helen~M Byrne.
\newblock Quantification of spatial and phenotypic heterogeneity in an
  agent-based model of tumour-macrophage interactions.
\newblock {\em PLoS Comput Biol}, 19(3):e1010994, Mar 2023.

\bibitem{Hickey.CellSyst.2024}
John~W Hickey, Eran Agmon, Nina Horowitz, Tze-Kai Tan, Matthew Lamore, John~B
  Sunwoo, Markus~W Covert, and Garry~P Nolan.
\newblock Integrating multiplexed imaging and multiscale modeling identifies
  tumor phenotype conversion as a critical component of therapeutic {T} cell
  efficacy.
\newblock {\em Cell Syst}, 15(4):322--338, Apr 2024.

\bibitem{Issa:2021aa}
Naiem~T Issa, Vasileios Stathias, Stephan Sch{\"u}rer, and Sivanesan
  Dakshanamurthy.
\newblock {Machine and deep learning approaches for cancer drug repurposing.}
\newblock {\em Semin Cancer Biol}, 68:132--142, Jan 2021.

\bibitem{Stephan:2024aa}
Simon Stephan, St{\'e}phane Galland, Ouassila Labbani~Narsis, Kenji Shoji,
  S{\'e}bastien Vachenc, St{\'e}phane Gerart, and Christophe Nicolle.
\newblock {Agent-based approaches for biological modeling in oncology: A
  literature review.}
\newblock {\em Artif Intell Med}, 152:102884, Jun 2024.

\bibitem{Moore.JTheorBiol.2004}
Helen Moore and Natasha~K Li.
\newblock A mathematical model for chronic myelogenous leukemia {(CML) and T}
  cell interaction.
\newblock {\em J Theor Biol}, 227(4):513--523, Apr 2004.

\bibitem{Kogan.SIAMJApplMath.2010}
Yuri Kogan, Urszula Fory{\'s}, Ofir Shukron, Natalie Kronik, and Zvia Agur.
\newblock {Cellular immunotherapy for high grade gliomas: Mathematical analysis
  deriving efficacious infusion rates based on patient requirements.}
\newblock {\em SIAM J Appl Math}, 70(6):1953--1976, Jan 2010.

\bibitem{Bunimovich-Mendrazitsky.BullMathBiol.2007}
Svetlana Bunimovich-Mendrazitsky, Eliezer Shochat, and Lewi Stone.
\newblock Mathematical model of {BCG} immunotherapy in superficial bladder
  cancer.
\newblock {\em Bull Math Biol}, 69(6):1847--1870, Aug 2007.

\bibitem{Bunimovich-Mendrazitsky.JTheorBiol.2011}
Svetlana Bunimovich-Mendrazitsky, Jean Claude~Gluckman, and Joel Chaskalovic.
\newblock A mathematical model of combined bacillus {Calmette-Guerin (BCG)} and
  interleukin {(IL)-2} immunotherapy of superficial bladder cancer.
\newblock {\em J Theor Biol}, 277(1):27--40, May 2011.

\bibitem{Okuneye.ComputSystOncol.2021}
Kamaldeen Okuneye, Daniel Bergman, Jeffrey~C Bloodworth, Alexander~T Pearson,
  Randy~F Sweis, and Trachette~L Jackson.
\newblock A validated mathematical model of {FGFR3-mediated} tumor growth
  reveals pathways to harness the benefits of combination targeted therapy and
  immunotherapy in bladder cancer.
\newblock {\em Comput Syst Oncol}, 1(2):e1019, Jun 2021.

\bibitem{Li.BullMathBiol.2024}
Chenghang Li, Zonghang Ren, Guiyu Yang, and Jinzhi Lei.
\newblock Mathematical modeling of tumor immune interactions: {The role of
  Anti-FGFR and Anti-PD-1 in the combination therapy}.
\newblock {\em Bull Math Biol}, 86(9):116, Aug 2024.

\bibitem{Ramaj.MathBiosci.2023}
Tedi Ramaj and Xingfu Zou.
\newblock On the treatment of melanoma: {A} mathematical model of oncolytic
  virotherapy.
\newblock {\em Math Biosci}, 365:109073, Sep 2023.

\bibitem{Valle.ApplMathModel.2021}
Paul~A Valle, Luis~N Coria, and Karla~D Carballo.
\newblock {Chemoimmunotherapy for the treatment of prostate cancer: Insights
  from mathematical modelling.}
\newblock {\em Appl Math Model}, 90:682--702, Feb 2021.

\bibitem{Kogan.CancerRes.2012}
Yuri Kogan, Karin Halevi-Tobias, Moran Elishmereni, Stanimir Vuk-Pavlovi{\'c},
  and Zvia Agur.
\newblock Reconsidering the paradigm of cancer immunotherapy by computationally
  aided real-time personalization.
\newblock {\em Cancer Res}, 72(9):2218--2227, May 2012.

\bibitem{Ji.PLoSComputBiol.2019}
Zhiwei Ji, Weiling Zhao, Hui-Kuan Lin, and Xiaobo Zhou.
\newblock Systematically understanding the immunity leading to {CRPC}
  progression.
\newblock {\em PLoS Comput Biol}, 15(9):e1007344, Sep 2019.

\bibitem{Bitsouni.JTheorBiol.2022}
Vasiliki Bitsouni and Vasilis Tsilidis.
\newblock {Mathematical modeling of tumor-immune system interactions: The
  effect of rituximab on breast cancer immune response.}
\newblock {\em J Theor Biol}, 539:111001, Apr 2022.

\bibitem{Nazari.PLoSComputBiol.2018}
Fereshteh Nazari, Alexander~T Pearson, Jacques~Eduardo N{\"o}r, and Trachette~L
  Jackson.
\newblock A mathematical model for {IL-6-mediated}, stem cell driven tumor
  growth and targeted treatment.
\newblock {\em PLoS Comput Biol}, 14(1):e1005920, Jan 2018.

\bibitem{Pang.ApplMathModel.2021}
Liuyong Pang, Sanhong Liu, Fang Liu, Xinan Zhang, and Tianhai Tian.
\newblock Mathematical modeling and analysis of tumor-volume variation during
  radiotherapy.
\newblock {\em Appl Math Model}, 89:1074--1089, Jan 2021.

\bibitem{Louzoun.JTheorBiol.2014}
Yoram Louzoun, Chuan Xue, Gregory~B Lesinski, and Avner Friedman.
\newblock A mathematical model for pancreatic cancer growth and treatments.
\newblock {\em J Theor Biol}, 351:74--82, Jun 2014.

\bibitem{Lourenco.JBiolSyst.2023}
Edgard Lourenco~Jr, Diego~S Rodrigues, Maria~E Antunes, Paulo~Fa Mancera, and
  Guilherme Rodrigues.
\newblock A simple mathematical model of non-small cell lung cancer involving
  macrophages and {CD8+ T} cells.
\newblock {\em J Biol Syst}, 31(04):1407--1431, Oct 2023.

\bibitem{Fletcher.JTheorBiol.2012}
Alexander~G Fletcher, Christopher J~W Breward, and S~Jonathan~Chapman.
\newblock Mathematical modeling of monoclonal conversion in the colonic crypt.
\newblock {\em J Theor Biol}, 300:118--133, May 2012.

\bibitem{Sameen.JTheorBiol.2016}
Sheema Sameen, Roberto Barbuti, Paolo Milazzo, Antonio Cerone, Marzia Del~Re,
  and Romano Danesi.
\newblock Mathematical modeling of drug resistance due to {KRAS} mutation in
  colorectal cancer.
\newblock {\em J Theor Biol}, 389:263--273, Jan 2016.

\bibitem{Lo.JTheorBiol.2013}
Wing-Cheong Lo, Edward W~Jr Martin, Charles~L Hitchcock, and Avner Friedman.
\newblock Mathematical model of colitis-associated colon cancer.
\newblock {\em J Theor Biol}, 317:20--29, Jan 2013.

\bibitem{Mohammad-Mirzaei.iScience.2023}
Navid Mohammad~Mirzaei, Wenrui Hao, and Leili Shahriyari.
\newblock Investigating the spatial interaction of immune cells in colon
  cancer.
\newblock {\em iScience}, 26(5):106596, May 2023.

\bibitem{Koenders.JTheorBiol.2016}
M~A Koenders and R~Saso.
\newblock A mathematical model of cell equilibrium and joint cell formation in
  multiple myeloma.
\newblock {\em J Theor Biol}, 390:73--79, Feb 2016.

\bibitem{Gallaher.JTheorBiol.2018}
Jill Gallaher, Kamila Larripa, Marissa Renardy, Blerta Shtylla, Nessy Tania,
  Diana White, Karen Wood, Li~Zhu, Chaitali Passey, Michael Robbins, Natalie
  Bezman, Suresh Shelat, Hearn Jay~Cho, and Helen Moore.
\newblock Methods for determining key components in a mathematical model for
  tumor-immune dynamics in multiple myeloma.
\newblock {\em J Theor Biol}, 458:31--46, Dec 2018.

\bibitem{Bouchnita.JTheorBiol.2024}
Anass Bouchnita and Vitaly Volpert.
\newblock Phenotype-structured model of intra-clonal heterogeneity and drug
  resistance in multiple myeloma.
\newblock {\em J Theor Biol}, 576:111652, Jan 2024.

\bibitem{Da.JBiolSyst.2020}
Jairo~Gomes Da~Silva, Rafael~Martins De~Morais, Izabel Cristina~Rodrigues
  Da~Silva, Mostafa Adimy, and Paulo~Fernando De~Arruda~Mancera.
\newblock A mathematical model for treatment of papillary thyroid cancer using
  the {Allee} effect.
\newblock {\em J Biol Syst}, 28(03):701--718, Feb 2020.

\bibitem{Delitala.JTheorBiol.2012}
Marcello Delitala and Tommaso Lorenzi.
\newblock A mathematical model for the dynamics of cancer hepatocytes under
  therapeutic actions.
\newblock {\em J Theor Biol}, 297:88--102, Mar 2012.

\bibitem{Barazzuol.JTheorBiol.2010}
Lara Barazzuol, Neil~G Burnet, Raj Jena, Bleddyn Jones, Sarah~J Jefferies, and
  Norman~F Kirkby.
\newblock A mathematical model of brain tumour response to radiotherapy and
  chemotherapy considering radiobiological aspects.
\newblock {\em J Theor Biol}, 262(3):553--565, Feb 2010.

\bibitem{Serre.CancerRes.2016}
Raphael Serre, Sebastien Benzekry, Laetitia Padovani, Christophe Meille,
  Nicolas Andr{\'e}, Joseph Ciccolini, Fabrice Barlesi, Xavier Muracciole, and
  Dominique Barbolosi.
\newblock Mathematical modeling of cancer immunotherapy and its synergy with
  radiotherapy.
\newblock {\em Cancer Res}, 76(17):4931--4940, Sep 2016.

\bibitem{Labrie:2022aa}
Marilyne Labrie, Joan~S Brugge, Gordon~B Mills, and Ioannis~K Zervantonakis.
\newblock {Therapy resistance: opportunities created by adaptive responses to
  targeted therapies in cancer.}
\newblock {\em Nat Rev Cancer}, 22(6):323--339, Jun 2022.

\bibitem{Waarts:2022aa}
Michael~R Waarts, Aaron~J Stonestrom, Young~C Park, and Ross~L Levine.
\newblock {Targeting mutations in cancer.}
\newblock {\em J Clin Invest}, 132(8), Apr 2022.

\bibitem{Hutchinson.JTheorBiol.2016}
L~G Hutchinson, E~A Gaffney, P~K Maini, J~Wagg, A~Phipps, and H~M Byrne.
\newblock {Vascular phenotype identification and anti-angiogenic treatment
  recommendation: A pseudo-multiscale mathematical model of angiogenesis.}
\newblock {\em J Theor Biol}, 398:162--180, Jun 2016.

\bibitem{He.BullMathBiol.2018}
Yixuan He, Anita Kodali, and Dorothy~I Wallace.
\newblock Predictive modeling of neuroblastoma growth dynamics in xenograft
  model after bevacizumab {anti-VEGF} therapy.
\newblock {\em Bull Math Biol}, 80(8):2026--2048, Aug 2018.

\bibitem{Zheng.JMathBiol.2018}
Xiaoming Zheng and Mohye Sweidan.
\newblock {A mathematical model of angiogenesis and tumor growth: Analysis and
  application in anti-angiogenesis therapy.}
\newblock {\em J Math Biol}, 77(5):1589--1622, Nov 2018.

\bibitem{Druker:2001aa}
B~J Druker, M~Talpaz, D~J Resta, B~Peng, E~Buchdunger, J~M Ford, N~B Lydon,
  H~Kantarjian, R~Capdeville, S~Ohno-Jones, and C~L Sawyers.
\newblock {Efficacy and safety of a specific inhibitor of the BCR-ABL tyrosine
  kinase in chronic myeloid leukemia.}
\newblock {\em N Engl J Med}, 344(14):1031--1037, Apr 2001.

\bibitem{Breccia:2010aa}
Massimo Breccia and Giuliana Alimena.
\newblock {Nilotinib: a second-generation tyrosine kinase inhibitor for chronic
  myeloid leukemia.}
\newblock {\em Leuk Res}, 34(2):129--134, Feb 2010.

\bibitem{Johnson:2022aa}
Melissa Johnson, Marina~Chiara Garassino, Tony Mok, and Tetsuya Mitsudomi.
\newblock {Treatment strategies and outcomes for patients with EGFR-mutant
  non-small cell lung cancer resistant to EGFR tyrosine kinase inhibitors:
  Focus on novel therapies.}
\newblock {\em Lung Cancer}, 170:41--51, Aug 2022.

\bibitem{Rodriguez:2023aa}
Jonathan Rodriguez, Abdon Iniguez, Nilamani Jena, Prasanthi Tata, Zhong-Ying
  Liu, Arthur~D Lander, John Lowengrub, and Richard~A Van~Etten.
\newblock {Predictive nonlinear modeling of malignant myelopoiesis and tyrosine
  kinase inhibitor therapy.}
\newblock {\em eLife}, 12:e84149, Apr 2023.

\bibitem{Bouchnita:2020aa}
Anass Bouchnita, Vitaly Volpert, Mark~J Koury, and Andreas Hellander.
\newblock {A multiscale model to design therapeutic strategies that overcome
  drug resistance to tyrosine kinase inhibitors in multiple myeloma.}
\newblock {\em Math Biosci}, 319:108293, Jan 2020.

\bibitem{Perini:2018aa}
Guilherme~Fleury Perini, Glaciano~Nogueira Ribeiro, Jorge~Vaz Pinto~Neto,
  Laura~Tojeiro Campos, and Nelson Hamerschlak.
\newblock {BCL-2 as therapeutic target for hematological malignancies.}
\newblock {\em J Hematol Oncol}, 11(1):65, May 2018.

\bibitem{Lee:2021aa}
Jonggul Lee, Donggu Lee, and Yangjin Kim.
\newblock {Mathematical model of STAT signalling pathways in cancer development
  and optimal control approaches.}
\newblock {\em R Soc Open Sci}, 8(9):210594, Sep 2021.

\bibitem{Ballesta:2013aa}
Annabelle Ballesta, Jonathan Lopez, Nikolay Popgeorgiev, Philippe Gonzalo,
  Marie Doumic, and Germain Gillet.
\newblock {Data-driven modeling of SRC control on the mitochondrial pathway of
  apoptosis: Implication for anticancer therapy optimization.}
\newblock {\em PLoS Comput Biol}, 9(4):e1003011, Apr 2013.

\bibitem{Mukherjee:2024aa}
Arnab Mukherjee, Suzanna Abraham, Akshita Singh, S~Balaji, and K~S Mukunthan.
\newblock {From Data to Cure: A Comprehensive Exploration of Multi-omics Data
  Analysis for Targeted Therapies.}
\newblock {\em Mol Biotechnol}, Apr 2024.

\bibitem{Galluzzi:2020aa}
Lorenzo Galluzzi, Juliette Humeau, Aitziber Buqu{\'e}, Laurence Zitvogel, and
  Guido Kroemer.
\newblock {Immunostimulation with chemotherapy in the era of immune checkpoint
  inhibitors.}
\newblock {\em Nat Rev Clin Oncol}, 17(12):725--741, Dec 2020.

\bibitem{Zouein:2022aa}
Joseph Zouein, Fady~G Haddad, Roland Eid, and Hampig~R Kourie.
\newblock {The combination of immune checkpoint inhibitors and chemotherapy in
  advanced non-small-cell lung cancer: the rational choice.}
\newblock {\em Immunotherapy}, 14(2):155--167, Feb 2022.

\bibitem{Liao:2024aa}
Kang-Ling Liao, Adam~J Wieler, and Pedro M~Lopez Gascon.
\newblock {Mathematical modeling and analysis of cancer treatment with
  radiation and anti-PD-L1.}
\newblock {\em Math Biosci}, 374:109218, Aug 2024.

\bibitem{Lee:2020aa}
Won~Suk Lee, Hannah Yang, Hong~Jae Chon, and Chan Kim.
\newblock {Combination of anti-angiogenic therapy and immune checkpoint
  blockade normalizes vascular-immune crosstalk to potentiate cancer immunity.}
\newblock {\em Exp Mol Med}, 52(9):1475--1485, Sep 2020.

\bibitem{Li2024.04.30.591845}
Chenghang Li, Yongchang Wei, and Jinzhi Lei.
\newblock {Quantitative cancer-immunity cycle modeling for predicting disease
  progression in advanced metastatic colorectal cancer}.
\newblock {\em bioRxiv}, page doi: https://doi.org/10.1101/2024.04.30.591845,
  2024.

\bibitem{Yang.MathComputSimulat.2015}
Jin Yang, Sanyi Tang, and Robert~A Cheke.
\newblock Modelling pulsed immunotherapy of tumour--immune interaction.
\newblock {\em Math Comput Simulat}, 109:92--112, Mar 2015.

\bibitem{June:2018aa}
Carl~H June, Roddy~S O'Connor, Omkar~U Kawalekar, Saba Ghassemi, and Michael~C
  Milone.
\newblock {CAR T cell immunotherapy for human cancer.}
\newblock {\em Science}, 359(6382):1361--1365, Mar 2018.

\bibitem{Owens.BullMathBiol.2021}
Katherine Owens and Ivana Bozic.
\newblock {Modeling {CAR T}-Cell Therapy with Patient Preconditioning.}
\newblock {\em Bull Math Biol}, 83(5):42, Mar 2021.

\bibitem{Adhikarla:2024aa}
Vikram Adhikarla, Dennis Awuah, Enrico Caserta, Megan Minnix, Maxim Kuznetsov,
  Amrita Krishnan, Jefferey Y~C Wong, John~E Shively, Xiuli Wang, Flavia
  Pichiorri, and Russell~C Rockne.
\newblock {Designing combination therapies for cancer treatment: Application of
  a mathematical framework combining CAR T-cell immunotherapy and targeted
  radionuclide therapy.}
\newblock {\em Front Immunol}, 15:1358478, 2024.

\bibitem{Kara:2024aa}
Erdi Kara, Trachette~L Jackson, Chartese Jones, Rockford Sison, and Reginald~L
  McGee~Ii.
\newblock {Mathematical modeling insights into improving CAR T cell therapy for
  solid tumors with bystander effects.}
\newblock {\em NPJ Syst Biol Appl}, 10(1):105, Sep 2024.

\bibitem{Giorgadze:2022aa}
Tina Giorgadze, Henning Fischel, Ansel Tessier, and Kerri-Ann Norton.
\newblock {Investigating Two Modes of Cancer-Associated Antigen Heterogeneity
  in an Agent-Based Model of Chimeric Antigen Receptor T-Cell Therapy.}
\newblock {\em Cells}, 11(19), Oct 2022.

\bibitem{Sahoo:2020aa}
Prativa Sahoo, Xin Yang, Daniel Abler, Davide Maestrini, Vikram Adhikarla,
  David Frankhouser, Heyrim Cho, Vanessa Machuca, Dongrui Wang, Michael Barish,
  Margarita Gutova, Sergio Branciamore, Christine~E Brown, and Russell~C
  Rockne.
\newblock {Mathematical deconvolution of CAR T-cell proliferation and
  exhaustion from real-time killing assay data.}
\newblock {\em J R Soc Interface}, 17(162):20190734, Jan 2020.

\bibitem{Prybutok:2022aa}
Alexis~N Prybutok, Jessica~S Yu, Joshua~N Leonard, and Neda Bagheri.
\newblock {Mapping CAR T-Cell Design Space Using Agent-Based Models.}
\newblock {\em Front Mol Biosci}, 9:849363, 2022.

\bibitem{Kirouac.NatBiotechnol.2023}
Daniel~C Kirouac, Cole Zmurchok, Avisek Deyati, Jordan Sicherman, Chris Bond,
  and Peter~W Zandstra.
\newblock Deconvolution of clinical variance in {CAR-T} cell pharmacology and
  response.
\newblock {\em Nat Biotechnol}, 41(11):1606--1617, Nov 2023.

\bibitem{Ratajczyk.MathBiosciEng.2017}
Elzbieta Ratajczyk, Urszula Ledzewicz, Maciej Leszczynski, and Avner Friedman.
\newblock {The role of TNF-$\alpha$ inhibitor in glioma virotherapy: A
  mathematical model.}
\newblock {\em Math Biosci Eng}, 14(1):305--319, Feb 2017.

\bibitem{Marino.JTheorBiol.2008}
Simeone Marino, Ian~B Hogue, Christian~J Ray, and Denise~E Kirschner.
\newblock A methodology for performing global uncertainty and sensitivity
  analysis in systems biology.
\newblock {\em J Theor Biol}, 254(1):178--196, Sep 2008.

\bibitem{Lillacci.PLoSComputBiol.2010}
Gabriele Lillacci and Mustafa Khammash.
\newblock Parameter estimation and model selection in computational biology.
\newblock {\em PLoS Comput Biol}, 6(3):e1000696, Mar 2010.

\bibitem{Mitra.NatCommun.2018}
Eshan~D Mitra, Raquel Dias, Richard~G Posner, and William~S Hlavacek.
\newblock Using both qualitative and quantitative data in parameter
  identification for systems biology models.
\newblock {\em Nat Commun}, 9(1):3901, Sep 2018.

\bibitem{Linden.PLoSComputBiol.2022}
Nathaniel~J Linden, Boris Kramer, and Padmini Rangamani.
\newblock Bayesian parameter estimation for dynamical models in systems
  biology.
\newblock {\em PLoS Comput Biol}, 18(10):e1010651, Oct 2022.

\bibitem{Liepe.NatProtoc.2014}
Juliane Liepe, Paul Kirk, Sarah Filippi, Tina Toni, Chris~P Barnes, and Michael
  P~H Stumpf.
\newblock A framework for parameter estimation and model selection from
  experimental data in systems biology using approximate bayesian computation.
\newblock {\em Nat Protoc}, 9(2):439--456, Feb 2014.

\bibitem{Kramer.BMCBioinformatics.2014}
Andrei Kramer, Ben Calderhead, and Nicole Radde.
\newblock {Hamiltonian Monte Carlo methods for efficient parameter estimation
  in steady state dynamical systems.}
\newblock {\em BMC Bioinformatics}, 15(1):253, Jul 2014.

\bibitem{Schmiester.JMathBiol.2020}
Leonard Schmiester, Daniel Weindl, and Jan Hasenauer.
\newblock Parameterization of mechanistic models from qualitative data using an
  efficient optimal scaling approach.
\newblock {\em J Math Biol}, 81(2):603--623, Aug 2020.

\bibitem{Giampiccolo.NPJSystBiolAppl.2024}
Stefano Giampiccolo, Federico Reali, Anna Fochesato, Giovanni Iacca, and Luca
  Marchetti.
\newblock Robust parameter estimation and identifiability analysis with hybrid
  neural ordinary differential equations in computational biology.
\newblock {\em NPJ Syst Biol Appl}, 10(1):139, Nov 2024.

\bibitem{Rodriguez.Bioinformatics.2010}
Maria Rodriguez-Fernandez and Julio~R. Banga.
\newblock {SensSB: a software toolbox for the development and sensitivity
  analysis of systems biology models.}
\newblock {\em Bioinformatics}, 26(13):1675--1676, May 2010.

\bibitem{Glont.Bioinformatics.2020}
Mihai Glont, Chinmay Arankalle, Krishna Tiwari, Tung V~N Nguyen, Henning
  Hermjakob, and Rahuman~S Malik-Sheriff.
\newblock Biomodels parameters: a treasure trove of parameter values from
  published systems biology models.
\newblock {\em Bioinformatics}, 36(17):4649--4654, Nov 2020.

\bibitem{Schalte.Bioinformatics.2023}
Yannik Sch{\"a}lte, Fabian Fr{\"o}hlich, Paul~J Jost, Jakob Vanhoefer, Dilan
  Pathirana, Paul Stapor, Polina Lakrisenko, Dantong Wang, Elba Raim{\'u}ndez,
  Simon Merkt, Leonard Schmiester, Philipp St{\"a}dter, Stephan Grein, Erika
  Dudkin, Domagoj Doresic, Daniel Weindl, and Jan Hasenauer.
\newblock {pyPESTO: A modular and scalable tool for parameter estimation for
  dynamic models.}
\newblock {\em Bioinformatics}, 39(11), Nov 2023.

\bibitem{Masoudi-Nejad.SeminCancerBiol.2015}
Ali Masoudi-Nejad, Gholamreza Bidkhori, Saman Hosseini~Ashtiani, Ali Najafi,
  Joseph~H Bozorgmehr, and Edwin Wang.
\newblock {Cancer systems biology and modeling: Microscopic scale and
  multiscale approaches.}
\newblock {\em Semin Cancer Biol}, 30:60--69, Feb 2015.

\bibitem{Meyer.CellSyst.2021}
Pablo Meyer and Julio Saez-Rodriguez.
\newblock Advances in systems biology modeling: 10 years of crowdsourcing
  {DREAM} challenges.
\newblock {\em Cell Syst}, 12(6):636--653, Jun 2021.

\bibitem{Yue.NPJSystBiolAppl.2022}
Rongting Yue and Abhishek Dutta.
\newblock Computational systems biology in disease modeling and control, review
  and perspectives.
\newblock {\em NPJ Syst Biol Appl}, 8(1):37, Oct 2022.

\bibitem{Chen.MathBiosci.2022}
Yu~Chen and Xiulan Lai.
\newblock Modeling the effect of gut microbiome on therapeutic efficacy of
  immune checkpoint inhibitors against cancer.
\newblock {\em Math Biosci}, 350:108868, Aug 2022.

\bibitem{Heath.ComputSciRev.2009}
Allison~P Heath and Lydia~E Kavraki.
\newblock Computational challenges in systems biology.
\newblock {\em Comput Sci Rev}, 3(1):1--17, Feb 2009.

\bibitem{Walpole.AnnuRevBiomedEng.2013}
Joseph Walpole, Jason~A Papin, and Shayn~M Peirce.
\newblock Multiscale computational models of complex biological systems.
\newblock {\em Annu Rev Biomed Eng}, 15:137--154, Apr 2013.

\bibitem{Fletcher.WIREsMechDis.2022}
Alexander~G Fletcher and James~M Osborne.
\newblock Seven challenges in the multiscale modeling of multicellular tissues.
\newblock {\em WIREs Mech Dis}, 14(1):e1527, Jan 2022.

\bibitem{Cappuccio.BriefBioinform.2016}
Antonio Cappuccio, Paolo Tieri, and Filippo Castiglione.
\newblock {Multiscale modelling in immunology: A review.}
\newblock {\em Brief Bioinform}, 17(3):408--418, May 2016.

\bibitem{Kazerouni.iScience.2020}
Anum~S Kazerouni, Manasa Gadde, Andrea Gardner, David A~2nd Hormuth, Angela~M
  Jarrett, Kaitlyn~E Johnson, Ernesto A B~F Lima, Guillermo Lorenzo, Caleb
  Phillips, Amy Brock, and Thomas~E Yankeelov.
\newblock Integrating quantitative assays with biologically based mathematical
  modeling for predictive oncology.
\newblock {\em iScience}, 23(12):101807, Dec 2020.

\bibitem{Zhao.InformationFusion.2024}
Yuxin Zhao, Xiaobo Li, Changjun Zhou, Hao Pen, Zhonglong Zheng, Jun Chen, and
  Weiping Ding.
\newblock A review of cancer data fusion methods based on deep learning.
\newblock {\em Information Fusion}, 108:102361, Aug 2024.

\bibitem{Boehm.NatRevCancer.2022}
Kevin~M Boehm, Pegah Khosravi, Rami Vanguri, Jianjiong Gao, and Sohrab~P Shah.
\newblock Harnessing multimodal data integration to advance precision oncology.
\newblock {\em Nat Rev Cancer}, 22(2):114--126, Feb 2022.

\bibitem{Lorenzo.AnnuRevBiomedEng.2024}
Guillermo Lorenzo, Syed~Rakin Ahmed, David~A Hormuth~Ii, Brenna Vaughn,
  Jayashree Kalpathy-Cramer, Luis Solorio, Thomas~E Yankeelov, and Hector
  Gomez.
\newblock Patient-specific, mechanistic models of tumor growth incorporating
  artificial intelligence and big data.
\newblock {\em Annu Rev Biomed Eng}, 26:529--560, Apr 2024.

\bibitem{Alber.NPJDigitMed.2019}
Mark Alber, Adrian Buganza~Tepole, William~R Cannon, Suvranu De, Salvador
  Dura-Bernal, Krishna Garikipati, George Karniadakis, William~W Lytton, Paris
  Perdikaris, Linda Petzold, and Ellen Kuhl.
\newblock Integrating machine learning and multiscale modeling-perspectives,
  challenges, and opportunities in the biological, biomedical, and behavioral
  sciences.
\newblock {\em NPJ Digit Med}, 2:115, Nov 2019.

\bibitem{Kozowska.PLoSComputBiol.2020}
Emilia Koz{\l}owska, Rafa{\l} Suwi{\'n}ski, Monika Giglok, Andrzej
  {\'S}wierniak, and Marek Kimmel.
\newblock Mathematical model predicts response to chemotherapy in advanced
  non-resectable non-small cell lung cancer patients treated with
  platinum-based doublet.
\newblock {\em PLoS Comput Biol}, 16(10):e1008234, Oct 2020.

\bibitem{Metzcar.FrontImmunol.2024}
John Metzcar, Catherine~R Jutzeler, Paul Macklin, Alvaro K{\"o}hn-Luque, and
  Sarah~C Br{\"u}ningk.
\newblock {A review of mechanistic learning in mathematical oncology.}
\newblock {\em Front Immunol}, 15:1363144, Mar 2024.

\bibitem{Perez-Lopez.NatRevCancer.2024}
Raquel Perez-Lopez, Narmin Ghaffari~Laleh, Faisal Mahmood, and Jakob~Nikolas
  Kather.
\newblock A guide to artificial intelligence for cancer researchers.
\newblock {\em Nat Rev Cancer}, 24(6):427--441, Jun 2024.

\end{thebibliography}

\end{document}